\documentstyle[12pt,epsfig]{article}

\newcommand{\Preprint}{\vspace*{-1.5cm} \noindent hep-ph/9601202 \hfill 
  FTUV/95-50 \\ \mbox{}\hfill IFIC/95-52 \\  \mbox{}\hfill
  December 1995\vspace*{0.5cm}}
\catcode`\@=11
\long\def\@makefntext#1{
\protect\noindent \hbox to 3.2pt {\hskip-.9pt  
$^{{\ninerm\@thefnmark}}$\hfil}#1\hfill}		

\def\@makefnmark{\hbox to 0pt{$^{\@thefnmark}$\hss}}  
	
\def\ps@myheadings{\let\@mkboth\@gobbletwo
\def\@oddhead{\hbox{}
\rightmark\hfil\ninerm\thepage}   
\def\@oddfoot{}\def\@evenhead{\ninerm\thepage\hfil
\leftmark\hbox{}}\def\@evenfoot{}
\def\sectionmark##1{}\def\subsectionmark##1{}}

\renewcommand{\thefootnote}{\fnsymbol{footnote}}

\newcounter{sectionc}\newcounter{subsectionc}\newcounter{subsubsectionc}
\renewcommand{\section}[1] {\vspace*{0.6cm}  
        \refstepcounter{sectionc}
\setcounter{subsectionc}{0}\setcounter{subsubsectionc}{0}\noindent 
	{\normalsize\bf\thesectionc. #1}\par\vspace*{0.4cm}}
\renewcommand{\subsection}[1] {\vspace*{0.6cm}\addtocounter{subsectionc}{1} 
	\setcounter{subsubsectionc}{0}\noindent 
	{\normalsize\it\thesectionc.\thesubsectionc. #1}\par\vspace*{0.4cm}}
\renewcommand{\subsubsection}[1]
{\vspace*{0.6cm}\addtocounter{subsubsectionc}{1}
	\noindent {\normalsize\rm\thesectionc.\thesubsectionc.\thesubsubsectionc. 
	#1}\par\vspace*{0.4cm}}

\newcounter{appendixc}
\newcounter{subappendixc}[appendixc]
\newcounter{subsubappendixc}[subappendixc]

\renewcommand{\appendix}[1] {\vspace*{0.6cm}
        \refstepcounter{appendixc}
        \setcounter{figure}{0}
        \setcounter{table}{0}
        \setcounter{equation}{0}
        \renewcommand{\thefigure}{\Alph{appendixc}.\arabic{figure}}
        \renewcommand{\thetable}{\Alph{appendixc}.\arabic{table}}
        \renewcommand{\theappendixc}{\Alph{appendixc}}
        \renewcommand{\theequation}{\Alph{appendixc}.\arabic{equation}}
        \noindent{\bf Appendix \theappendixc #1}\par\vspace*{0.4cm}}

\def\abstracts#1{{
	\centering{\begin{minipage}{12.2truecm}\footnotesize\baselineskip=12pt\noindent
	\centerline{\footnotesize ABSTRACT}\vspace*{0.3cm}
	\parindent=0pt #1
	\end{minipage}}\par}} 

\renewenvironment{thebibliography}[1]
	{\begin{list}{\arabic{enumi}.}
	{\usecounter{enumi}\setlength{\parsep}{0pt}
\setlength{\leftmargin 1.25cm}{\rightmargin 0pt}
	 \setlength{\itemsep}{0pt} \settowidth
	{\labelwidth}{#1.}\sloppy}}{\end{list}}

\newcounter{itemlistc}
\newcounter{romanlistc}
\newcounter{alphlistc}
\newcounter{arabiclistc}

\newcommand{\fcaption}[1]{
        \refstepcounter{figure}
        \setbox\@tempboxa = \hbox{\footnotesize Fig.~\thefigure. #1}
        \ifdim \wd\@tempboxa > 6in
           {\begin{center}
        \parbox{6in}{\footnotesize\baselineskip=12pt Fig.~\thefigure. #1}
            \end{center}}
        \else
             {\begin{center}
             {\footnotesize Fig.~\thefigure. #1}
              \end{center}}
        \fi}

\newcommand{\tcaption}[1]{
        \refstepcounter{table}
        \setbox\@tempboxa = \hbox{\footnotesize Table~\thetable. #1}
        \ifdim \wd\@tempboxa > 6in
           {\begin{center}
        \parbox{6in}{\footnotesize\baselineskip=12pt Table~\thetable. #1}
            \end{center}}
        \else
             {\begin{center}
             {\footnotesize Table~\thetable. #1}
              \end{center}}
        \fi}

\def\@cite#1#2{\unskip\nobreak\relax
    \def\@tempa{$\m@th^{\hbox{\the\scriptfont0 #1}}$}%
    \futurelet\@tempc\@citexx}
\def\@citexx{\ifx.\@tempc\let\@tempd=\@citepunct\else
    \ifx,\@tempc\let\@tempd=\@citepunct\else
    \let\@tempd=\@tempa\fi\fi\@tempd}
\def\@citepunct{\@tempc\edef\@sf{\spacefactor=\the\spacefactor\relax}\@tempa
    \@sf\@gobble}

\def\citenum#1{{\def\@cite##1##2{##1}\cite{#1}}}
\def\citea#1{\@cite{#1}{}}

\newcount\@tempcntc
\def\@citex[#1]#2{\if@filesw\immediate\write\@auxout{\string\citation{#2}}\fi
  \@tempcnta\z@\@tempcntb\m@ne\def\@citea{}\@cite{\@for\@citeb:=#2\do
    {\@ifundefined
       {b@\@citeb}{\@citeo\@tempcntb\m@ne\@citea\def\@citea{,}{\bf ?}\@warning
       {Citation `\@citeb' on page \thepage \space undefined}}%
    {\setbox\z@\hbox{\global\@tempcntc0\csname b@\@citeb\endcsname\relax}%
     \ifnum\@tempcntc=\z@ \@citeo\@tempcntb\m@ne
       \@citea\def\@citea{,}\hbox{\csname b@\@citeb\endcsname}%
     \else
      \advance\@tempcntb\@ne
      \ifnum\@tempcntb=\@tempcntc
      \else\advance\@tempcntb\m@ne\@citeo
      \@tempcnta\@tempcntc\@tempcntb\@tempcntc\fi\fi}}\@citeo}{#1}}
\def\@citeo{\ifnum\@tempcnta>\@tempcntb\else\@citea\def\@citea{,}%
  \ifnum\@tempcnta=\@tempcntb\the\@tempcnta\else
   {\advance\@tempcnta\@ne\ifnum\@tempcnta=\@tempcntb \else \def\@citea{--}\fi
    \advance\@tempcnta\m@ne\the\@tempcnta\@citea\the\@tempcntb}\fi\fi}

 1
 1
 1

\font\ninerm=cmr9

\renewcommand{\theequation}{\arabic{sectionc}.\arabic{equation}}
\def\refjl#1#2#3#4#5#6{\bibitem{#1} #2, {\it #3} {\bf #4} (#5) #6.}
\def\refbk#1#2#3#4{\bibitem{#1} #2, {\it #3}, #4.}
\def\etal{{\it et al}\ }
\def\NP{Nucl. Phys.}
\def\NPPS{Nucl. Phys. B (Proc. Suppl.)}
\def\PL{Phys. Lett.}
\def\PRL{Phys. Rev. Lett.}
\def\PR{Phys. Rev.}
\def\PRep{Phys. Rep.}
\def\ZP{Z. Phys.}

\def\APNY{Ann. Phys., NY}
\def\NC{Nuovo Cimento}
\def\RMP{Rev. Mod. Phys.}
\def\RPP{Rep. Prog. Phys.}
\def\ARNPS{Ann. Rev. Nucl. Part. Sci.}
\def\PTP{Progr. Theor. Phys.}
\def\SJNP{Sov. J. Nucl. Phys.}
\def\PPNP{Prog. Part. Nucl. Phys.}
\def\slashchar#1{\setbox0=\hbox{$#1$}\dimen0=\wd0%
\setbox1=\hbox{/}\dimen1=\wd1%
\ifdim\dimen0>\dimen1%
\rlap{\hbox to
\dimen0{\hfil/\hfil}}#1\else                                        
\rlap{\hbox to \dimen1{\hfil$#1$\hfil}}/\fi}
\newcommand{\eqn}[1]{(\ref{#1})}
\newcommand{\be}{\begin{equation}}
\newcommand{\ee}{\end{equation}}
\newcommand{\no}{\nonumber}
\newcommand{\bel}[1]{\be\label{#1}}
\newcommand{\ba}{\begin{array}{c}}
\newcommand{\bat}{\begin{array}{cc}}
\newcommand{\ea}{\end{array}}
\newcommand{\beqn}{\begin{eqnarray}}
\newcommand{\eeqn}{\end{eqnarray}}

\newcommand{\bi}{\begin{itemize}}
\newcommand{\ei}{\end{itemize}}

\newcommand{\rms}{\rm\scriptsize}
\newcommand{\rmf}{\rm\footnotesize}

\def\BBB{\stackrel{\scriptscriptstyle
    {(-)\hphantom{\scriptstyle 0}}}{B^0}}
\def\ffb{\stackrel{\scriptscriptstyle (-)}{f}}

\newcommand{\toU}{\stackrel{\mbox{\rms U(1)}}{\longrightarrow}}
\newcommand{\toG}{\stackrel{\mbox{\rms G}}{\longrightarrow}}
\newcommand{\ssb}{\stackrel{\mbox{\rms SSB}}{\longrightarrow}}

\newcommand{\toLow}{\stackrel{q^2 \ll M_W^2}{\,\longrightarrow\,}}

\newcommand{\e}{\mbox{\rm e}}

\newcommand{\bV}{\mbox{\boldmath $V$}}
\newcommand{\bM}{\mbox{\boldmath $M$}}

\newcommand{\cL}{{\cal L}}

\newcommand{\cM}{{\cal M}}
\newcommand{\cO}{{\cal O}}
\newcommand{\cP}{{\cal P}}
\newcommand{\cQ}{{\cal Q}}
\newcommand{\cH}{{\cal H}}

\newcommand{\cI}{{\cal I}}
\newcommand{\cF}{{\cal F}}
\newcommand{\cC}{{\cal C}}
\newcommand{\cJ}{{\cal J}}
\begin{document}
\Preprint

\centerline{\normalsize\bf FLAVOURDYNAMICS\footnote{\bf
Lectures given at the XXIII International Meeting on Fundamental Physics,
Comillas (Spain) 22--26 May 1995}}
\baselineskip=22pt
\centerline{\footnotesize A. PICH}
\baselineskip=13pt
\centerline{\footnotesize\it  Departament de F\'{\i}sica Te\`orica
and IFIC, Universitat de Val\`encia -- CSIC,}
\baselineskip=12pt
\centerline{\footnotesize\it  Dr. Moliner 50, E-46100 Burjassot,
Val\`encia, Spain}
\vspace*{0.3cm}

\vspace*{0.9cm}
\abstracts{These lectures provide an introductory overview of the
dynamics of flavour-changing transitions.
The main emphasis is put on present tests of
the quark-mixing matrix structure and the phenomenological determination
of its parameters.
The interplay of strong interactions in weak decays and
the important role of flavour symmetries for controlling the size of QCD
corrections to some semileptonic transitions are discussed.}

\normalsize\baselineskip=15pt
\setcounter{footnote}{0}
\renewcommand{\thefootnote}{\alph{footnote}}

\section{Flavour Structure of the Standard Model}
\label{sec:introduction}

The Standard Model (SM)
is a gauge theory, based on the group
%
$SU(3)_C \otimes SU(2)_L \otimes U(1)_Y$,
%
which describes strong, weak and electromagnetic interactions,
via the exchange of the corresponding spin--1 gauge fields:
8 massless gluons and 1 massless photon for the strong and
electromagnetic interactions, respectively,
and 3 massive bosons, $W^\pm$ and $Z$, for the weak interaction.
The fermionic matter content is given by the known
leptons and quarks, which are organized in a 3--fold
family structure:
\bel{eq:families}
\left[\bat \nu_e & u \\  e^- & d \ea \right] \, , \qquad\quad
\left[\bat \nu_\mu & c \\  \mu^- & s \ea \right] \, , \qquad\quad
\left[\bat \nu_\tau & t \\  \tau^- & b \ea \right] \, , 
\ee
where
(each quark appears in 3 different {\it colours}) 
\bel{eq:structure}
\left[\bat \nu_l & q_u \\  l^- & q_d \ea \right] \,\,\equiv\,\,
\left(\ba \nu_l \\ l^- \ea \right)_{\! L} , \quad
\left(\ba q_u \\ q_d \ea \right)_{\! L} , \quad l^-_R , 
\quad (q_u)_R , \quad
(q_d)_R ,
\ee
plus the corresponding antiparticles.
Thus, the left-handed fields are $SU(2)_L$ doublets, while
their right-handed partners transform as $SU(2)_L$ singlets.
The 3 fermionic families in \eqn{eq:families} appear
to have identical properties (gauge interactions); they only
differ by their mass and their flavour quantum number.

The gauge symmetry is broken by the vacuum,
which triggers the Spontaneous Symmetry Breaking (SSB)
of the electroweak group to the electromagnetic subgroup:
\bel{eq:ssb}
SU(3)_C \otimes SU(2)_L \otimes U(1)_Y \, \ssb\,
SU(3)_C \otimes U(1)_{QED} \, .
\ee
The SSB mechanism generates the masses of the weak gauge bosons,
and gives rise to the appearance of
a physical scalar particle in the model, the so-called {\it Higgs}.
The fermion masses and mixings are also generated through the 
SSB mechanism.  

The SM constitutes one of the most successful achievements
in modern physics. It provides a very elegant theoretical
framework, which is able to describe all known experimental
facts in particle physics.
A detailed description of the SM and its present
phenomenological status can be found in Refs.~\citenum{jaca:94}
and \citenum{sorrento:94}, which discuss the electroweak and strong
sectors, respectively.

In spite of its enormous phenomenological success, the SM leaves too many
unanswered questions to be considered as a complete description of the
fundamental forces.
We do not understand yet why fermions are replicated in three
(and only three)
nearly identical copies? Why the pattern of masses and mixings
is what it is?  Are the masses the only difference among the three
families? What is the origin of the SM flavour structure?
Which dynamics is responsible for the observed CP violation?

The fermionic flavour is the main source of
arbitrary free parameters in the SM: 9 fermion masses,
3 mixing angles and 1 complex phase (assuming the neutrinos to be
massless).
The problem of fermion--mass
generation is deeply related with the mechanism responsible for the SSB.
Thus, the origin of these parameters lies in the most obscure part of
the SM Lagrangian: the scalar sector. 
Clearly, the dynamics of flavour appears to be ``terra incognita''
which deserves a careful investigation.

The flavour structure looks richer in the quark sector,
where mixing phenomena among the different families occurs
(leptons would also mix if neutrino masses were non-vanishing).
A precise measurement of the quark mixings would allow to test
their predicted unitarity structure, and could give some hints
about the unknown underlying dynamics.
Since quarks are confined within hadrons,
an accurate determination of their mixing parameters requires
first a good understanding of hadronization effects
in flavour--changing transitions.
The interplay of strong interactions in weak decays
plays a crucial role, which,
unfortunately, is rather difficult to control due to the
non-perturbative character of QCD at long distances.

The purpose of these lectures is to provide an introductory overview
of our present knowledge on the quark--mixing couplings
and the prospects for further improvements.
I will try to emphasize those theoretical aspects which
are more relevant for our understanding of the flavour--changing dynamics.
Further experimental considerations are discussed elsewhere in these
proceedings.

\subsection{Charged--Current Interactions}

In the SM flavour--changing transitios occur only in the charged--current
sector:
\bel{eq:cc_mixing}
\cL_{\mbox{\rms CC}}\, = \, {g\over 2\sqrt{2}}\,\left\{
W^\dagger_\mu\,\left[\sum_{ij}\,
\bar u_i\gamma^\mu(1-\gamma_5) \bV_{\!\! ij} d_j 
\, +\,\sum_l\, \bar\nu_l\gamma^\mu(1-\gamma_5) l
\right]\, + \, \mbox{\rm h.c.}\right\}\, .
\ee
The so-called Cabibbo\cite{CA:63}--Kobayashi--Maskawa\cite{KM:73} (CKM)
matrix {\boldmath $V$} couples any {\it up--type} quark with all
{\it down--type} quarks.
It originates from the same Yukawa couplings giving rise to the
quark masses.

Before SSB, there is no mixing among the different quarks.
In order to understand the origin of
the matrix {\boldmath $V$},
let us consider the general case of $N_G$ generations of fermions, 
and denote $\nu_j'$, $l'_j$, $u'_j$, $d'_j$ the members of the weak
family $j$\  ($j=1,\ldots,N_G$),
with definite transformation properties under the gauge group.
The $W$ boson couples to these fields as in Eq.~\eqn{eq:cc_mixing},
but without any mixing matrix {\boldmath $V$}
(i.e., with {\boldmath $V$}$_{\!\! ij} = \delta_{ij}$).

The SSB mechanism generates fermion masses proportional to the vacuum
expectation value of the scalar field,
$\langle\emptyset |\phi^{(0)}|\emptyset \rangle\equiv v/\sqrt{2}$.
The resulting quark--mass
eigenstates are however not the same as the eigenstates of the weak
interactions.
The most general Yukawa Lagrangian,
\beqn\label{eq:N_Yukawa}
\cL_Y &=&\sum_{jk}\,\left\{
\left(\bar u'_j , \bar d'_j\right)_L \left[ c^{(d)}_{jk} 
\left(\ba \phi^{(+)}\\ \phi^{(0)}\ea\right)\, d'_{kR}  + 
c^{(u)}_{jk}
\left(\ba \phi^{(0)\dagger}\\ -\phi^{(+)\dagger}\ea\right)\, u'_{kR}
\right] 
\right.\no\\ && \quad\,\,\,\left. + 
\left(\bar \nu'_j , \bar l'_j\right)_L c^{(l)}_{jk}
\left(\ba \phi^{(+)}\\ \phi^{(0)}\ea\right)\, l'_{kR}\right\} 
\, +\, \mbox{\rm h.c.},
\eeqn
is not diagonal in quark flavour,
since this condition is not required by gauge invariance.
Thus, the couplings
$c^{(d)}_{jk}$, $c^{(u)}_{jk}$ and $c^{(l)}_{jk}$
are arbitrary constants.

After SSB, the Yukawa Lagrangian can be written as
\bel{eq:N_Yuka}
\cL_Y = - \left(1 + {H\over v}\right)\,\left\{
\overline{\mbox{\boldmath $d$}}'_L \mbox{\boldmath $M$}_d' 
\mbox{\boldmath $d$}'_R \, + \, 
\overline{\mbox{\boldmath $u$}}'_L \mbox{\boldmath $M$}_u' 
\mbox{\boldmath $u$}'_R
\, + \,
\overline{\mbox{\boldmath $l$}}'_L \mbox{\boldmath $M$}'_l 
\mbox{\boldmath $l$}'_R \, + 
\mbox{\rm h.c.}\right\} .
\ee
Here, $\mbox{\boldmath $d$}'$, $\mbox{\boldmath $u$}'$
and $\mbox{\boldmath $l$}'$ denote vectors in flavour 
space, and the corresponding mass matrices are given by
\bel{eq:M_c_relation}
(\mbox{\boldmath $M$}'_d)_{ij}\,\equiv\, 
- c^{(d)}_{ij}\, v/\sqrt{2}\, , \quad
(\mbox{\boldmath $M$}'_u)_{ij}\,\equiv\, 
- c^{(u)}_{ij}\, v/\sqrt{2}\, , \quad
(\mbox{\boldmath $M$}'_l)_{ij}\,\equiv\,
 - c^{(l)}_{ij}\, v/\sqrt{2}\, .
\ee
The diagonalizacion of these matrices determines the mass
eigenstates $d_j$, $u_j$ and $l_j$.

The matrix $\mbox{\boldmath $M$}_d'$ can be decomposed as
$\mbox{\boldmath $M$}_d'=\mbox{\boldmath $H$}_d
\mbox{\boldmath $U$}_d=\mbox{\boldmath $S$}_d^\dagger 
\mbox{\boldmath $\cM$}_d \mbox{\boldmath $S$}_d 
\mbox{\boldmath $U$}_d$, where
$\mbox{\boldmath $H$}_d\equiv
\sqrt{\mbox{\boldmath $M$}_d'{\mbox{\boldmath $M$}_d'}^\dagger}$ 
is an hermitian positive--definite matrix,
while $\mbox{\boldmath $U$}_d$ is unitary. 
$\mbox{\boldmath $H$}_d$ can be diagonalized
by a unitary matrix $\mbox{\boldmath $S$}_d$; the resulting matrix 
$\mbox{\boldmath $\cM$}_d$ is diagonal,
hermitian and positive definite.
Similarly,
one has
$\mbox{\boldmath $M$}_u'= \mbox{\boldmath $H$}_u 
\mbox{\boldmath $U$}_u= \mbox{\boldmath $S$}_u^\dagger 
\mbox{\boldmath $\cM$}_u \mbox{\boldmath $S$}_u 
\mbox{\boldmath $U$}_u$ and 
$\mbox{\boldmath $M$}_l'= \mbox{\boldmath $H$}_l 
\mbox{\boldmath $U$}_l= \mbox{\boldmath $S$}_l^\dagger 
\mbox{\boldmath $\cM$}_l \mbox{\boldmath $S$}_l 
\mbox{\boldmath $U$}_l$.
In terms of the diagonal mass matrices, 
$\mbox{\boldmath $\cM$}_d=\mbox{\rm diag}(m_d,m_s,m_b,\ldots)$,
$\mbox{\boldmath $\cM$}_u=\mbox{\rm diag}(m_u,m_c,m_t,\ldots)$,
$\mbox{\boldmath $\cM$}_l=\mbox{\rm diag}(m_e,m_\mu,m_\tau,\ldots)$,
the Yukawa Lagrangian takes the
simpler form
\bel{eq:N_Yuk_diag}
\cL_Y = - \left(1 + {H\over v}\right)\,\left\{
\overline{\mbox{\boldmath $d$}} \mbox{\boldmath $\cM$}_d 
\mbox{\boldmath $d$} \, + \, 
\overline{\mbox{\boldmath $u$}} \mbox{\boldmath $\cM$}_u 
\mbox{\boldmath $u$} \, + \,
\overline{\mbox{\boldmath $l$}} \mbox{\boldmath $\cM$}_l 
\mbox{\boldmath $l$} \right\} ,
\ee
where the mass eigenstates are defined by
\beqn\label{eq:S_matrices}
\mbox{\boldmath $d$}_L &\!\!\!\!\equiv &\!\!\!\! 
\mbox{\boldmath $S$}_d\, \mbox{\boldmath $d$}'_L \, , 
\qquad\,\,\,\,\,\,\,\,\,
\mbox{\boldmath $u$}_L \equiv \mbox{\boldmath $S$}_u \, 
\mbox{\boldmath $u$}'_L \, , 
\qquad\,\,\,\,\,\,\,\,\,
\mbox{\boldmath $l$}_L \equiv \mbox{\boldmath $S$}_l \, 
\mbox{\boldmath $l$}'_L \, ,
\no\\
\mbox{\boldmath $d$}_R &\!\!\!\!\equiv &\!\!\!\! 
\mbox{\boldmath $S$}_d \mbox{\boldmath $U$}_d\,  
\mbox{\boldmath $d$}'_R \, , \qquad
\mbox{\boldmath $u$}_R \equiv \mbox{\boldmath $S$}_u 
\mbox{\boldmath $U$}_u\, \mbox{\boldmath $u$}'_R \, , \qquad
\mbox{\boldmath $l$}_R \equiv \mbox{\boldmath $S$}_l 
\mbox{\boldmath $U$}_l \, \mbox{\boldmath $l$}'_R \, .
\eeqn

Since, $\overline{\mbox{\boldmath $f$}'}_L \mbox{\boldmath $f$}'_L =
\overline{\mbox{\boldmath $f$}}_L \mbox{\boldmath $f$}_L$ and 
$\overline{\mbox{\boldmath $f$}'}_R \mbox{\boldmath $f$}'_R =
\overline{\mbox{\boldmath $f$}}_R \mbox{\boldmath $f$}_R\,$
($f=d,u,l$), the form of the neutral--current part of the 
$SU(2)_L\otimes U(1)_Y$ Lagrangian does not change when expressed
in terms of mass eigenstates. Therefore, there are no
flavour--changing neutral currents in the SM
(GIM mechanism \cite{GIM:70}).
This is a consequence of treating all equal--charge fermions
on the same footing.
However, $\overline{\mbox{\boldmath $u$}}'_L \mbox{\boldmath $d$}'_L = 
\overline{\mbox{\boldmath $u$}}_L \mbox{\boldmath $S$}_u 
\mbox{\boldmath $S$}_d^\dagger 
\mbox{\boldmath $d$}_L\equiv 
\overline{\mbox{\boldmath $u$}}_L \mbox{\boldmath $V$} 
\mbox{\boldmath $d$}_L$. In general, 
$\mbox{\boldmath $S$}_u\not= \mbox{\boldmath $S$}_d$; thus
a $N_G\times N_G$ unitary mixing matrix $\bV$
appears in the quark charged--current sector, and one gets the
Lagrangian in Eq.~\eqn{eq:cc_mixing}.

We can
redefine the neutrino flavours in such a way as to eliminate
the analogous mixing in the lepton sector:
$\overline{\mbox{\boldmath $\nu$}}_L' \mbox{\boldmath $l$}'_L = 
\overline{\mbox{\boldmath $\nu$}}_L' \mbox{\boldmath $S$}^\dagger_l 
\mbox{\boldmath $l$}_L
\equiv
\overline{\mbox{\boldmath $\nu$}}_l \mbox{\boldmath $l$}_L$.
The lepton flavour is then conserved in the minimal SM
without right-handed neutrinos.
   
The fermion masses and the quark--mixing matrix $\bV$
are all determined by the Yukawa couplings in Eq.~\eqn{eq:N_Yukawa}. 
However, the Yukawas are not known; therefore we have a bunch of
arbitrary parameters.
A general $N_G\times N_G$ unitary matrix contains $N_G^2$ real
parameters [$N_G (N_G-1)/2$ moduli and $N_G (N_G+1)/2$ phases].
In the case of {\boldmath $V$}, many of these parameters are
irrelevant, because we can always choose arbitrary
quark phases.
Under the phase redefinitions $u_i\to \e^{i\phi_i} u_i$ and
$d_j\to\e^{i\theta_j} d_j$, the mixing matrix changes as
$\mbox{\boldmath $V$}_{\!\! ij}\to 
\mbox{\boldmath $V$}_{\!\! ij}\,\e^{i(\theta_j-\phi_i)}$;
thus, $2 N_G-1$ phases are unobservable.
The number of physical free parameters in the quark--mixing matrix
gets then reduced to $(N_G-1)^2$:
$N_G(N_G-1)/2$ moduli and $(N_G-1)(N_G-2)/2$ phases.

In the simpler case of two generations, {\boldmath $V$}
is determined by a single parameter, the so-called
Cabibbo angle \cite{CA:63}:
\bel{eq:cabibbo}
\mbox{\boldmath $V$}\, = \, \left[
\bat \phantom{-}\cos{\theta_C} &\sin{\theta_C} \\ 
-\sin{\theta_C}& \cos{\theta_C}\ea
\right]\, .
\ee
With $N_G=3$, the CKM matrix is described by 3 angles and 1 phase
\cite{KM:73}.
Different (but equivalent) representations can be found in the literature.
The Particle Data Group \cite{pdg:94} advocates the use of the
following one as the {\it standard} CKM parametrization:
\be\label{eq:CKM_pdg}
\mbox{\boldmath $V$}\, = \, \left[
\begin{array}{ccc}
c_{12}c_{13}  &  s_{12} c_{13}  &  s_{13} e^{-i\delta_{13}} \\
-s_{12} c_{23}-c_{12}s_{23}s_{13}e^{i\delta_{13}} &
c_{12} c_{23}- s_{12} s_{23} s_{13} e^{i\delta_{13}} &
s_{23} c_{13}  \\
s_{12} s_{23}-c_{12}c_{23}s_{13}e^{i\delta_{13}} &
-c_{12} s_{23}- s_{12} c_{23} s_{13} e^{i\delta_{13}} &
c_{23} c_{13}  
\ea
\right] .
\ee
Here $c_{ij} \equiv \cos{\theta_{ij}}$ and
$s_{ij} \equiv \sin{\theta_{ij}}$, with $i$ and $j$ being
generation labels ($i,j=1,2,3$).
The real angles $\theta_{12}$, $\theta_{23}$ and $\theta_{13}$
can all be made to lie in the first quadrant, by an appropriate 
redefinition of quark field phases; then,
$c_{ij}\geq 0$, $s_{ij}\geq 0$ and $0\leq \delta_{13}\leq 2\pi$.

$\delta_{13}$ is the only complex phase in the SM
Lagrangian; thus, it is a unique source of CP--violation.
In fact, it was for this reason that the third generation
was assumed to exist \cite{KM:73}, 
before the discovery of the $\tau$ and the $b$ .
With two generations, the SM could not explain the observed
CP--violation in the $K$ system.

\goodbreak

\mbox{}\vspace*{-1.1cm}

\subsection{Fermion Mass Spectrum}

The measured lepton masses \cite{pdg:94,bes:95},
\bel{eq:l_mass}
\ba
m_e \, = \,
 (0.51099906\pm0.00000015) \;\mbox{\rm MeV} , \\
m_\mu \, = \,  (105.658389 \pm 0.000034) \;\mbox{\rm MeV} ,
\\
m_\tau \, =\, (1777.0 \pm 0.3) \;\mbox{\rm MeV} , 
\ea
\ee
are very different. This indicates a hierarchy of the original Yukawa
couplings, which increase from one generation to the other.
A similar pattern is found in the quark spectrum:
\cite{GL:82,LE:90,BPR:94,JM:94,NA:95,CH:95,AL:94,DA:94,TY:94,CDF:95,D0:95}
\be\label{eq:q_masses}
\begin{array}{ccc}
\overline{m}_d(1 \,\mbox{\rm GeV}) \, =\, (8.5 \pm 2.5) \;\mbox{\rm MeV} , 
& \qquad\; &
\overline{m}_u(1 \,\mbox{\rm GeV})\, =\, (5.0 \pm 2.5) \;\mbox{\rm MeV} , 
\\ 
\overline{m}_s(1 \,\mbox{\rm GeV}) \, = \, (180\pm 25) \;\mbox{\rm MeV}  , 
&&
\overline{m}_c \, =\,  (1.25\pm 0.05) \;\mbox{\rm GeV} , 
\\
\overline{m}_b \, =\,  (4.25\pm 0.10) \;\mbox{\rm GeV} , 
&&
m_t \, =\,  (180\pm 12) \;\mbox{\rm GeV} . \ea
\ee
Since $m_d > m_u$, the first quark family behaves analogously to
the lepton ones, where the heavier states (the charged leptons)
correspond to the $T_3=-1/2$ members of the weak doublets.
However, the second and third quark generations show the opposite
behaviour:
the up--type quarks are heavier than their
doublet partnerts.

The present experimental bounds \cite{pdg:94,aleph:95} on the 
neutrino masses are:
\bel{eq:nu_masses}
m_{\nu_e} < 7.0\;\mbox{\rm eV}  \;\; (95\%\,\mbox{\rm CL}), \quad
m_{\nu_\mu} < 0.27\;\mbox{\rm MeV}  \;\; (90\%\,\mbox{\rm CL}), \quad
m_{\nu_\tau} < 24\;\mbox{\rm MeV}  \;\; (95\%\,\mbox{\rm CL}). 
\ee

At this point, one should make a word of caution concerning quark
masses. Since quarks are confined within hadrons, the kinematical
concept of ({\it on-shell}) quark mass, corresponding to a free asymptotic
state, is meaningless.
What we call quark masses are the parameters appearing
in the mass term of the QCD Lagrangian (the so-called {\it current} 
quark masses).
Like any other coupling, these parameters need to be appropriately
defined within some renormalization scheme. Thus, they depend on
the chosen renormalization scheme 
(we use the $\overline{MS}$ scheme)
and are functions of the
renormalization scale; i.e., masses are {\it running} parameters
which we denote by $\overline{m}(\mu)$.

One can adopt the value of the
running mass at the scale $\mu = m_q$
to characterize the quark mass, i.e.
$\overline{m}_q\equiv\overline{m}_q(\overline{m}_q)$.
This is not possible for the light flavours because our
perturbative QCD formulae are no loger valid at such low scales; this 
is why the up, down and strange quark masses in \eqn{eq:q_masses} 
have been normalized at a common reference scale
$\mu=$ 1 GeV. The phenomenological determination of these masses
suffers from rather large uncertainties. 

Using the chiral symmetry properties of the light--quark sector
\cite{chpt:95,EC:95},
one can fix more precisely the  ratios \cite{GL:82,LE:90,chpt:95}
\bel{eq:m_ratios}
{2 m_s\over m_d + m_u} = 22.6\pm 3.3 \, , \qquad\quad
{m_d - m_u \over m_d + m_u} = 0.25 \pm 0.04 \, ,
\ee
which do not have any scale or scheme dependence (QCD is flavour blind).

For heavy quarks, one can also define the mass as the pole of the
quark propagator. This quantity would correspond to the kinematical
{\it on-shell} mass measured in the leptonic case. The only problem
is that there should not be any pole for confined quarks.
Nevertheless, the definition of {\it pole} mass still makes sense
within perturbation theory. The relation between the perturbative
{\it pole} mass and the {\it running} mass is: \cite{TA:81,GBGS:90}
\bel{eq:pole_mass}
m_q^{\mbox{\rms pole}} = \overline{m}_q(m_q^{\mbox{\rms pole}}) \,
\left\{ 1 + {4\over 3} {\alpha_s(m_q^{\mbox{\rms pole}})\over \pi}
  + O\left(\alpha_s^2\right)\right\} \, .
\ee
The numerical difference between these two definitions is quite
important. Even at the top--mass scale where $\alpha_s$ is small
one has
$m_t^{\mbox{\rms pole}} - \overline{m}_t \sim 7$ GeV, which is
of the same size as the present experimental uncertainty.
Thus, when quoting any numerical values for quark masses,
it is very important to specify the exact definition to which they
refer.
The measured value of $m_t$ given in \eqn{eq:q_masses} seems
to correspond to a {\it pole} mass (we will assume that in the
following); however, the adopted definition has not been
given in the published experimental papers \cite{CDF:95,D0:95}.

\section{Weak Decays}
\label{sec:decays}\setcounter{equation}{0}
\vspace*{-0.6cm}
\subsection{$\mu^-\to e^-\bar\nu_e\nu_\mu$}

The simplest flavour--changing process is the leptonic
decay of the muon, which proceeds through the $W$--exchange
diagram shown in Fig.~\ref{fig:mu_decay}.
The momentum transfer carried by the intermediate $W$ is very small
compared to $M_W$. Therefore, the vector--boson propagator reduces
to a contact interaction,
\bel{eq:low_energy}
{-g_{\mu\nu} + q_\mu q_\nu/M_W^2 \over q^2-M_W^2}\quad\;
 \toLow\quad\; {g_{\mu\nu}\over M_W^2}\, .
\ee
The decay can then be described through an effective local
4--fermion Hamiltonian,
\bel{eq:mu_v_a}
\cH_{\mbox{\rms eff}}\, = \, {G_F \over\sqrt{2}}
\left[\bar e\gamma^\alpha (1-\gamma_5) \nu_e\right]\,
\left[ \bar\nu_\mu\gamma_\alpha (1-\gamma_5)\mu\right]\, , 
\ee
where
\bel{eq:G_F}
{G_F\over\sqrt{2}} = {g^2\over 8 M_W^2}
\ee
is called the Fermi coupling constant.
$G_F$ 
is fixed by the total decay width,
\bel{eq:mu_lifetime}
{1\over\tau_\mu}\, = \, \Gamma(\mu^-\to e^-\bar\nu_e\nu_\mu)
\, = \, {G_F^2 m_\mu^5\over 192 \pi^3}\,
\left( 1 + \delta_{\mbox{\rms RC}}\right) \, 
f\left(m_e^2/m_\mu^2\right) \, ,
\ee
where
$\, f(x) = 1-8x+8x^3-x^4-12x^2\ln{x}$,
and
\bel{eq:qed_corr}
(1+\delta_{\mbox{\rms RC}})  =  
\left[1+{\alpha\over 2\pi}\left({25\over 4}-\pi^2\right)\right]\,
\left[ 1 +{3\over 5}{m_\mu^2\over M_W^2} - 2 {m_e^2\over M_W^2}\right]
=  0.9958 \, 
\ee
takes into account the leading higher-order corrections \cite{KS:59}.
The measured lifetime \cite{pdg:94},
$\tau_\mu=(2.19703\pm 0.00004)\times 10^{-6}$ s,
implies the value
\bel{eq:gf}
G_F\, = \, (1.16639\pm 0.00002)\times 10^{-5} \:\mbox{\rm GeV}^{-2}
\,\approx\, {1\over (293 \:\mbox{\rm GeV})^2} \, .
\ee
%

\begin{figure}[bth]
\vfill
\centerline{
\begin{minipage}[t]{.4\linewidth}\centering
\epsfig{file=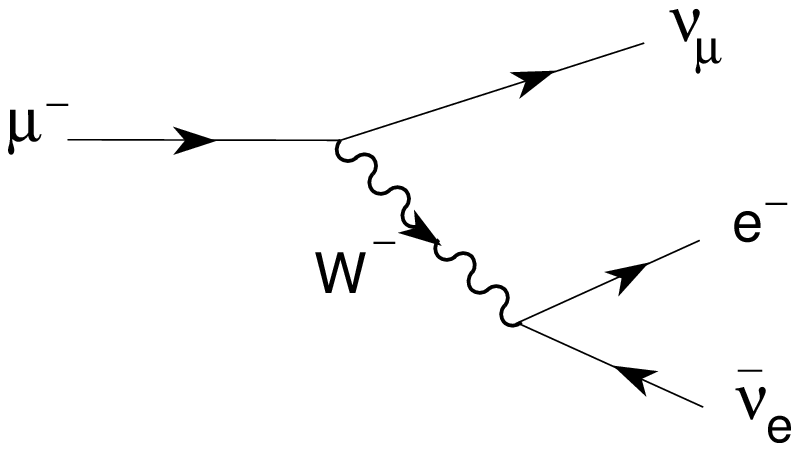,height=35mm}
\caption{$\mu$-decay diagram.}
\label{fig:mu_decay}
\end{minipage}
\hspace{0.4cm}
\begin{minipage}[t]{.54\linewidth}\centering
\epsfig{file=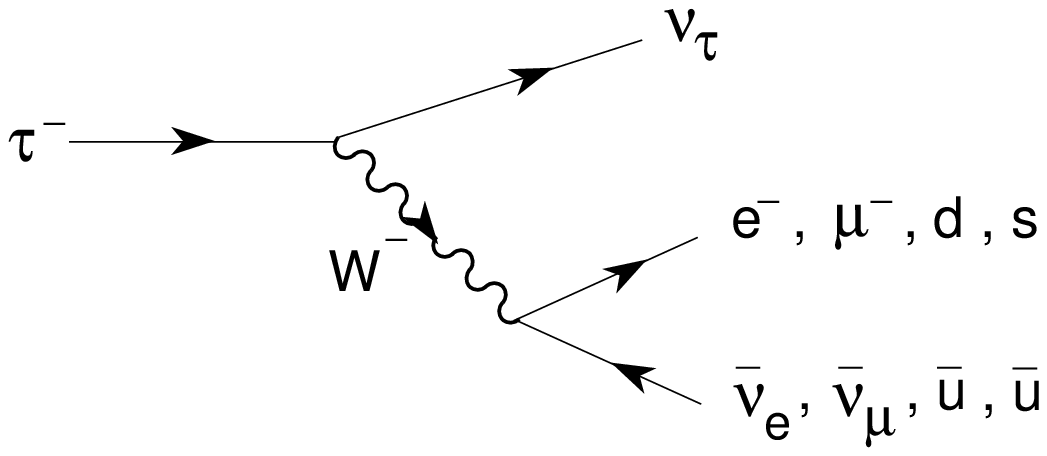,height=35mm}
\caption{$\tau$-decay diagram.}
\end{minipage}
}
\vfill
\end{figure}

\subsection{$\tau$ Decay}

The decays of the $\tau$ lepton proceed through the same
$W$--exchange mechanism as the leptonic $\mu$ decay.
The only difference is that several final states
are kinematically allowed:
$\tau^-\to\nu_\tau e^-\bar\nu_e$,
$\tau^-\to\nu_\tau\mu^-\bar\nu_\mu$,
$\tau^-\to\nu_\tau d\bar u$ and $\tau^-\to\nu_\tau s\bar u$.
Owing to the universality of the $W$--couplings, all these
decay modes have equal amplitudes (if final fermion masses and
QCD interactions are neglected), except for an additional
$N_C |\bV_{\!\! ui}|^2$ factor ($i=d,s$) in the semileptonic
channels, where $N_C=3$ is the number of quark colours. 
Making trivial kinematical changes in Eq.~\eqn{eq:mu_lifetime},
one easily gets the lowest--order prediction for the total
$\tau$ decay width:
\bel{eq:tau_decay_width}
{1\over\tau_\tau}\equiv\Gamma(\tau) \approx
\Gamma(\mu) \left({m_\tau\over m_\mu}\right)^5
\left\{ 2 + N_C 
\left( |\bV_{\!\! ud}|^2 + |\bV_{\!\! us}|^2\right)\right\}
\approx {5\over\tau_\mu}\left({m_\tau\over m_\mu}\right)^5
  ,
\ee
where we have used the CKM unitarity relation
$|\bV_{\!\! ud}|^2 + |\bV_{\!\! us}|^2 = 1 - |\bV_{\!\! ub}|^2
\approx 1$ 
(we will see later that this is an excellent approximation).
From the measured muon lifetime, one has then
$\tau_\tau\approx 3.3\times 10^{-13}$ s, to be compared
with the experimental value \cite{taurep}
$\tau_\tau^{\mbox{\rms exp}} = (2.916\pm 0.016)\times 10^{-13}$ s.

The branching ratios into the different decay modes are
predicted to be:
\beqn\label{eq:tau_br}
\mbox{\rm Br}(\tau^-\to\nu_\tau e^-\bar\nu_e) \approx {1\over 5} = 20\% 
\qquad
\qquad\quad && [\mbox{\rm exp:}\; (17.79\pm 0.09)\%] \, ,
\no\\
{\mbox{\rm Br}(\tau^-\to\nu_\tau \mu^-\bar\nu_\mu)\over
\mbox{\rm Br}(\tau^-\to\nu_\tau e^-\bar\nu_e)}\approx
{f(m_\mu^2/m_\tau^2)\over f(m_e^2/m_\tau^2)} = 0.97256 
\qquad && \;\; [\mbox{\rm exp:}\; 0.974\pm 0.007] \, ,
\\
R_\tau\equiv {\Gamma(\tau\to\nu_\tau + \mbox{\rm Hadrons})\over 
\Gamma(\tau^-\to\nu_\tau e^-\bar\nu_e)} \approx N_C \quad
\qquad && \;\; [\mbox{\rm exp:}\; 3.647\pm 0.024] \, ,
\no
\eeqn
in good agreement with the measured numbers \cite{taurep}, indicated
on the right. Our naive predictions only deviate
from the experimental results by about 20\%.
A much better agreement is obtained for the absolute value of the
leptonic decay widths:
\bel{eq:tau_to_e}
\Gamma(\tau^-\to\nu_\tau e^-\bar\nu_e) 
\approx {1\over\tau_\mu}\left({m_\tau\over m_\mu}\right)^5
= \, 6.12\times 10^{11}\;\mbox{\rm s}^{-1} \qquad
[\mbox{\rm exp:}\; (6.10\pm 0.05)\times 10^{11}\;\mbox{\rm s}^{-1}] \, .
\ee
The reason why this prediction works
much better can be easily understood.
The branching ratios are more sensitive to the corrections
induced by strong interactions, which we are completely neglecting.
These corrections are generated by gluonic exchanges between the
final quarks and only affect the semileptonic decay modes.
As indicated by the measured $R_\tau$ value, they amount
to a 20\%  effect.

\subsection{Charm Decays}

%
\begin{figure}[htb]
\centering       
\epsfig{file=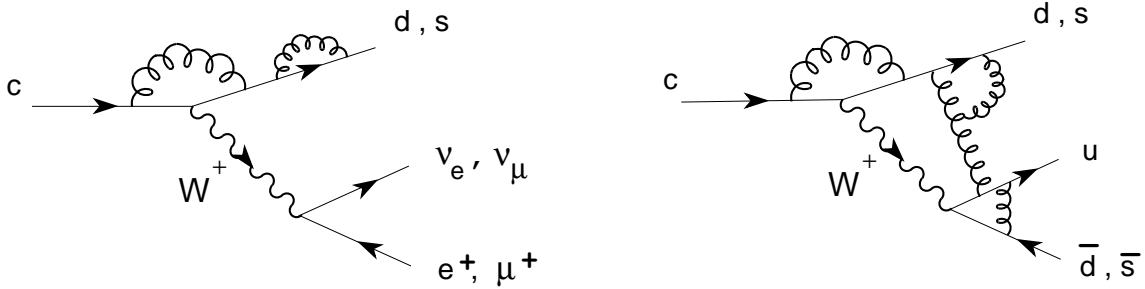,width=6.0in}
\caption{Feynman diagrams for semileptonic and non-leptonic $c$ decays.}
\label{fig:c_decay}
\end{figure}
%

At lowest order, the decay of the charm quark is described
by the same Feynman diagram as the $\mu$ and $\tau$ decays.
The initial $c$ quark converts into a $d$ or  $s$ flavour
through the emission of a $W^+$; the corresponding vertices
contain a $\bV_{\!\! ci}^*$ ($i=d,s$) mixing factor.
The emitted $W^+$ can give rise to several fermonic final states 
which are
kinematically allowed: $\nu_e e^+$, $\nu_\mu\mu^+$, $u\bar d$
and $u\bar s$. Adding all possible decay modes, the total charm
decay width is predicted to be
\bel{eq:c_decay}
\Gamma(c)\approx 
{1\over\tau_\mu} \left( {m_c\over m_\mu}\right)^5
\left( |\bV_{\!\! cd}|^2 + |\bV_{\!\! cs}|^2 \right) \,
\left\{ 2 + N_C \left( |\bV_{\!\! ud}|^2 + |\bV_{\!\! us}|^2\right)
\right\} \, .
\ee
Using the approximate relations (they would be exact with
only two fermion generations)
$|\bV_{\!\! cd}|^2 + |\bV_{\!\! cs}|^2 = 1 - |\bV_{\!\! cb}|^2
\approx 1$ and 
$|\bV_{\!\! ud}|^2 + |\bV_{\!\! us}|^2 = 1 - |\bV_{\!\! ub}|^2
\approx 1$, one gets a charm lifetime
\be
\tau_c \approx {1\over 5} \,\tau_\mu \left( {m_\mu\over m_c}\right)^5 
\approx 7.6\times 10^{-13} \:\mbox{\rm s} \, ,
\ee
where we have taken $m_c\sim 1.5$ GeV.
This estimate should be compared with the measured lifetimes
of the different charm hadrons: \cite{pdg:94}
\bel{eq:c_had_life}\begin{array}{lll}
\tau_{D^+} \, = \, (10.57\pm 0.15) \times 10^{-13}\:\mbox{\rm s}  ,
& \qquad &
\tau_{\Lambda_c^+}\, =\, 
(2.00\, {}^{+0.11}_{-0.10})\times 10^{-13}\:\mbox{\rm s},
\\
\tau_{D^0} \, = \, (4.15\pm 0.04) \times 10^{-13}\:\mbox{\rm s} ,
&&
\tau_{\Xi_c^+} \, = \, (3.5\, {}^{+0.7}_{-0.4}) \times 10^{-13}\:\mbox{\rm s},
\\
\tau_{D_s^+} \, = \, (4.67\pm 0.17)\times 10^{-13}\:\mbox{\rm s} ,
&&
\tau_{\Xi_c^0} \, = \, 
(0.98\, {}^{+0.23}_{-0.15}) \times 10^{-13}\:\mbox{\rm s}.
\ea\ee
Obviously, changing the numerical value of $m_c$ one could trivially
fit any of those measurements. However, there is no way to understand
the large splittings among the different hadronic lifetimes.
Our naive estimate assumes a similar lifetime $\tau_c$ for
all charm hadrons,
whereas one measures \cite{pdg:94}
$\tau(D^+)/\tau(D^0)\approx 2.5$ or
$\tau(D^+)/\tau(\Xi^0_c)\approx 11$.

The prediction works substantially better for the semileptonic
decay widths:
\bel{eq:c_eX}
\Gamma(c\to X l^+\nu_l) \equiv {\mbox{\rm Br}(c\to X l^+\nu_l)\over \tau_c}
\approx {1\over\tau_\mu}
\left( {m_c\over m_\mu}\right)^5 \approx
2.6\times 10^{11}\:\mbox{\rm s}^{-1} \, ,
\ee
to be compared with the experimental values \cite{pdg:94,RB:95,CLEO:95d}
\beqn\label{eq:c_semileptonic}
\Gamma(D^+\to  X^0 l^+\nu_l) & = & 
(1.63\pm 0.18)\times 10^{11}\:\mbox{\rm s}^{-1} \, ,
\no\\
\Gamma(D^0\to X^- l^+\nu_l) & = & 
(1.65\pm 0.08)\times 10^{11}\:\mbox{\rm s}^{-1} \, ,
\\
\Gamma(\Lambda_c^+\to  X^0 l^+\nu_l) & = & \phantom{0}
(2.3\pm 0.9)\phantom{0}\times 10^{11}\:\mbox{\rm s}^{-1} \, . \no
\eeqn

The problem with our lifetime prediction has to do with
the strong interactions that we have ignored. In the semileptonic
decays, gluons can only couple to a single $c\to d,s$ hadronic current;
they produce then a sizeable but not too large correction. 
However, the hadronic decay modes contain two different quark currents
and gluons can couple everywhere. The corrections induced by those
gluons exchanged from one quark current to the other appear to be
crucial to understand the hadronic charm decays.
Notice, that QCD is more important in the semileptonic
$c$--decay modes than in the inclusive hadronic $\tau$--decay.
This should be expected, because now we are considering decays
such as $D^+\to X^0 l^+\nu_l$ which refer to a given charm hadron;
i.e., these decays, although inclusive in the final state, are in
fact exclusive with respect to the initial hadron.

\subsection{Bottom Decays}

Applying again the same naive arguments, one gets
\bel{eq:b_decay}
{1\over\tau_b}\equiv
\Gamma(b)\approx {1\over\tau_\mu} \left( {m_b\over m_\mu}\right)^5
\chi^{\mbox{\rms CKM}} \, N_{\mbox{\rmf eff}} \, ,
\ee
where
\beqn\label{eq:b_fac}
&&\qquad \chi^{\mbox{\rms CKM}}  \,\equiv \,
f_c |\bV_{\!\! cb}|^2 + |\bV_{\!\! ub}|^2 \, , 
\no\\
N_{\mbox{\rmf eff}} &\!\! \equiv &\!\!
2 + f_\tau + N_C \left[
f_c' \left( |\bV_{\!\! cd}|^2 + |\bV_{\!\! cs}|^2 \right)
+ \left( |\bV_{\!\! ud}|^2 + |\bV_{\!\! us}|^2 \right) \right] \, .
\eeqn
Owing to the larger mass of the $b$, heavier final states
involving charmed hadrons or the $\tau$ lepton can be now produced.
Therefore, one can no longer neglect the kinematical effects 
induced by the non-zero final masses,
which have been parametrized through the factors $f_c$,
$f_\tau$ and $f_c'$.
Taking $m_c\sim 1.5$ GeV and $m_b\sim 4.5$ GeV, we get the
rough estimate $f_\tau\sim f_c\sim f_c'\sim f(m_c^2/m_b^2) \sim 0.5$,
which implies
$N_{\mbox{\rmf eff}}\approx 7$.
We have then
\bel{eq:tau_b}
\tau_b\approx {2.2\times 10^{-15}\:\mbox{\rm s}\over \chi^{\mbox{\rms CKM}}}
\,\times\,\left({7\over N_{\mbox{\rmf eff}}}\right) \, .
\ee
%

%
\begin{figure}[htb]
\centering       
\epsfig{file=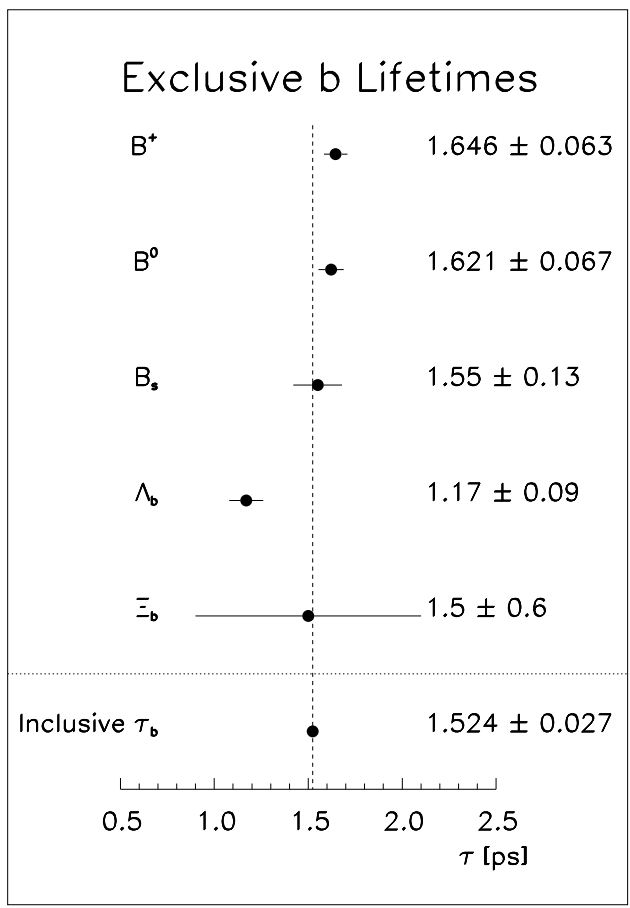,height=120mm}
\caption{Measured lifetimes of the $b$ hadrons \protect\cite{BH:95}.}
\label{fig:b_lifetimes}
\end{figure}
%

The experimental lifetimes of the known $b$ hadrons are plotted in
Fig.~\ref{fig:b_lifetimes}.
Except for $\tau_{\Lambda_b}$ which is slightly smaller, all hadrons turn out
to have a similar lifetime. This provides some support to our 
approximation \eqn{eq:b_decay}, which ignores hadronization effects.
Using Eq.~\eqn{eq:tau_b},
the inclusive measurement $\tau_b = (1.524\pm 0.027)\times 10^{-12}$ s
implies $\chi^{\mbox{\rms CKM}}\sim 10^{-3}$.
Thus, the mixing of the bottom with the up and charm quarks is very small.
The unitarity of the CKM matrix requires then $|\bV_{\!\! tb}|\approx 1$.

Our prediction for the $b$ semileptonic branching ratio is
\bel{eq:b_semileptonic}
\Gamma(b\to X l^-\bar\nu_l) \equiv
{\mbox{\rm Br}(b\to X l^-\bar\nu_l) \over\tau_b} \approx
{1\over\tau_\mu} \left({m_b\over m_\mu}\right)^5 \chi^{\mbox{\rms CKM}}
\approx 6.4\times 10^{13}\,\chi^{\mbox{\rms CKM}}\:\mbox{\rm s}^{-1} \, ,
\ee
to be compared with the experimental measurements: \cite{RB:95,BH:95}
\bel{eq:semi_exp}\ba
\Gamma(b\to X l^-\bar\nu_l) \; = \;
(7.3\pm 0.3)\times 10^{10}\:\mbox{\rm s}^{-1}  ,
\\
\Gamma(\bar B^0\to X^+ l^-\bar\nu_l) \, = \, 
(6.3\pm 0.7)\times 10^{10}\:\mbox{\rm s}^{-1}  ,
\\
\Gamma(B^-\to X^0 l^-\bar\nu_l) \, = \, 
(6.1\pm 1.4)\times 10^{10}\:\mbox{\rm s}^{-1} ,
\\
\Gamma(B\to X\tau^-\bar\nu_\tau) \, = \, 
(1.7\pm 0.3)\times 10^{10}\:\mbox{\rm s}^{-1}  .
\ea\ee
These numbers can be easily understood with a mixing factor
$\chi^{\mbox{\rms CKM}}\sim 10^{-3}$.
The unknown factor $\chi^{\mbox{\rms CKM}}$
cancels out in the branching ratio, which is expected to be
\bel{eq:b_br}
\mbox{\rm Br}(b\to l^-\bar\nu_l X) \sim {1\over N_{\mbox{\rmf eff}}} \sim 
14\%\, ,
\ee
in reasonable agreement with the measured values: \cite{RB:95,BH:95}
\bel{eq:b_br_exp}\ba
\mbox{\rm Br}(b\to X l^-\bar\nu_l) \; = \;
 (11.2\pm 0.4)\% \, ,
\\
\mbox{\rm Br}(\bar B^0\to X^+ l^-\bar\nu_l) \, = \, 
(10.2\pm1.0)\% \, ,
\\
\mbox{\rm Br}(B^-\to X^0 l^-\bar\nu_l) \, = \, 
(10.1\pm2.3)\% \, . 
\ea\ee

  Thus, our naive description of the quark decay process seems to 
work much better
for the bottom than for the charm. Owing to the larger mass of the b quark,
the strength of the QCD coupling is smaller, implying that the missing 
corrections induced by the strong interactions are not so
crucial as in  the charm system.

\subsection{Strange Decays}

Let us try to analyze the decay width of the quark $s$ in the same way.
The only kinematically allowed decays are
$s\to u \, e^-\bar\nu_e$, $s\to u \,\mu^-\bar\nu_\mu$
and $s\to u \, d\bar u$. Therefore,
\bel{eq:s_width}
{1\over\tau_s}\approx {1\over\tau_\mu} \left({m_s\over m_\mu}\right)^5
|\bV_{\!\! us}|^2 \left\{ 2 + N_C |\bV_{\!\! ud}|^2 \right\} 
\approx {5 \sin^2{\theta_C}\over\tau_\mu}\left({m_s\over m_\mu}\right)^5
\approx {1\over 4\times 10^{-7}\:\mbox{\rm s}} \, ,
\ee
where we have taken 
$|\bV_{\!\! us}|\approx\sin{\theta_C}\approx 0.22$
and $m_s\approx 200$ MeV. 
The measured lifetimes of strange hadrons deviate strongly from this estimate:
\cite{pdg:94}
\bel{eq:K_life}\begin{array}{lll}
\tau(K_S) \, = \, (0.8926\pm 0.0012)\times 10^{-10}\:\mbox{\rm s} , 
&\quad &
\tau(K_L) \, = \,  (5.17\pm 0.04)\times 10^{-8}\:\mbox{\rm s} , 
\\
\tau(K^+) \, = \, (1.2371\pm 0.0029)\times 10^{-8}\:\mbox{\rm s} , 
&&
\tau(\Lambda) \, = \, (2.632\pm 0.020)\times 10^{-10}\:\mbox{\rm s} .  
\ea\ee
This time we need to explain lifetimes
which differ by two orders of magnitude:
$\tau(K^+)/\tau(K_S) = 139$!

The predicted semileptonic decay width,
\bel{eq:s_semi}
\Gamma(s\to e^-\bar\nu_e X) \approx {1\over\tau_\mu} 
\left({m_s\over m_\mu}\right)^5
|\bV_{\!\! us}|^2 \approx 5.4\times 10^5\: \mbox{\rm s}^{-1} ,
\ee
does not agree either with the measurements: \cite{pdg:94}
\beqn\label{eq:s_semi_exp}
\Gamma(K^-\to\pi^0 e^-\bar\nu_e) &\!\! = & \!\! 
(3.90\pm 0.05)\times 10^{6}\:\mbox{\rm s}^{-1}  , 
\no\\
\Gamma(\bar K^0 \to\pi^+ e^-\bar\nu_e) &\!\! = & \!\! 
(7.49\pm 0.11)\times 10^{6}\:\mbox{\rm s}^{-1}  , 
\\
\Gamma(\Lambda \to p\, e^-\bar\nu_e) &\!\! = & \!\! 
(3.16\pm 0.06)\times 10^{6}\:\mbox{\rm s}^{-1}  . 
\no
\eeqn

Thus, the strange decays cannot be understood with our naive
description, which ignores the strong interactions.
QCD is a crucial ingredient at the low scales relevant for $s$ decays.
The dramatic effect of the gluonic interactions is cleary
shown by the famous enhancement of $\Delta I=1/2$ transitions
observed in $K$ decays: \cite{pdg:94}
\bel{eq:DI=1/2}
{\Gamma(K_S\to\pi^+\pi^-)\over\Gamma(K^+\to\pi^+\pi^0)} =
{\mbox{\rm Br}(K_S\to\pi^+\pi^-)\,\tau(K^+)\over
\mbox{\rm Br}(K^+\to\pi^+\pi^0)\,\tau(K_S)} = 449 \, .
\ee
This huge ratio is predicted to be four in the absence of 
QCD {\it corrections}!

\section{General Analysis of Semileptonic Decays}
\label{sec:semileptonic}\setcounter{equation}{0}

Several important lessons can be extracted from our 
previous discussion of weak decays:
\begin{itemize}
\item Whenever hadrons (quarks) are involved, gluonic corrections play
an important role in the decay amplitude.
\item The interplay of QCD is stronger at lower
scales. Gluonic effects are moderate in $b$ decay, but quite sizeable
(sometimes 100\%) in the charm system.
The strenght of the strong interactions is so large at the $m_s$ scale,
that there is no way to understand the decays of
strange hadrons with a free--quark
description.
\item  The dynamical effect of the strong interaction is more important
in non-leptonic transitions, where there are two different quark currents
and gluons can couple everywhere.
In semileptonic decays, gluons can only be exchanged within a single
quark current; their contribution is then much smaller.
\item Inclusive transitions are less sensitive to QCD than exclusive
hadronic decays where hadronization effects are obviously crucial.
\end{itemize}

Thus, we need to worry about QCD if we want to study
the quark--mixing structure of flavour--changing transitions. 
Unfortunately, the long--distance regime of the strong interactions
is still not well understood.
While the SM Lagrangian is formulated in terms of quarks and gluons,
the measurable weak decays involve hadronic bound states of these
fundamental constituents.
Some insight into the confinement dynamics which binds quarks into hadrons
is then needed, to make possible an accurate determination of the SM
parameters from experimental data.

The best way to proceed is to select those processes 
with smaller gluonic corrections, where we have some chance to control the
QCD effects.
Let us consider the semileptonic weak decay
$H\to H' l^- \bar\nu_l$, associated with the corresponding quark transition
$d_j\to u_i l^-\bar\nu_l$.
Since quarks are confined within hadrons, the decay amplitude
\bel{eq:T_decay}
T[H\to H' l \bar\nu_l]\,\approx\, {G_F\over\sqrt{2}} \,\bV_{\!\! ij}\,
\,\langle H'| \bar u_i \gamma^\mu (1-\gamma_5) d_j | H\rangle \,\,
\bar l \gamma_\mu (1-\gamma_5) \nu_l
\ee
always involves an hadronic matrix element of the weak left current.
The evaluation of this matrix element is a non-perturbative QCD problem
and, therefore, introduces unavoidable theoretical uncertainties.

Usually, one looks for a semileptonic transition
where the matrix element can be fixed at some kinematical point, by
a symmetry principle (this will be discussed in the next section).
This has the virtue of reducing the theoretical
uncertainties to the level of symmetry--breaking corrections and
kinematical extrapolations.
The standard example is a $0^-\to 0^-$ decay such as $K\to\pi l\nu$,
$D\to K l\nu$ or $B\to D l\nu$.
Only the vector current can contribute in this case:
\bel{eq:vector_me}
\langle P'(k')| \bar u_i \gamma^\mu d_j | P(k)\rangle \, = \, C_{PP'}\,
\left\{ (k+k')^\mu f_+(q^2) + (k-k')^\mu f_-(q^2)\right\} .
\ee
Here, $C_{PP'}$ is a Clebsh--Gordan factor and $q^2=(k-k')^2$
the momentum transfer carried by the intermediate $W$.
The unknown strong dynamics is fully contained in the two form factors
$f_\pm(q^2)$.

Since
$(k-k')^\mu\,\bar l\gamma_\mu (1-\gamma_5)\nu_l\sim m_l$, the contribution
of $f_-(q^2)$ is kinematically suppressed in the $e$ and $\mu$ modes.
Moreover, since $f_-(q^2)\approx (m_{u_i}-m_{d_j})$ 
[see Eq.~\eqn{eq:der_curr}],
there is an additional strong suppression of the $f_-(q^2)$ term
in the light--flavour case.
The decay width can then be written as
\bel{eq:decay_width}
\Gamma(P\to P' l \nu) 
\, \approx\, {G_F^2 M_P^5\over 192\pi^3}\, |\bV_{\!\! ij}|^2\, C_{PP'}^2\,
|f_+(0)|^2\, \cI\, \left(1+\delta_{\mbox{\rms RC}}\right) ,
\ee
where $\delta_{\mbox{\rms RC}}$ is an electroweak radiative
correction factor and $\cI$ denotes a phase--space integral, 
which in the $m_l=0$ limit takes
the form
\bel{eq:ps_integral}
\cI\,\approx\,\int_0^{(M_P-M_{P'})^2}\! {dq^2\over M_P^8}\;
\lambda^{3/2}(q^2,M_P^2,M_{P'}^2)\, 
\left| {f_+(q^2)\over f_+(0)}\right|^2 .
\ee

The usual procedure to determine $|\bV_{\!\! ij}|$ involves three steps:
\begin{enumerate}
\item Measure the shape of the $q^2$ distribution. This fixes the ratio
$|f_+(q^2)/f_+(0)|$ and therefore determines $\cI$.
\item Measure the total decay width $\Gamma$. Since $G_F$ is already known
from $\mu$ decay, one gets then an
experimental value for the product $|f_+(0)|\, |\bV_{\!\! ij}|$.
\item Get a theoretical prediction for the normalization $f_+(0)$.
\end{enumerate}
The important point to realize is that theoretical input is
always needed. Thus, the accuracy of the $|\bV_{\!\! ij}|$
determination is limited by our ability to  calculate the relevant
hadronic input.

\section{Conserved Vector Current}
\label{sec:cvc}\setcounter{equation}{0}

Symmetries are a powerful tool to
derive general constraints, without entering into
the detailed dynamics.
Since we have not been able to solve QCD  in the
difficult non-perturbative regime, we would like
to get at least some handle on the strong interactions
through symmetry considerations.

To simplify the discussion, let us consider
the electromagnetic interaction of a fermion with
charge $Q$:
\bel{eq:QED}
\cL_{\mbox{\rms QED}} = 
 i \,\overline{\Psi}(x)\gamma^\mu\partial_\mu\Psi(x)
\, - \, m\, \overline{\Psi}(x)\Psi(x) 
+  e Q \, A_\mu(x) \,
\overline{\Psi}(x)\gamma^\mu\Psi(x) \, .
\ee
The strength of the QED interaction is
proportional to the electric charge; i.e.,
$Q_u=2/3$, $Q_d=-1/3$ and $Q_e=-1$.
What is not so trivial is the fact that 
the hadronic charges are exactly equal to the
sum of charges of their constituents quarks.
For instance, the electric charge of the
proton is just given by 
$Q_p = 2 Q_u + Q_d = 1$.
That protons and electrons have the same
(up to a sign) charge has been experimentally verify
to a very good precision: \cite{pdg:94}
\bel{eq:charges}
\left|{Q_p\over Q_e}\right| - 1 < 1.0\times 10^{-21}\, .
\ee
However, one would naively expect that the interaction 
of the photon with a bound hadronic state such as the proton 
would be quite different from the one with an elementary electron. 
Fig.~\ref{fig:proton_charge} shows the effective
photon couplings of the proton and the electron,
once higher--order quantum corrections are taken into account.
While the electron vertex only gets higher--order QED contributions,
the quark constituents within the proton are affected by all kinds
of gluonic exchanges.
The fact that all these complicated QCD interactions finally
reduce to $Q_p=-Q_e$ looks somewhat miraculous.

%
\begin{figure}[htb]
\centering       
\epsfig{file=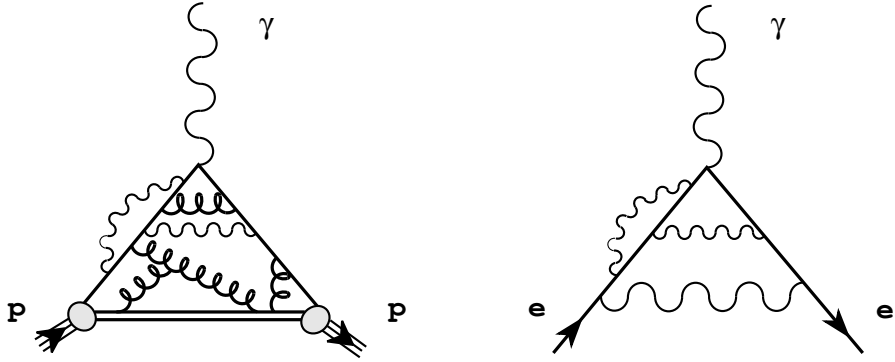,height=5cm}
\caption{Electromagnetic couplings of the proton and the electron.}
\label{fig:proton_charge}
\end{figure}

The quantum loop corrections generate an electromagnetic form
factor; i.e., they change the tree--level coupling
$eQ$ to $e Q\, F_1(q^2)$, where $q^2$ is the squared
quadrimomentum of the photon.
The explicit calculation shows that the tree--level coupling
of an {\it on-shell} photon does not get modified by higher--order
contributions.
Although individual Feynman diagrams do generate non-zero
corrections to $F_1(0)$, a complete cancellation occurs among the
different contributions in such a way that the
tree--level result $F_1(0)=1$ is recovered.
This is why the electric charges of the electron and the proton
are finally the same (in modulus), in spite of the very different
dynamics involved.

Magic cancellations of this kind usually originate from some
dynamical symmetry. In QED, the symmetry at work 
is the invariance under phase redefinitions of the fermion
fields:
\bel{eq:global}
\Psi(x) \quad\toU\quad \Psi'(x)\,\equiv\,\exp{\{i Q \theta\}}\,\Psi(x) \, ,
\ee
where $\theta$ is an arbitrary real constant.

A global symmetry of the Lagrangian always implies \cite{noether}
the existence of a {\it conserved current}, satisfying
$\partial_\mu J^\mu = 0$,  
and an associated {\it conserved charge}, 
$\cQ(t) \equiv\int d^3x\, J^0(\vec{x},t)$,
which  is a constant of motion, i.e. $d\cQ(t)/dt = 0$. 
In the QED case, the conserved quantities are the electromagnetic 
vector current,
\bel{eq:current}
J^\mu_{\mbox{\rms em}} =  Q	\,\overline{\Psi}(x)\gamma^\mu \Psi(x) \, ,
\ee
and the associated electromagnetic charge
\bel{eq:charge}
\cQ_{\mbox{\rms em}}(t) \equiv\int d^3x\, 
J^0_{\mbox{\rms em}}(\vec{x},t) = 
Q \,\int d^3x\, |\Psi(x)|^2 = Q \, . 
\ee

It is very easy to see how the conservation of the electromagnetic
current guarantees that $F_1(0)=1$, both for the electron and the
proton.
In the proton case, the photon couples to the electromagnetic
quark current
$J^\mu_{\mbox{\rms em}} = \sum_q Q_q\, \bar q \gamma^\mu q$.
The proton  electromagnetic vertex is then given by
the hadronic matrix element
\beqn\label{eq:proton_em}
\langle p(k')\, |\, J^\mu_{\mbox{\rms em}}(x)\, |\, p(k)\rangle 
&\!\! =&\!\!
\langle p(k')\, |\,\e^{iPx}\, J^\mu_{\mbox{\rms em}}(0)\,\,\e^{-iPx}\, 
|\, p(k)\rangle
\no\\ &\!\! =&\!\!
\e^{-iqx}\, Q_p\; \bar p(k') \left[ F_1(q^2) \,\gamma^\mu
- i F_2(q^2) \,\sigma^{\mu\nu} q_\nu \right] p(k) \, ,\qquad
\eeqn
where $q^\mu=(k-k')^\mu$ is the photon momentum,
$p(k)$ and $\bar p(k')$ the Dirac spinors of the incoming and
outgoing protons,
and $Q_p =\sum_q Q_q$ the conserved proton charge.
We have first used translation invariance to extract the $x$--dependence,
and then a general Lorentz decomposition of the resulting amplitude,
obeying the current conservation constraint
$q_\mu J^\mu_{\mbox{\rms em}} = 0$,
has been performed. Thus, the strong dynamics
is parametrized by two form factors.
Taking $\mu=0$ and integrating over $d^3x$, one has
\beqn\label{eq:F1=1}
\langle p(k')\, |\, \cQ_{\mbox{\rms em}}(t)\, |\, p(k)\rangle 
&\!\! =&\!\!
Q_p \,\langle p(k')\, |\, p(k)\rangle 
\no\\ &\!\! =&\!\!
\left[ (2\pi)^3\delta^{(3)}(q)\, p(k')^\dagger p(k)\right]
\, Q_p \, F_1(0) \, ,
\eeqn
where the first and second lines are obtained from the left- and 
right-hand sides of Eq.~\eqn{eq:proton_em}, respectively.
Since $\langle p(k')\, |\, p(k)\rangle =
(2\pi)^3\delta^{(3)}(q)\, p(k')^\dagger p(k)$, this
proofs the wanted result $F_1(0)=1$.
The same derivation applies to the electron case
(without the sum over constituent quarks!).

\subsection{Chiral Symmetry}

In the absence of quark masses, the QCD Lagrangian splits into
two independent quark sectors,
\bel{eq:LR_sectors}
\cL_{\mbox{\rms QCD}}^{(m=0)}
\,=\, -{1\over 4}\, G^{\mu\nu}_aG_{\mu\nu}^a
+ i\bar q_L \gamma^\mu D_\mu q_L  +
i\bar q_R \gamma^\mu D_\mu q_R \ .
\ee
Here, $q$ denotes the flavour (and colour) vector \ 
$q = \mbox{\rm column}(u,d,\ldots)$,  and
$D_\mu q$ the corresponding QCD covariant derivative.
Thus, $\cL_{\mbox{\rms QCD}}^{(m=0)}$
is invariant under
independent {\it global} $G\equiv SU(N_f)_L\otimes SU(N_f)_R$
transformations of the left- and
right-handed quarks in flavour space,
where $N_f$ denotes the number of quark flavours:
\bel{eq:chiral_sym}
q_L \, \toG \, g_L \, q_L \, , \qquad\qquad
q_R \, \toG \, g_R \, q_R \, , \qquad\qquad
g_{L,R} \in SU(N_f)_{L,R} \, .
\ee
The associated conserved currents are
$L^\mu = V^\mu - A^\mu$ and $R^\mu = V^\mu + A^\mu$, with
\bel{eq:V_A_currents}
V^\mu_{ij} \equiv \bar q_j\gamma^\mu q_i \, , \qquad\qquad
A^\mu_{ij} \equiv \bar q_j\gamma^\mu\gamma_5 q_i \, ,
\ee
the $SU(N_f)$ multiplets of vector and axial currents.
All these currents would be conserved in a massless quark world.
However, the chiral symmetry is explicitly broken by the quark 
mass term which communicates the left- and right-handed sectors
[$\cL_{\cM} = -(\bar q_R \cM q_L + \bar q_L \cM^\dagger q_R)$].
The current divergences can be easily obtained,
using the QCD equations of motion:
\bel{eq:der_curr}
\partial_\mu V^\mu_{ij} = i\, (m_{q_j}-m_{q_i}) \,\, \bar q_j q_i 
\, , \qquad\qquad
\partial_\mu A^\mu_{ij} = i\, (m_{q_j}+m_{q_i})\,\,\bar q_j\gamma_5 q_i 
\, .
\ee
Notice, that the vector currents are still conserved for non-zero but
equal quark masses [$SU(N_f)_V$ symmetry].

The light quark masses ($m_u$, $m_d$, $m_s$) 
are quite small compared with a typical hadronic scale of about
1 GeV.
We have then an approximate
$SU(3)_L\otimes SU(3)_R$
symmetry, leading to useful constraints \cite{sorrento:94,chpt:95,EC:95}. 
For instance, adapting the derivation given before for the
electromagnetic case, it is straightforward to proof that 
the conservation of the vector current implies:
\bel{eq:CVC}
\langle p\, | \,\bar u\gamma^\mu d\, |\, n\rangle \, = \,
\bar p \gamma^\mu n \qquad\qquad (q^2=0) \, .
\ee
Thus, in the isospin limit ($m_u=m_d$), strong interactions do not
change the normalization of this hadronic matrix element at $q^2=0$.

A similar statement does not hold for the axial currents because
chirality is not respected by the QCD vacuum
[$\langle\emptyset| \bar q q | \emptyset\rangle =
\langle\emptyset| (\bar q_L q_R + \bar q_R q_L) | \emptyset\rangle
\not= 0$].
The chiral symmetry of the Lagrangian is spontaneously broken to its
vectorial subgroup:
\bel{eq:SSB}
SU(3)_L\otimes SU(3)_R \ssb SU(3)_V \, ,
\ee
and, according to Goldstone's theorem \cite{GO:61},
an octet of massless pseudoscalars ($\pi$, $K$, $\eta$) appears
in the hadronic spectrum.
The Goldstone nature of the pseudoscalar octet leads to many
interesting implications, which go beyond the scope of these lectures  
(a detailed discussion can be found in Refs.~\citenum{sorrento:94},
\citenum{chpt:95} and \citenum{EC:95}).
For our present purposes, the relevant thing is that the massless
pion couples to the axial current, giving rise to a pole at
$q^2=0$ which modifies the free--quark normalization:
\bel{eq:PCAC}
\langle p\, | \,\bar u\gamma^\mu\gamma_5 d\, |\, n\rangle \, = \,
g_A \;\bar p \gamma^\mu n \qquad\qquad (q^2=0) \, ,
\ee
where \cite{pdg:94} $g_A = 1.2573\pm 0.0028 \not= 1$.

\section{Determination of the CKM mixings for light quarks}
\label{sec:CKM_light}\setcounter{equation}{0}

The previous two sections have provided all the needed ingredients
to allow an accurate determination of the CKM mixings among the
up, down and strange quarks.

\subsection{V$_{ud}$}
\label{subsec:Vud}

The most accurate measurement of $\bV_{\!\! ud}$ is done with
superallowed nuclear $\beta$ decays of the Fermi type [$0^+\to 0^+$],
where the nuclear matrix element 
$\langle N'|\bar u\gamma^\mu d|N\rangle$
can be fixed by vector--current conservation.
The CKM factor is obtained through the relation \cite{MA:91},
\bel{eq:ft_value}
|\bV_{\!\! ud}|^2\, =\, 
{\pi^3\ln{2}\over ft\, G_F^2 m_e^5\, (1+\delta_{\mbox{\rms RC}})}
\, = \, {(2984.4\pm 0.1)\,\mbox{\rm s}\over ft\,(1+\delta_{\mbox{\rms RC}})} ,
\ee
where the factor $ft$ denotes a {\it comparative half--life} corrected
for phase--space and Coulomb effects \cite{HA:90,TO:92,BA:92}.
In order to obtain $|\bV_{\!\! ud}|$, one needs to perform a careful 
analysis
of radiative corrections \cite{MS:86,SZ:86,SI:87,JR:87},
including both short--distance contributions
$\Delta_{\mbox{\rms inner}}=0.0234\pm 0.0012$, and 
nucleus--dependent corrections $\Delta_{\mbox{\rms outer}}\equiv
\delta_{\mbox{\rms RC}}-\Delta_{\mbox{\rms inner}}$.
These radiative corrections are quite large,
$\delta_{\mbox{\rms RC}}\sim 3$--4\%, and have a crucial role in order
to bring the results from different nuclei into good agreement.
Table ~\ref{tab:Vud} shows the values of $|\bV_{\!\! ud}|$ obtained
from various superallowed $\beta$ transitions.
The final result quoted by the Particle Data Group \cite{pdg:94} is
\bel{eq:Vud}
|\bV_{\!\! ud}|\, =\, 0.9736\pm 0.0010 \, .
\ee
%

\begin{table}[bht]
\centering
\caption{$Ft\equiv ft \, (1+\Delta_{\mbox{\protect\rms outer}})$ and
$|\protect\bV_{\!\! ud}|$ for various superallowed $\beta$ decays.
\protect\cite{MA:91}}
\label{tab:Vud}
\vspace{0.2cm}
\begin{tabular}{|c|c|c|}
\hline Nucleus & $Ft$ & $|\bV_{\!\! ud}|$ \\ \hline
${}^{14}$O & $3067.9\pm 2.4$ s & $0.9750\pm 0.0007$ \\
${}^{26m}$Al & $3071.1\pm 2.6$ s & $0.9744\pm 0.0007$ \\
${}^{34}$Cl & $3074.2\pm 3.1$ s & $0.9740\pm 0.0008$ \\
${}^{38m}$K & $3071.5\pm 3.2$ s & $0.9744\pm 0.0008$ \\
${}^{42}$Sc & $3077.1\pm 2.9$ s & $0.9735\pm 0.0008$ \\
${}^{46}$V & $3078.7\pm 3.2$ s & $0.9732\pm 0.0008$ \\
${}^{50}$Mn & $3073.2\pm 5.2$ s & $0.9741\pm 0.0010$ \\
${}^{54}$Co & $3075.1\pm 3.7$ s & $0.9738\pm 0.0008$ \\
\hline
\end{tabular}
\end{table}

An independent determination can be obtained from neutron decay,
$n\to p\, e^-\bar\nu_e$.
The axial current also contributes in this case; therefore, one needs
to use the experimental value of the axial coupling $g_A$.
The measured neutron lifetime \cite{pdg:94},
$\tau_n = 887.0\pm 2.0$~s, implies \cite{MA:91}:
\bel{eq:n_decay}
|\bV_{\!\! ud}|\, =\, 
\left\{ {(4904.0\pm 5.0)\,\mbox{\rm s}\over \tau_n\, (1+3 g_A^2)}\right\}^{1/2}
\, =\, 0.981\pm 0.002 \, ,
\ee
which is bigger than 
\eqn{eq:Vud}. 
Thus, a better measurement of $g_A$ and $\tau_n$ is needed.

The pion $\beta$ decay $\pi^+\to\pi^0 e^+\nu_e$ offers a 
cleaner way to measure
$|\bV_{\!\! ud}|$. It is a pure vector transition, with very small theoretical
uncertainties. At $q^2=0$, the relevant hadronic matrix element
does not receive \cite{GL:84} isospin breaking contributions of first order in
$m_d-m_u$; i.e.,
$f_+(0)= 1 + \cO[(m_d-m_u)^2]$.
Moreover, the small available phase--space makes it possible to theoretically
control the form factor with high accuracy over the entire kinematical
domain.
Unfortunately, owing to the kinematical suppression, this decay mode
has a small branching fraction.
The present experimental value is not very precise,
Br$(\pi^+\to\pi^0 e^+\nu_e)=(1.025\pm 0.034)\times 10^{-8}$; it implies
$|\bV_{\!\! ud}| = 0.968\pm 0.018$.
An accurate measurement of this transition would be very valuable.

\goodbreak

\mbox{}\vspace*{-1.1cm}

\subsection{V$_{us}$}
\label{subsec:Vus}

The decays $K^+\to\pi^0 l^+\nu_l$ and $K^0\to\pi^-l^+\nu_l$ 
are ideal for measuring $|\bV_{\!\! us}|$, because the relevant hadronic
form factors are well understood. 
$SU(3)$--breaking corrections are very suppressed
and isospin violations
can be easily taken into account.
For $K^0\to\pi^-l^+\nu_l$,
the $SU(3)$ symmetry relation $f_+(0)=1$ does not get any correction
linear in the quark mass differences  \cite{AG:64,BS:60}.
The $K^+\to\pi^0 l^+\nu_l$ form factor gets, however, a calculable
contribution proportional to $(m_d-m_u)$, induced by $\pi^0$--$\eta$
mixing: \cite{chpt:95}
\bel{eq:f_chpt}
f_+^{K^0\pi^-}(0) = 1 + \cO[(m_s-m_u)^2]\, ; \qquad 
f_+^{K^+\pi^0}(0) = 1 + {3 (m_d-m_u)\over 4m_s-2m_u-2m_d} +\cdots
\ee
Using Chiral Perturbation Theory methods \cite{chpt:95,EC:95},
the leading higher--order corrections to \eqn{eq:f_chpt}
and the low--momentum behaviour of the $f_\pm(q^2)$ form factors
can be rigorously computed.
The resulting  values \cite{GL:85},
\bel{eq:fp_chpt}
f_+^{K^0\pi^-}(0)=0.977 \, , \qquad\quad
f_+^{K^+\pi^0}(0)/f_+^{K^0\pi^-}(0)=1.022 \, , 
\ee
should be compared with
the experimental ratio \cite{pdg:94}
$|f_+^{K^+\pi^0}(0)/f_+^{K^0\pi^-}(0)|=1.028\pm 0.010$.
The accurate calculation of these quantities allows to extract
\cite{LR:84} a precise determination of $|\bV_{\!\! us}|$:
\bel{eq:Vus}
|\bV_{\!\! us}|\, =\, 0.2196\pm 0.0023 \, .
\ee

The analysis of semileptonic hyperon decay data can also
provide information on $|\bV_{\!\! us}|$. However,  the theoretical
uncertainties are larger, owing to the first--order
$SU(3)$--breaking effects in the axial--vector couplings \cite{DHK:87}.
The Particle Data Group \cite{pdg:94} quotes the result
$|\bV_{\!\! us}|\, =\, 0.222\pm 0.003$.
The average with \eqn{eq:Vus} gives the final value:
\bel{eq:V_us}
|\bV_{\!\! us}|\, =\, 0.2205\pm 0.0018 \, .
\ee

\section{V$_{\! cd}$ and V$_{\! cs}$}
\label{sec:Vcd}\setcounter{equation}{0}

$|\bV_{\!\! cd}|$ is deduced
from deep inelastic $\nu_\mu$ and $\bar\nu_\mu$ scattering data,
by measuring the dimuon production rates off valence $d$ quarks;
i.e., $\nu_\mu d\to\mu^- c$ with the charm quark detected
through $c\to \mu^+\nu_\mu d$ or $\mu^+\nu_\mu s$.
One gets in this way, the product \cite{pdg:94}
$\overline{B}_c\,|\bV_{\!\! cd}|^2 = (0.49\pm 0.05)\times 10^{-2}$,
where $\overline{B}_c$ is the average semileptonic branching fraction
of the produced charmed hadrons.
Using  \cite{pdg:94} $\overline{B}_c=0.099\pm 0.012$, yields
\bel{eq:Vcd}
|\bV_{\!\! cd}|\, =\, 0.224\pm 0.016 \, .
\ee

Similarly, one could extract $|\bV_{\!\! cs}|$ from 
$\nu_\mu s\to\mu^- c$ data. The resulting values depend, however, 
on assumptions about the strange quark density in the parton sea.
Assuming that the strange quark sea does not
exceed the value corresponding to an $SU(3)$ symmetric sea, leads
to the conservative lower bound  \cite{AB:82} $|\bV_{\!\! cs}|>0.59$.

Better information is obtained from the decays $D\to\bar K l^+\nu_l$.
The measured $q^2$ distribution \cite{pdg:94} 
can be fitted with the pole parametrization
$f_+^D(q^2)/f_+^D(0) = M^2/(M^2-q^2)$ and $M=M_{D^*}\approx 2.1$ GeV,
which corresponds to the assumption that the form factor is dominated by
the lightest intermediate meson with the right quantum numbers.
This determines the corresponding integral $\cI$, implying
\bel{eq:D_K_width}
\Gamma(D\to\bar K l^+\nu_l)\, = \, |f_+^D(0)|^2\, |\bV_{\!\! cs}|^2 \;
(1.54\times 10^{11}\: s^{-1}) \, .
\ee
Using \cite{RB:95}
$\Gamma(D\to\bar K l^+\nu_l)=(8.4\pm 0.4)\times 10^{10}$
s$^{-1}$, one gets then:
\bel{eq:Vcs}
|f_+^D(0)|\, |\bV_{\!\! cs}|\, =\, 0.74\pm 0.02 \pm 0.02\, ,
\ee
where the second error is from the uncertainty in the $q^2$ dependence
\cite{pdg:94}.

%
\begin{figure}[htb]
\centering       
\epsfig{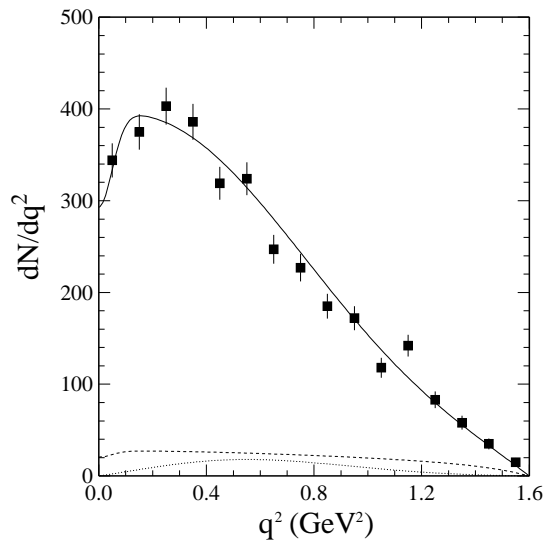}
\caption{Measured $q^2$--distribution for $D\to\bar K l^+\nu_l$
(CLEO II \protect\cite{cleo:93}).
The solid curve is a fit to the data with a pole form factor, which gives
$M=2.00\pm 0.12 \pm 0.18$.}
\label{fig:D_K_decay}
\end{figure}
%

The status of our theoretical understanding of charm form factors
is quite crude. The symmetry arguments are not very helpful here,
because the charm--quark mass is too heavy for using the $SU(4)$
limit, and, at the same time, is too light to obtain accurate
results from the opposite limit $m_c\to\infty$.
Symmetry--breaking corrections are very important.
The conservative assumption $|f_+^D(0)|<1$, implies
$|\bV_{\!\! cs}|>0.66$.
Taking the range
$|f_+^D(0)|= 0.75\pm 0.15$, which covers the main part of the existing
calculations, one gets:
\bel{eq:V_cs}
|\bV_{\!\! cs}|\, =\, 0.99\pm 0.20 \, .
\ee

Theoretical uncertainties are largely avoided by taking 
decay--width ratios,
such as $\Gamma(D\to\pi l^+\nu_l)/\Gamma(D\to \bar K l^+\nu_l)$,
where the form--factor uncertainty is reduced to the level of
$SU(3)$ breaking. The recent CLEO II measurements \cite{cleo:93b} give
\bel{eq:cleo_ratio}
\left|{f_+^{D\pi}(0)\over f_+^{DK}(0)}\right|
\left|{\bV_{\!\! cd}\over\bV_{\!\! cs}}\right|
= 0.25\pm 0.03 \, .
\ee
Taking the conservative range
$f_+^{D\pi}(0)/ f_+^{DK}(0) = 1.0\pm 0.2$, this implies
\bel{eq:cd_cs}
\left|{\bV_{\!\! cd}\over\bV_{\!\! cs}}\right|
= 0.25\pm 0.06 \, .
\ee

The present measurements of $D\to\pi l^+\nu_l$
and $D\to \bar K l^+\nu_l$ 
are still too poor to provide an accurate determination
of the CKM factors. 
A 1\% measurement
of these semileptonic ratios seems possible \cite{marbella:93} at a future
tau--charm factory; this would allow a precise determination of
$|\bV_{\!\! cd}|/|\bV_{\!\! cs}|$.
For this to be the case, however, a better theoretical
understanding of $SU(3)$--breaking effects is mandatory.
The prospects for extracting the absolute values of
$|\bV_{\!\! cd}|$ and $|\bV_{\!\! cs}|$ with a similar
accuracy are not so good; we need first to improve  in a 
significative way our ability to control QCD effects.

\section{Heavy Quark Symmetry}
\label{eq:HQET}\setcounter{equation}{0}

The chiral symmetries of massless QCD are not relevant for
heavy quarks. This is the reason why we have not been able to
pin down the charm CKM factors with a precision comparable to
the one achieved for light quarks.
There is, however, another approximate limit of QCD
which turns out to be rather useful: the infinite--mass limit.

The dynamical simplifications which occur in the heavy--mass
limit can be easily understood by looking back to the more
familiar atomic physics.
The quantum mechanical properties of an electron in the
Coulomb potential of an atomic nucleus
are regulated by the reduced mass
$m_e M/(m_e + M)\approx m_e << M$, where $M$ is the heavy
nuclear mass.
Therefore, different isotopes ($M\not=M'$) of the same
atom ($Z=Z'$) have the same chemical properties to a very
good approximation (isotopic symmetry).
Moreover, atoms with nuclear spin $S$ are $(2S+1)$ degenerate,
in the limit $M\to\infty$ (spin symmetry).

The QCD analog is slightly more complicated, but the general idea
is the same.
The quarks confined inside hadrons exchange momentum of a magnitude
of about $\Lambda\sim M_p/3 \approx 300$ MeV. The scale $\Lambda$
characterize the typical amount  by which quarks are off-shell;
it also determines the  hadronic size
$R_{\mbox{\rms had}}\sim 1/\Lambda$.
If we consider a {\it heavy--light} hadron composed of one heavy quark $Q$
and any
number of light constituents, the light quark(s) is (are) very far
off-shell by an amount of order $\Lambda$.
However, if $M_Q>>\Lambda$, the heavy quark is almost on-shell
and its Compton wavelength $\lambda_Q\sim 1/M_Q$
is much smaller than the hadronic size $R_{\mbox{\rms had}}$.

Although the quark interactions change the momentum
of $Q$ by $\delta P_Q\sim\Lambda$, its velocity only changes
by a negligible amount,
$\delta v_Q\sim\Lambda/M_Q << 1$.
Thus, $Q$ moves approximately with constant velocity.
 In the hadron rest frame, the heavy quark is almost at rest
and acts as a static source of gluons. It is surrounded by a 
complicated, strongly interacting cloud of light quarks,
antiquarks and gluons, sometimes referred to as the
{\it brown muck}.
To resolve the quantum numbers of the heavy quark would
require a hard probe with $Q^2\geq M_Q^2$; however,
the soft gluons coupled to the {\it brown muck} can only resolve
larger distances of order $R_{\mbox{\rms had}}$.
The light hadronic constituents are blind to the
flavour and spin orientation of the heavy quark;
they only feel its colour field which extends over large distances
because of confinement.
Thus, in the infinite $M_Q$ limit,
the properties of heavy--light hadrons are independent
of the mass ({\it flavour} symmetry) and spin 
({\it spin} symmetry) of the heavy source of colour \cite{IW:89}.

In order to put this qualitative arguments 
within a more formal framework, let us write the heavy quark
momentum as
\bel{eq:P_Q}
P_Q^\mu \equiv M_Q v^\mu + k^\mu \, ,
\ee
where $v^\mu$ is the hadron four-velocity 
($v^2=1$) and $k^\mu$ the
{\it residual} momentum of order $\Lambda$.
In the limit $M_Q\to\infty$ with $v^\mu$ kept fixed
\cite{IW:89},
 the QCD Feynman rules simplify
considerably \cite{GR:90}. The heavy quark propagator becomes
\bel{eq:Q_propagator}
{i\over \slashchar{P}_Q - M_Q} \, = \, {i\over v\cdot k}\, 
{1+ \slashchar{v}\over 2} + \cO(k/M_Q) \, .
\ee
The factors $P_\pm\equiv (1\pm \slashchar{v})/2$ are
energy projectors ($P_\pm^2=P_\pm$, $P_\pm P_\mp =0$).
Thus, the propagator is independent of $M_Q$ and only the 
positive energy projection of the
heavy quark field propagates. Moreover, since
$P_+\gamma^\mu P_+ = P_+ v^\mu P_+$, the quark--gluon
vertex reduces to
\bel{eq:Q_vertex}
i g \left({\lambda^a\over 2}\right) \gamma^\mu
\quad\longrightarrow\quad
i g \left({\lambda^a\over 2}\right) v^\mu \, .
\ee
The resulting interaction is then independent of the heavy--quark
spin.

These Feynman rules can be easily incorporated into an effective
Lagrangian. Making  the field redefinition
\bel{eq:h_def}
Q(x)\approx \e^{-iM_Q v\cdot x}\, h_v^{(Q)}(x) \, ,
\ee
where 
$h_v^{(Q)}=P_+ h_v^{(Q)} = \slashchar{v} h_v^{(Q)}$
(i.e., we are only considering the positive--energy projection
of the heavy--quark spinor),
the heavy--quark Lagrangian becomes \cite{EH:90,GE:90}
\bel{eq:L_Q}
\cL_{\mbox{\rms QCD}}^{(Q)} \, =\, \bar Q \, (i\,\slashchar{D} - M_Q)\, Q
\,\approx\, \bar h_v^{(Q)} \, i\, (v\cdot D) h_v^{(Q)} \, ,
\ee
showing explicitly that the interaction is independent of
the mass and spin of the heavy quark.
The corresponding equation of motion is:
\bel{eq:motion}
i \,\slashchar{D} Q = M_Q Q \quad\longrightarrow\quad
i \, (v\cdot D) h_v^{(Q)} =0 \, .
\ee
The phase factor in \eqn{eq:h_def} has removed the {\it kinetic}
piece $M_Q v^\mu$ from the heavy quark momentum, so that in 
momentum space a derivative acting on $h_v^{(Q)}$ just
produces the {\it residual} momentum $k^\mu$.
Notice that $h_v^{(Q)}$ is a two--component spinor,
which destroys a quark $Q$ but does not create the corresponding
antiquark; pair creation does not occur in the
heavy quark effective theory (HQET).

\subsection{Spectroscopic Implications}

Let us denote $s_l$ the total spin of the light degrees of
freedom in a hadron containing a single heavy quark $Q$.
In the $M_Q\to\infty$ limit, the dynamics is independent of
the heavy--quark spin. Therefore, there will be two
degenerate hadronic states with $J=s_l\pm\frac{1}{2}$.
For $Q\bar q$ mesons the ground state has negative parity
and $s_l=1/2$, giving a doublet  of degenerate spin--zero and
spin--one mesons.
The measured charm and bottom spectrum shows indeed that
this is true to a quite good approximation: \cite{pdg:94}
\bel{D_B_spectrum}\begin{array}{lll}
M_{D^*} - M_D = (142.12\pm0.07)\,\mbox{\rm MeV} , &\qquad &
(M_{D^*} - M_D)/ M_D \approx 8\%  ,
\\
M_{B^*} - M_B = (46.0\pm 0.6)\,\mbox{\rm MeV} , &&
(M_{B^*} - M_B)/ M_B \approx 0.9\%  .
\ea\ee
The infinite--mass limit works much better for the
bottom, although the result is also good in the charm case.
We expect these mass splittings to get corrections of the
form
$M_{P^*}-M_P\approx a/M_Q$; this gives
the refined prediction
$M_{B^*}^2 - M_B^2 \approx M_{D^*}^2 - M_D^2$,
which is in very good agreement with the data: \cite{pdg:94}
\bel{eq:D_B_masses}
M_{D^*}^2 - M_D^2 \approx 0.53 \:\mbox{\rm GeV}^2 , \qquad\quad
M_{B^*}^2 - M_B^2 \approx 0.49 \:\mbox{\rm GeV}^2 .
\ee

\subsection{Weak Decay Form Factors}

Let us consider the semileptonic decay $B\to D l \bar\nu_l$. The decay
amplitude involves the hadronic matrix element
$\langle D | \bar c\gamma^\mu b | B\rangle$,
which depends on two general form factors $f_+(q^2)$ and $f_-(q^2)$
[see Eq.~\eqn{eq:vector_me}].
If the masses of the bottom and charm quarks are taken to be heavy, we can use
the HQET formalism to analyze this matrix element.
It is convenient to work with a mass--independent normalization for the
meson states; i.e., to redefine the hadronic states as
\bel{eq:had_def}
|\widetilde M(v)\rangle \equiv {1\over \sqrt{M_P}} \, |M(p)\rangle \, ,
\ee
with the normalization
$\langle\widetilde M(v')|\widetilde M(v)\rangle = 2 v^0 (2\pi)^3
\delta^{(3)}(\vec{p}-\vec{p}^{\,\prime})$.

In the heavy--quark theory, the 
wanted matrix element of the vector current takes the form \cite{IW:89}
\bel{eq:V_HQET}
\langle \widetilde D(v_D) |\, \bar h^{(c)}_{v_D} \gamma^\mu h^{(b)}_{v_B} 
\, | \widetilde B(v_B) \rangle
\, =\, \xi(v_D\!\cdot\! v_B)\; (v_D+ v_B)^\mu \, ,
\ee
where $\xi(v_D\!\cdot\! v_B)$ is an unknown form factor. 
That there is no term proportional to $(v_D - v_B)^\mu$
can be seen by contracting the matrix element with $(v_D - v_B)^\mu$
and using $\slashchar{v}_B h^{(b)}_{v_B} = h^{(b)}_{v_B}$
and $\bar h^{(c)}_{v_D} \slashchar{v}_D = \bar h^{(c)}_{v_D}$.
Thus, the non-perturbative problem has been reduced to a single form factor
which only depends on the relative velocity
$(v_B-v_D)^2 = 2 (1 - v_B\!\cdot\! v_D)$.

The physical picture behind \eqn{eq:V_HQET} is quite easy to understand.
The $B\to D$ transition is induced by the action of an external vector
current coupled to the heavy quark. Before the action of the current,
the non-perturbative {\it brown muck} orbits around the heavy quark $b$ which
acts as a (static in the rest frame)
colour source; the whole system moves with a velocity $v_B$.
The effect of the current is to replace instantaneously the quark $b$ by
a quark $c$ moving with velocity $v_D$. If $v_B=v_D$ nothing happens;
the light quarks are unable to
realize that a heavy--quark transition has taken place,
because the interaction is flavour independent. However,
if $v_B\not=v_D$ the {\it brown muck} suddenly feels itself interacting
with a moving coulour source. The soft--gluon exchanges needed to rearrange
the light degrees of freedom into a final meson moving with velocity
$v_D$ generate a form factor suppression $\xi(v_D\!\cdot\! v_B)$,
which can only depend on the Lorentz boost $\gamma = v_D\!\cdot\! v_B$
connecting the rest frames of the initial and final mesons.
The flavour symmetry guarantees that this form factor is a universal
function independent of the heavy mass (i.e., it is the same
for $B\to B$, and $B\to D$ transitions).

When $v_B=v_D\equiv v$, the vector current 
$J^\mu = \bar h^{(c)}_{v_D} \gamma^\mu h^{(b)}_{v_B}=
\bar h^{(c)}_{v} v^\mu h^{(b)}_{v}$
is conserved:
\bel{eq:conv_j_v}
\partial_\mu J^\mu \, = \,\bar h^{(c)}_{v} (v\cdot D) h^{(b)}_{v} +
\bar h^{(c)}_{v} (v\cdot\stackrel{\leftarrow}{D}) h^{(b)}_{v} = 0 \, ,
\ee
since $(v\cdot D) h^{(c,b)}_{v}=0$ by the equations of motion.
The associated conserved charge
\bel{eq:H_charge}
N_{cb}\,\equiv\,\int d^3x\, J^0(x) \, =\,
\int d^3x\, \bar h^{(c)\dagger}_{v}\, h^{(b)}_{v}
\ee
is a generator of the flavour symmetry. Acting over a $B$ meson, it
replaces a quark $b$ by a quark $c$: \
$N_{cb} | \widetilde B(v) \rangle = | \widetilde D(v) \rangle$. 
Therefore, it satisfies
\bel{eq:N_norm}
\langle \widetilde D(v) | N_{cb} | \widetilde B(v) \rangle =
\langle \widetilde D(v) |  \widetilde D(v) \rangle =
2 v^0 (2\pi)^3 \delta^{(3)}(\vec{0}) \, .
\ee
Comparing with Eq.~\eqn{eq:V_HQET}
[the integration over $d^3x$ is the same as in Eq.~\eqn{eq:F1=1}],
one gets the important result:
\bel{eq:IW_norm}
\xi(1) = 1 \, .
\ee
This is the formal statement corresponding to the fact that the
{\it brown muck} does not feel any change if $v_D=v_B$.

Notice, that the light-- and heavy--quark symmetries
allow us to pin down the normalization of the corresponding
form factors at rather different kinematical points.
For massless (or equal--mass) quarks, the conservation of the vector current
fixes $f_+(q^2)$ at zero momentum transfer.
The heavy--quark limit, however, provides information on the point
of zero recoil for the $D$ meson. Since 
\bel{eq:vv_kinem}
v_B\!\cdot\! v_D = {M_B^2 + M_D^2 - q^2 \over 2 M_B M_D} \, ,
\ee
the equal velocity
regime corresponds to the maximum momentum transfer to the final leptons:
$q^2_{\mbox{\rms max}} = (M_B-M_D)^2$.

Up to now, we have only used the flavour symmetry associated with the
infinite--mass limit. There is in addition a useful spin symmetry
relating the $B\to D$ and $B\to D^*$ transitions.
Owing to the spin--1 character of the $D^*$, the decay $B\to D^*l\bar\nu_l$
gets contributions from both the vector and the axial--vector currents.
A general Lorentz parametrization would involve four independent
form factors: \cite{BSW:85}
\beqn\label{eq:BtoD*}
\lefteqn{\langle D^*(p')| \bar c \gamma^\mu (1-\gamma_5) b | B(p)\rangle
= 
{2i \over M_B + M_{D^*}} \varepsilon^{\mu\nu\alpha\beta}
\epsilon^*_\nu p'_\alpha p_\beta \, V(q^2)
 - 2 M_{D^*} {\epsilon^*\!\cdot\! q\over q^2} q^\mu\, A_0(q^2)
 }\no\\ && \!\!
    \mbox{} - (M_B + M_{D^*}) \epsilon^{*\mu}\, A_1(q^2) +
{\epsilon^*\!\cdot\! q \over M_B + M_{D^*}} (p + p')^\mu\, A_2(q^2)
+ 2 M_{D^*} {\epsilon^*\!\cdot\! q\over q^2} q^\mu\, A_3(q^2) \, , 
\no\\
\eeqn
where
\bel{eq:A_rel}
A_3(q^2) = {(M_B + M_{D^*}) \over 2 M_{D^*}}\, A_1(q^2) -
{(M_B - M_{D^*}) \over 2 M_{D^*}}\, A_2(q^2) \, ; \qquad
A_3(0) = A_0(0) \, .
\ee
In the infinite--mass limit,
this matrix element reduces to the simpler expression:\cite{IW:89}
\beqn\label{eq:BtoD*_HQET}
\langle \widetilde{D}^*(v')\, |\, \bar h^{(c)}_{v'} \gamma^\mu (1-\gamma_5)
 h^{(b)}_{v}\, | \,\widetilde{B}(v) \rangle 
&\!\! = &\!\!
i \varepsilon^{\mu\nu\alpha\beta} \epsilon^*_\nu 
v'_\alpha v_\beta\,\xi(v\!\cdot\! v')
\qquad\qquad \no\\ &\!\! &\!\!\!\!\!\!\!\!\!\!\!
\mbox{} -\left\{\epsilon^{*\mu}\, (1 +  v\!\cdot\! v') - 
v'^\mu (\epsilon^*\!\cdot\! v) \right\}\,\xi(v\!\cdot\! v')
\, , \qquad\quad
\eeqn
which depends on a single unknown form factor.
Moreover, this form factor is precisely the same appearing in \eqn{eq:V_HQET}.
The spin symmetry implies \cite{IW:89}
that all $B\to D$ and
$B\to D^*$ form factors are given
in terms of the universal function $\xi(v\!\cdot\! v')$.

The infinite--mass limit is the starting point for a systematic expansion
in powers of $E/M_Q$, which allows to analyze the properties of hadrons
containing a heavy quark. Further details on the HQET and many other
phenomenological applications can be found in Refs.~\citenum{GR:92}, 
\citenum{NE:94} and \citenum{WI:94}.


\section{CKM mixings of the $b$ quark}
\label{sec:CKM_b}\setcounter{equation}{0}
\vspace*{-0.6cm}
\subsection{V$_{cb}$}
\label{subsec:Vcb}

The cleanest determination of $|\bV_{\!\! cb}|$ uses the decay
$B\to D^* l \bar\nu_l$, 
where the relevant hadronic form factor
can be controlled at the level of a few per cent, close
to the zero--recoil region \cite{NE:91}.
The decay $B\to D^* l \bar\nu_l$ has the largest branching fraction of any
exclusive $B$ decay. In addition, the relevant kinematical variable
$v_B\!\cdot\! v_{D^*}$ can only vary over a small range, 1 to 1.5, where
the variation of the form factors is expected to be soft and HQET
techniques can be applied.
Compared with the analogous decay into a pseudoscalar meson,
$B\to D l \bar\nu_l$, the vector mode has two important advantages:
1) The $B\to D^*$ matrix element does not get any $1/M_Q$ 
correction  \cite{LU:90} at zero recoil;
corrections to the infinite--mass limit are then of order
$1/M_Q^2$.
2) Whereas $\Gamma(B\to D l \bar\nu_l)$ has a suppression factor
$|\vec{p}_D|^3$ at  $|\vec{p}_D|=0$ [see Eq.~\eqn{eq:ps_integral}],
such a suppression is not present in the $B\to D^* l \bar\nu_l$
decay mode.

The differential decay distribution is proportional to 
$|\bV_{\!\! cb}|^2\, |\cF(v_B\!\cdot\! v_{D^*})|^2$, where
the form factor $\cF(y)$ coincides with
$\xi(y)$, up to symmetry--breaking corrections of order
$\alpha_s(M_Q)$ and $\Lambda^2/M_Q^2$.
The calculated short--distance QCD corrections 
\cite{PG:83,CGP:84,VS:87,PW:88,FG:90,JM:91,BG:91,NE:92}
and the present estimates
of the $1/M_Q^2$ contributions \cite{FN:93,MA:94,SUV:95,NE:94b}
result in \cite{NE:95}
\bel{eq:F(1)}
\cF(1) = 0.91\pm 0.04\, .
\ee
%

%
\begin{figure}[htb]
\centering       
\epsfig{file=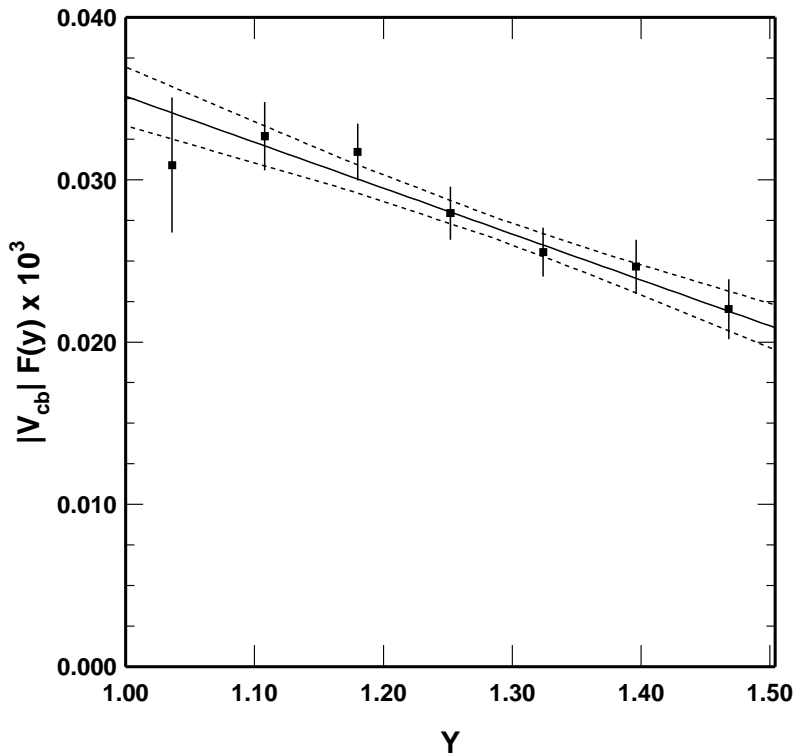,height=10cm}
\caption{Measured CLEO II distribution \protect\cite{BH:95,CLEO:95} of 
$|\protect\bV_{\!\! cb}|\,\cF(y)\times 10^3$ ($y\equiv v_B\!\cdot\! v_{D^*}$).
The curves show a linear fit to $\cF(y)$ and the
$\pm 1\sigma$ variations in the fit parameters.}
\label{fig:B_D*_distrib}
\end{figure}
%

The measurement of the $D^*$ recoil spectrum has been performed by several
experiments \cite{ARGUS:93,CLEO:95,ALEPH:95,DELPHI:95}.
Extrapolating the data to the zero--recoil point, the present world average
gives \cite{NE:95}
\bel{eq:F_V_exp}
|\bV_{\!\! cb}|\, |\cF(1)| \, = \, 
(35.1\pm 1.7\, {}^{+1.4}_{-0.0})\times 10^{-3}\, .
\ee
Together with \eqn{eq:F(1)}, this implies \cite{NE:95}
a quite accurate determination of $\bV_{\!\! cb}$:
\bel{eq:V_cb_excl}
|\bV_{\!\! cb}|\, = \, 
(38.6\, {{}^{+2.4}_{-1.9}}_{\mbox{\rms exp}}\pm 1.7_{\mbox{\rms th}})
\times 10^{-3} \, .
\ee

Assuming that the inclusive semileptonic decay width of a bottom hadron
is given by the corresponding quark decay $b\to c\, l^-\bar\nu_l$, the
magnitude of $|\bV_{\!\! cb}|$ can be also determined from the ratio of
the measured semileptonic branching ratio and lifetime.
However, since $\Gamma(b\to c\, l^-\bar\nu_l)\propto m_b^5 \, f(m_c^2/m_b^2)$, 
this method 
is very sensitive to the not so well--known values of the bottom and
charm masses.
The mass dependence becomes milder if one chooses $m_b$ and
$\Delta m\equiv m_b-m_c$ as independent variables  \cite{SUV:95}
(this has the advantage that $\Delta m$ can be better constrained with HQET
methods \cite{FN:93}). 
Nevertheless, the predicted semileptonic decay width gets 
a large uncertainty of about 11\% from this source \cite{NE:95}.
The perturbative QCD corrections,
which are exactly known to $\cO(\alpha_s)$ only, are rather sizeable
\cite{NI:89,LSW:95,BBB:95}.
In order to properly include the effect of higher--order QCD corrections,
a careful analysis of the quark--mass definition is mandatory \cite{BN:94}.
Taking also into account the small non-perturbative contributions
\cite{BUV:92,BSUV:93,BKSV:94,MW:94}, one gets  \cite{NE:95}
\bel{eq:V_cb_incl}
|\bV_{\!\! cb}|\, = \, (39.8\pm 0.9_{\mbox{\rms exp}}\pm 4.0_{\mbox{\rms th}})
\times 10^{-3} \, .
\ee

The quoted experimental error includes additional theoretical
uncertainties.
The measurement of the inclusive semileptonic
branching ratio faces the difficulty of separating the contributions
of direct $b\to c\, l^-\bar\nu_l$ decays from the cascade process
$\bar b\to\bar c X$, $\bar c\to \bar s\, l^-\bar\nu_l$. 
This separation introduces a significant
model dependence because one needs to assume a theoretical prediction
for the shape of the primary spectrum.
The amount of model dependence has been significantly reduced using events
with two charged leptons from the combined process 
$e^+e^-\to b\bar b \to (c\, l^-\bar\nu_l) \, (\bar c\, l^+\nu_l)$.
In the absence of mixing, the primary decays give rise to a pair of
oppositely charged leptons, while a cascade process would flip the
lepton charge.

The good agreement between the exclusive and inclusive determinations
provides a good test of the theoretical approximations involved. Combining
\eqn{eq:V_cb_excl} and \eqn{eq:V_cb_incl}, one gets finally
\bel{eq:V_cb}
|\bV_{\!\! cb}|\, = \, (39\pm 2)\times 10^{-3} \, .
\ee

\subsection{V$_{ub}$}
\label{subsec:Vub}

The present determination of $|\bV_{\!\! ub}|$ is based on measurements
of the lepton momentum spectrum in inclusive $\bar B\to X_q l^-\bar\nu_l$
decays, where $X_q$ is any hadronic state containing a quark $q=c$ or $u$.
The experimental signature for inclusive $b\to u$ transitions is an excess
of leptons beyond the kinematic limit for the transition 
$b\to c\, l^-\bar\nu_l$.
The yield of leptons in this small portion of the Dalitz plot must be
extrapolated to the full allowed kinematic range and
the resulting fraction of $b\to u$ over $b\to c$ events
is then converted to the CKM ratio $|\bV_{\!\! ub}/\bV_{\!\! cb} |$.

%
\begin{figure}[bht]
\centering       
\epsfig{file=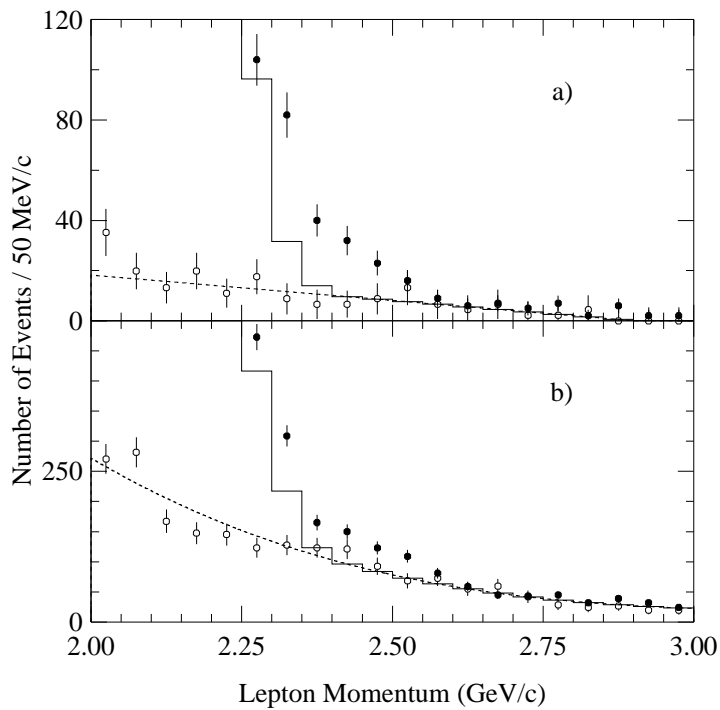,width=9.5cm}
\caption{Inclusive lepton spectrum in the endpoint region
  (CLEO II \protect\cite{CLEO:93}).
The two plots correspond to different experimental cuts on the same data.
The filled points represent the $\Upsilon(4S)$ data, whereas data taken below 
the resonance are indicated by open circles and fitted with the dashed line.
The solid histogram is a Monte Carlo simulation of $b\to c\, l^-\bar\nu_l$
processes. The excess of leptons between 2.4 and 2.6 GeV shows
the existence of $b\to u$ decays.}
\label{fig:b_to_u}
\end{figure}
%

This procedure is obviously
very sensitive to the assumed theoretical spectrum near the
kinematic limit for $\bar B\to D l^-\bar\nu_l$.
Using different models to estimate the systematic theoretical
uncertainties, the analyses of the experimental data 
\cite{CLEO:93,ARGUS:91}
give
\cite{pdg:94}
\bel{eq:Vub_Vcb}
\left| {\bV_{\!\! ub} / \bV_{\!\! cb}}\right|
\, =\,  0.08\pm 0.01_{\mbox{\rms exp}}\pm 0.02_{\mbox{\rms th}} \, . 
\ee
The large model dependence of this measurement is clearly reflected
by the size of the theoretical error (25\%). 
Together with \eqn{eq:V_cb}, this value implies
\bel{eq:V_ub}
|\bV_{\!\! ub}|\,  =\,  0.003\pm 0.001.
\ee

The CLEO Collaboration has recently reported \cite{CLEO:95b}
the first clear signal
for exclusive semileptonic decays of $B$ mesons into charmless final
states:
\beqn\label{eq:charmless}
\mbox{\rm Br}(\bar B\to\pi l\bar\nu_l) &\!\! = &\!\! \left\{
\bat (1.34\pm 0.45)\times 10^{-4}  \qquad & \mbox{\rm (ISGW)} \\
 (1.63\pm 0.57)\times 10^{-4}  \qquad & \mbox{\rm (BSW)} \ea\right. ;
\no\\ && \\
\mbox{\rm Br}(\bar B\to\rho l\bar\nu_l) &\!\! = &\!\! \left\{
\bat (2.28\, {}^{+0.69}_{-0.83})\times 10^{-4}  \qquad & \mbox{\rm (ISGW)}, \\
 (3.88\, {}^{+1.15}_{-1.39})\times 10^{-4} \qquad & \mbox{\rm (BSW)}, \ea\right. .
\no
\eeqn
Again, there is a significant model dependence  coming from the simulation
of reconstruction efficiencies. The two quoted results correspond to the 
theoretical models of Refs.~\citenum{ISGW:89} (ISGW) and 
\citenum{BSW:85} (BSW).

Unfortunately, heavy--quark symmetry does not help to fix the relevant 
form factors in heavy--to--light ($b\to u$) transitions. To extract
information on the CKM mixing factor, one has then to rely in 
model--dependent estimates of the hadronic matrix elements.
Depending on the chosen theoretical model,
the CLEO measurements imply values of $|\bV_{\!\! ub}|$  
which cover a broad range from 0.002 to 0.008 \cite{NE:95}.
Although this range is in good agreement with \eqn{eq:V_ub}, the
large theoretical uncertainty is rather disappointing.
Clearly, there is still large room for improvements here, both on the 
theoretical and experimental sides.
While more reliable methods to predict hadronic form factors should be 
developped, a good sample of measured exclusive $b\to u$ decays
would allow to discriminate among the different models and improve
our present theoretical tools.

\subsection{V$_{tb}$}
\label{subsec:Vtb}

The top quark has just been discovered recently \cite{CDF:95,D0:95}.
Thus, no direct measurement of $|\bV_{\!\! tb}|$
has been performed so far.
In fact, in order to identify top--quark events,
the CDF and D0 experiments have assumed that the top always decays
through $t\to b\, W^+$; i.e., $|\bV_{\!\! tb}|=1$.
This assumption is fully justified by the smallness of the measured
CKM mixings of the $b$ with the up and charm quarks.
Using the unitarity of the CKM matrix, the  experimental determinations
in \eqn{eq:V_cb} and \eqn{eq:V_ub} imply
\bel{eq:Vtb}
|\bV_{\!\! tb}|\, = \, 
\left\{ 1 - |\bV_{\!\! ub}|^2 - |\bV_{\!\! cb}|^2\right\}^{1/2}
\, > \, 0.999 \qquad (95\%\,\mbox{\rm CL}) .
\ee

Nevertheless, it would be nice to have a direct measurement of this
CKM factor, providing a test of the unitarity structure
of the SM quark mixings.
CDF has recently reported \cite{CDF:95b}
a preliminary value of the $t\to W b$ branching ratio:
Br$(t\to W b) = 0.87\, {}^{+0.13}_{-0.30}\, {}^{+0.13}_{-0.11}$.
The agreement with the theoretical expectation ($\sim 100\% $) shows indeed that
$|\bV_{\!\! tb}|\sim\cO(1)$; however, this determination has still a
rather large error.

\section{Unitarity Constraints on the CKM Matrix}
\label{sec:unitarity}\setcounter{equation}{0}

The present status of direct $\bV_{\!\! ij}$ determinations can
be easily summarized:
\bi
\item The light--quark mixings $|\bV_{\!\! ud}|$ and $|\bV_{\!\! us}|$
are rather well known (0.1\% and 0.8\% accuracy, respectively).
Moreover, since the theory is good, improved values could be obtained
with better data on semileptonic $\pi^+$ and $K$ decays.
\item $|\bV_{\!\! cd}|$ and $|\bV_{\!\! cs}|$  are very badly known
(7\% and 20\% accuracy, respectively). This could be largely improved
at a tau--charm factory. For this to be the case, however,
a better theoretical understanding of the strong dynamics is required.
\item $|\bV_{\!\! cb}|$ and $|\bV_{\!\! ub}|$ are also badly known
(5\% and 33\% accuracy, respectively). However, there are good
theoretical tools available. Thus, better determinations could be
easily performed at a $B$ factory.
\item Nothing is known about the CKM mixings involving the top quark,
except that $|\bV_{\!\! tb}|\sim\cO(1)$.
\ei

The entries of the first row are already accurate enough to perform
a sensible test of the unitarity of the CKM matrix:
\bel{eq:unitarity_test}
|\bV_{\!\! ud}|^2 + |\bV_{\!\! us}|^2 + |\bV_{\!\! ub}|^2 \, = \,
0.9965\pm 0.0021 \, .
\ee
It is important to notice that radiative corrections play here a
crucial role. If one uses $|\bV_{\!\! uj}|$ values determined without
radiative corrections, the result \eqn{eq:unitarity_test}
changes to $1.0384\pm 0.0027$, giving an apparent violation
of unitarity (by many $\sigma$'s) \cite{MA:91}.

Imposing the unitarity constraint $\bV \bV^\dagger = \bV^\dagger \bV =
\mbox{\boldmath $1$}$
(and assuming only three generations)
one can get a more precise picture of the CKM matrix.
The 90\% confidence limits on the magnitude of the CKM matrix elements are
then \cite{pdg:94}:

\bel{eq:CKM_values}
\bV \, = \, \left[
\begin{array}{ccc}
0.9745 \,\,\mbox{\rm to}\,\, 0.9757  & 0.219 \,\,\mbox{\rm to}\,\, 0.224 & 
0.002 \,\,\mbox{\rm to}\,\, 0.005 \\
0.218 \,\,\mbox{\rm to}\,\, 0.224 & 0.9736 \,\,\mbox{\rm to}\,\, 0.9750 & 
0.036 \,\,\mbox{\rm to}\,\, 0.046 \\
0.004 \,\,\mbox{\rm to}\,\, 0.014 & 0.034 \,\,\mbox{\rm to}\,\, 0.046 & 
0.9989 \,\,\mbox{\rm to}\,\, 0.9993
\end{array}
\right] ,
\ee
which correspond to
$s_{12} = 0.219$ to 0.223,
$s_{23} = 0.036$ to 0.046, and
$s_{13} = 0.002$ to 0.005.
The ranges given here are slightly different from (but consistent with)
the direct determinations mentioned before.
 
The CKM matrix shows a hierarchical pattern, with the
diagonal elements being very close to one, the ones connecting the
two first generations having a size
\bel{eq:lambda}
\lambda\equiv |\bV_{\!\! us}| = 0.2205\pm 0.0018 \, ,
\ee
the mixing between the second and third families being of order
$\lambda^2$, and the mixing between the first and third quark flavours
having a much smaller size of about $\lambda^3$.
It is then quite practical to use the 
approximate parametrization \cite{WO:83}:

\be\label{eq:wolfenstein}
\bV\, =\,
\left[ \matrix{\displaystyle \ 1- {\lambda^2 \over 2}
\hfill&
\displaystyle \ \ \ \ \ \ \lambda \hfill&
\displaystyle \ \ \ \ \ A\lambda^
3(\rho  - i\eta) \hfill \cr\displaystyle
\hfill& \displaystyle \hfill&
\displaystyle \hfill \cr\displaystyle \ \ \ -\lambda
\hfill& \displaystyle \ \
\ \ \ 1 -{\lambda^ 2 \over 2} \hfill& \displaystyle
\ \ \ \ \ A\lambda^ 2 \hfill
\cr\displaystyle \hfill& \displaystyle \hfill&
\displaystyle \hfill
\cr\displaystyle \ A\lambda^ 3(1-\rho -i\eta)
\hfill& \displaystyle \ \ \ \ \
-A\lambda^ 2 \hfill& \displaystyle
\ \ \ \ \ \ \ \ \ 1 \hfill \cr} \right]\
+\ O\left(\lambda^ 4 \right) \, ,
\ee
where 
\bel{eq:circle}
A= {|\bV_{\!\! cb}|\over\lambda^2} = 0.80\pm 0.04 \, , \qquad\quad
\sqrt{\rho^2+\eta^2} \, = \, 
\left|{\bV_{\!\! ub}\over \lambda \bV_{\!\! cb}}\right|
\, =\, 0.36\pm 0.10 \, .
\ee
Notice that when $|\bV_{\!\! ub}|$ is very small ($s_{13}\ll 1$)
the standard CKM parametrization in Eq.~\eqn{eq:CKM_pdg} only contains
complex phases in $\bV_{\!\! ub}$ and $\bV_{\!\! td}$; i.e., it follows
the same phase conventions than the matrix \eqn{eq:wolfenstein}.


\section{$B^0$--$\bar B^0$ Mixing}
\label{sec:BB_mixing}\setcounter{equation}{0}

Additional information on the CKM parameters can be obtained from
flavour--changing neutral--current transitions, occurring at the 1--loop
level. An important example is provided by 
the mixing between the $B^0$ meson and its antiparticle.
This process occurs through  the so--called box diagrams,
shown in Fig.~\ref{fig:boxdia}, where
two $W$ bosons are exchanged between a pair of quark lines.

\begin{figure}[bth]  
\centerline{\mbox{\epsfysize=4.0cm\epsffile{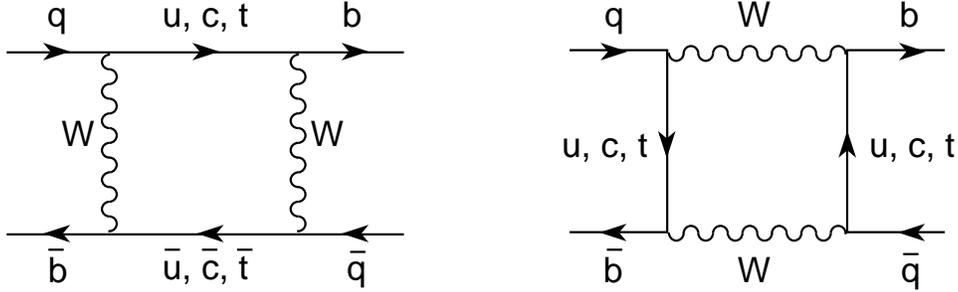}}}
\caption{$B^0$--$\bar B^0$ mixing diagrams.}
\label{fig:boxdia}
\end{figure}

The mixing amplitude is proportional to
\bel{eq:mixing}
\langle\bar B_d^0 | \cH_{\Delta B=2} |B_0\rangle\,\sim\,
\sum_{ij}\, \bV_{\!\! id}^{\phantom{*}}\bV_{\!\! ib}^*
\bV_{\!\! jd}^*\bV_{\!\! jb}^{\phantom{*}}\;
S(r_i,r_j) ,
\ee
where $S(r_i,r_j)$ is a loop function \cite{IL:81}
which depends on the masses
[$r_i\equiv \overline{m}_i^2/M_W^2$] 
of the up--type quarks running along the internal fermionic lines.
Owing to the unitarity of the CKM matrix, the mixing vanishes
for equal (up--type) quark masses (GIM mechanism \cite{GIM:70}); 
thus the effect
is proportional to the mass splittings between the $u$, $c$ and $t$ quarks.
Since the different CKM factors have all a similar size,
$\bV_{\!\! ud}^{\phantom{*}}\bV_{\!\! ub}^*\sim
\bV_{\!\! cd}^{\phantom{*}}\bV_{\!\! cb}^*\sim
\bV_{\!\! td}^{\phantom{*}}\bV_{\!\! tb}^*\sim A\lambda^3$, 
the final amplitude
is completely dominated by the top contribution:
\bel{eq:Vtd_mix}
\langle\bar B_d^0 | \cH_{\Delta B=2} |B_0\rangle\,\sim\, |\bV_{\!\! td}|^2 
S(r_t,r_t) \, .
\ee
This transition can then be used to perform
an indirect determination of $|\bV_{\!\! td}|$.

Notice that this determination has a qualitatively different character
than the ones obtained before from tree--level weak decays.
Now, we are going to test the structure of the electroweak theory at the
quantum level.
This flavour--changing transition could then be very sensitive to
{\it new physics} effects occurring at higher energy scales.
Moreover, the mixing amplitude crucially depends on the 
unitarity of the CKM matrix.
Without the GIM mechanism embodied in the CKM mixing structure, the
calculation of the analogous $K^0\to\bar K^0$ transition (replace
the $b$ quark by a $s$ in the box diagrams) would have failed
to explain the observed $K^0$--$\bar K^0$ mixing by several orders of
magnitude \cite{GL:74}.

\subsection{General Formalism for Meson--Antimeson Mixing}
\label{subsec:mixing_formalism}

The flavour quantum number is not conserved by weak interactions.
Thus a $P^0$ state ($P=K,D,B$)
can be transformed into its antiparticle $\bar P^0$.
As a consequence, the flavour eigenstates $P^0$ and $\bar P^0$ 
are not mass eigenstates and do not follow an exponential decay law.

Let us consider an arbitrary mixture of the two flavour states,
\bel{eq:mixed_state}
|\psi(t)\rangle \, =\, a(t) \, |P^0\rangle + b(t) \, |\bar P^0\rangle 
\,\equiv\, \left( \ba a(t) \\ b(t) \ea\right) \, .
\ee
The time evolution is governed by the equation
\bel{eq:t_eq}
i\, {d\over dt}\, |\psi(t)\rangle \, =\, \cM\, |\psi(t)\rangle \, ,
\ee
where $\cM$ is called the $P^0$--$\bar P^0$ mixing matrix.
Assuming CPT symmetry to hold,
this $2\times 2$ matrix can be written as
\be\label{eq:mass_matrix}
{\cal M} =
\left(   \begin{array}{cc} M & M_{12} \\ M_{12}^* & M \ea   \right)
- {i\over 2}
\left(   \begin{array}{cc} \Gamma & \Gamma_{12} \\
          \Gamma_{12}^* & \Gamma \ea   \right) .
\ee
The diagonal elements $M$ and $\Gamma$ are real parameters,
which would correspond to the mass and width of the neutral mesons
in the absence of mixing.
The off-diagonal entries contain the {\it dispersive} and
{\it absorptive} parts of the $\Delta P=2$ transition amplitude:
\beqn\label{eq:M12_G12}
M_{12} \!\!& = &\!\! {\langle P^0| \cH_{\Delta P=2} |\bar P^0\rangle\over 
2 M_K} + {1\over 2 M_K}\,
\cP\!\int ds\: {\sum_X \int dX\, 
\langle P^0| \cH_{\Delta P=1} | X\rangle\,
\langle X| \cH_{\Delta P=1} |\bar P^0\rangle
\over M_K^2 -s}\, ,
\no\\
\Gamma_{12} \!\!& = &\!\! {\pi\over M_K}\,
\sum_X \int dX\: \delta(M_K^2 -s)\:
\langle P^0| \cH_{\Delta P=1} | X\rangle\,
\langle X| \cH_{\Delta P=1} |\bar P^0\rangle \, .
\eeqn
The sum extends over all possible states $|X\rangle$ of invariant mass 
$\sqrt{s}$
to which the $|\bar P^0\rangle$ can decay; the symbol $dX$ denotes the
appropriate phase--space measure,
and $\cP$ stands for the principal value of the corresponding integral.
If CP were an exact symmetry, $M_{12}$ and $\Gamma_{12}$ would also be real.

The physical eigenstates of ${\cal M}$ are
\be\label{eq:eigenstates}
| P_\mp \rangle \, = \, {1\over\sqrt{|p|^2 + |q|^2}} \,
       \left[ p \, | P^0 \rangle \, \mp\, q \, | \bar P^0 \rangle \right] ,
\ee
with
\be\label{eq:q/p}
{q\over p} \, \equiv \, {1 - \bar\varepsilon \over
         1 + \bar\varepsilon} \, = \,
  \left( {M_{12}^* - {i\over 2}\Gamma_{12}^* \over
          M_{12} - {i\over 2}\Gamma_{12}} \right)^{1/2} .
\ee
If $M_{12}$ and $\Gamma_{12}$ were real, then $q/p = 1$ and
$| B_\mp \rangle $ would correspond to the
CP--even and CP--odd  states 
[we use the phase convention\footnote{
%
%
Since flavour is conserved by strong interactions, there is
some freedom in defining the phases of flavour eigenstates. 
In general, one could use
$\, |P^0_\zeta\rangle \equiv e^{-i\zeta} |P^0\rangle \, $ and 
$|\bar P^0_\zeta\rangle \equiv e^{i\zeta} |\bar P^0\rangle$,
which satisfy 
$\cC\cP\, |P^0_\zeta\rangle = - e^{-2i\zeta} |\bar P^0_\zeta\rangle$.
Both basis are trivially related:
$M_{12}^\zeta = e^{2i\zeta} M_{12}$,
$\Gamma_{12}^\zeta = e^{2i\zeta} \Gamma_{12}$ and
$(q/p)_\zeta = e^{-2i\zeta} (q/p)$.
Thus, $q/p\not=1$ does not necessarily imply CP violation.
CP is violated in the mixing matrix if $|q/p|\not=1$;
i.e., $\mbox{\rm Re}(\bar\varepsilon)\not=0$ and
$\langle P_- |P_+\rangle \not= 0$.
Note that 
$\langle P_- | P_+\rangle_\zeta =\langle P_- | P_+\rangle$.
Another phase--convention independent quantity is
$(q/p) \, (\bar A_f/A_f)$,
where $A_f\equiv A(P^0\!\to\! f)$ and
$\bar A_f\equiv A(\bar P^0\!\to\! f)$, for any final state $f$.} 
%
%
%
\ $\cC\cP |P^0\rangle = - |\bar P^0\rangle$]
\bel{eq:CP_states}
|P_{1,2}\rangle\equiv {1\over\sqrt{2}} \left( |P^0\rangle\mp 
|\bar P^0\rangle\right)\, , \qquad \qquad 
\cC\cP\, |P_{1,2}\rangle = \pm |P_{1,2}\rangle \, .
\ee

Note that if the $P^0$--$\bar P^0$ mixing violates CP, the two mass
eigenstates are no longer orthogonal:
\be \langle P_- | P_+\rangle = {|p|^2-|q|^2 \over |p|^2+|q|^2}
\approx 2 \,\mbox{\rm Re}\, (\bar\varepsilon) \, .
\ee

The time evolution of a state which was originally produced
as a $P^0$ or a  $\bar P^0$  is given by
\be\label{eq:evolution}
\left( \ba | P^0(t) \rangle  \\ | \bar P^0(t) \rangle \ea \right)
 =
\left( \begin{array}{cc} g_1(t)  & {q\over p} g_2(t) \\
     {p\over q} g_2(t) & g_1(t) \ea \right)
\left( \ba | P^0 \rangle  \\ | \bar P^0 \rangle \ea \right) \, ,
\ee
where
\be\label{eq:g}
\left( \ba g_1(t) \\ g_2(t) \ea \right) =
\e^{-iMt} \e^{-\Gamma t/2}
\left( \ba \cos{[(\Delta M - {i\over 2} \Delta\Gamma) t/2]} \\
   -i \sin{[(\Delta M - {i\over 2} \Delta\Gamma) t/2]} \ea \right) \, ,
\ee
with
\bel{eq:DM_DG}
\Delta M \equiv M_{P_+}-M_{P_-} \, ,\qquad\qquad
\Delta\Gamma\equiv\Gamma_{P_+}-\Gamma_{P_-}\, .
\ee

The main difference between the $K^0$--$\bar K^0$ and
$B^0$--$\bar B^0$ systems stems from the different kinematics involved.
The light kaon mass only allows the hadronic decay modes $K^0\to 2\pi$ and 
$K^0\to 3\pi$. Since $\cC\cP\, |\pi\pi\rangle = + |\pi\pi\rangle$, the
CP--even kaon state decays into $2\pi$ whereas the
CP--odd one decays into the phase--space suppressed $3\pi$ mode.
Therefore, there is a large lifetime difference and we have
a short--lived
$|K_S\rangle \equiv |K_-\rangle \approx 
|K_1\rangle + \bar\varepsilon_K |K_2\rangle $
and a long--lived
$|K_L\rangle \equiv |K_+\rangle \approx 
|K_2\rangle + \bar\varepsilon_K |K_1\rangle $
kaon,
with $\Gamma_{K_L}\ll\Gamma_{K_S}$.
One finds experimentally that
$\Delta\Gamma_{K^0}\approx -\Gamma_{K_S}\approx -2\Delta M_{K^0}$.

In the $B$ system, there are many open decay channels and a large part of them
are common to both mass eigenstates. Therefore, the $|B_\mp\rangle $ states
have a similar
lifetime; i.e., $\Delta\Gamma_{B^0}\ll\Gamma_{B^0}$.
Moreover, whereas the $B^0$--$\bar B^0$ transition
is dominated by the top box diagram, the decay amplitudes get
obviously their main contribution from the $b\to c$ transition. 
Thus, $\Delta\Gamma_{B^0} / \Delta M_{B^0} \sim m_b^2/ m_t^2 \ll 1$.
 
\subsection{Experimental Measurements}
\label{subsec:exp_mixing}

With $\Delta\Gamma_{B^0} / \Delta M_{B^0} \ll 1$,
the probability that a state
initially produced as $|B^0\rangle $ will become $|\bar B^0\rangle $
at time $t$ is given by
\bel{eq:mix_prob}
\mbox{\rm Prob}[B^0\to\bar B^0](t)
 \, = \, \frac{1}{2}\,\e^{-\Gamma_{B^0} t} \,
\left[ 1 - \cos{(\Delta M_{B^0} t)}\right] 
\,\equiv\, \frac{1}{2}\,\e^{-\tau}\,\left[1-\cos{(x\tau)}\right]
\, ,
\ee
where $\tau\equiv\Gamma_{B^0} t$ denotes the time measured in lifetime units
and
\bel{eq:x_def}
x \,\equiv\, {\Delta M_{B^0}\over\Gamma_{B^0}}
\ee
determines the frequency of the $B^0$--$\bar B^0$ mixing oscillations.
The time--integrated probability is given by
\bel{eq:mix_prob_integ}
\chi\,\equiv\, \mbox{\rm Prob}[B^0\to\bar B^0]\, = \, {x^2\over 2\, (1+x^2)}
\, .
\ee
Thus, $0\leq\chi<0.5$.

To experimentally measure the mixing transition requires the 
identification of the $B$--meson flavour
at both its production and decay time.
This can be done through flavour--specific decays such as
$B^0\to X l^+\nu_l$ and $\bar B^0\to X l^-\bar\nu_l$.
In general, mixing is measured by studying pairs of $B$ mesons so that
one $B$ can be used to {\it tag} the initial flavour of the other meson.
For instance, in $e^+e^-$ machines one looks into the pair 
production process
$e^+e^- \to B^0 \bar B^0 \to (X l\nu_l) \, (Y l \nu_l)$.
In the absence of mixing, the final leptons should have opposite charges.
The amount of like--sign leptons,
\bel{eq:N_ll_def}
R_{ll}\,\equiv\, {N(l^\pm l^\pm)\over N(l^\pm l^\mp) + N(l^\pm l^\pm)}\, ,
\ee
is then a clear signature of the
mixing transition.

At high--energy colliders a $B^\pm$ meson can be used to 
{\it tag} the flavour of the neutral $B$
[$b\bar b \to B^- B^0 X\to (Yl^-\bar\nu_l)\, (Zl^\pm\nu_l)\, X$,
 $b\bar b \to \bar B^0 B^+X\to (Yl^\mp\nu_l)\, (Zl^+\nu_l)\, X$];
then, $R_{ll} = \chi$.
The relation is slightly more complicated 
when the {\it tagging} is performed through another neutral $B$ which
also oscillates.
At LEP, where the two $B$ mesons are uncorrelated, $R_{ll}$ is just given by 
twice the probability that one $B$ oscillates times the probability that
the other $B$ does not change flavour.
The behaviour is quite different on the $\Upsilon(4S)$ resonance or
at the $B\bar B^*$ production threshold, because the $B^0\bar B^0$
pairs are produced coherently, i.e. in a state of definite orbital
angular momentum (odd/even at the $\Upsilon(4S)$/$B\bar B^*$ threshold).
Quantum statistics for spin zero particles requires then an
antisymmetric (symmetric) $B^0\bar B^0$ wave function for
odd (even) orbital angular momentum.
Taking this into account,
\bel{eq:R_ll_chi}
R_{ll} \, = \, \left\{ \begin{array}{lc}
x^2/[2\, (1+x^2)] \qquad\quad & [\Upsilon(4S)] \\
x^2\, (3+x^2)/[2\, (1+x^2)^2] \qquad & [B\bar B^*\,\mbox{\rm threshold}] \\
2\chi\, (1-\chi) & [\mbox{\rm LEP}]
\ea\right.\, .  
\ee

At the $\Upsilon(4S)$, the lepton like--sign fraction 
(corrected for leptons coming from $B^+B^-$ pairs)
directly measures the mixing transition $B_d^0\to\bar B_d^0$.
However, at higher energies both $B^0_d$ and $B^0_s$ are produced,
and one measures a combination of their mixing probabilities,
weighted by their production fractions:
$\bar\chi = f_{B^0_d}\chi_{B^0_d} + f_{B^0_s}\chi_{B^0_s}$.

Evidence for a large $B_d^0$--$\bar B_d^0$ mixing was first 
reported in 1987 by ARGUS \cite{ARGUS:87} and later confirmed by
CLEO \cite{CLEO:89}. This provided the first indication that the
top quark was very heavy. 
Since then, many experiments have analyzed
the mixing probabiliy
\cite{BH:95,FO:95}. The present world--average value of $\chi_{B^0_d}$
from threshold experiments is \cite{BH:95}
\bel{eq:chi_b}
\chi_{B^0_d} = 0.151 \pm 0.028 \, ,
\ee
which implies $x_{B^0_d} = 0.66 \pm 0.09$.
The high--energy measurements are compatible with this number and
together they indicate a maximal value for $\chi_{B^0_s}\sim 0.5$,
in agreement with the SM expectation
\bel{eq:x_s}
{x_{B^0_s}\over x_{B^0_d}} \sim {|\bV_{\!\! ts}|^2\over |\bV_{\!\! td}|^2}
\gg 1 \, .
\ee
Unfortunately, $\chi$ becomes insensitive to $x$ when mixing is
maximal. 
For instance, $\chi_{B^0_s}>0.4$ corresponds to the weak limit $x_{B^0_s}>2$.
Time integrated measurements are then not sensitive to
the rapid oscillations of the $B^0_s$ meson.

%
\begin{figure}[htb]
\centering       
\epsfig{file=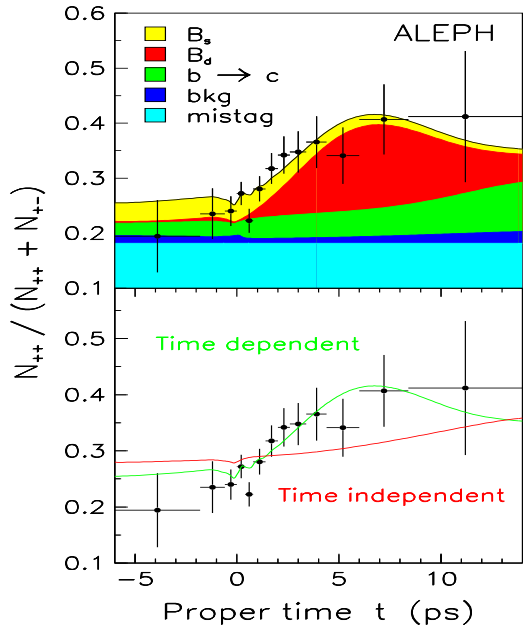,height=7.6cm,width=7cm}
\vspace*{0.5cm}
\caption{Dilepton like--sign fraction as a function of time
from ALEPH \protect\cite{ALEPH:94m}.}
\label{fig:mixing}
\end{figure}
%

The LEP experiments have performed explicit measurements of the
mixing probability $B^0_d\to\bar B^0_d$ as a function of time
\cite{ALEPH:94m,DELPHI:94m,OPAL:94m}. 
Fig.~\ref{fig:mixing} shows the time--dependent 
fraction of like--sign leptons
measured by ALEPH \cite{ALEPH:94m},
which provides clear evidence of the oscillatory behaviour.
A fit to the time dependence allows to extract $\Delta M_{B^0_d}$.
The present LEP average is  \cite{BH:95}
\bel{eq:DM_d_LEP}
\Delta M_{B^0_d} = 0.501 \pm 0.034\:\mbox{\rm ps}^{-1} \, .
\ee
Combined with \eqn{eq:chi_b} and the measured $B^0_d$ lifetime,
this gives the world average \cite{BH:95}
\bel{eq:DM_d_x_d}
\Delta M_{B^0_d} = 0.462 \pm 0.026\:\mbox{\rm ps}^{-1} \, ; \qquad\qquad
x_{B^0_d} = 0.76 \pm 0.05 \, .
\ee

The LEP experiments have also searched for a high--frequency component
in their fit to the proper--time distribution, trying to pin down the
$B^0_s$ contribution. The present upper limit on the
$B^0_s$--$\bar B^0_s$ mixing is \cite{OPAL:94m}
\bel{eq:DM_s_x_s}
\Delta M_{B^0_s} > 2.2\:\mbox{\rm ps}^{-1} \, , \qquad\quad
x_{B^0_s} > 3.0 \, , \qquad (95\%\,\mbox{\rm CL})\, .
\ee

\vspace*{-0.6cm}
\subsection{Mixing constraints on the CKM matrix}
\label{subsec:mixing_constraints}

The calculation of the short--distance box diagrams 
in Fig.~\ref{fig:boxdia}
is rather straightforward. Moreover, the leading and next--to--leading
gluonic corrections are already known \cite{BJW:90}. Unfortunately,
this is not enough to get an accurate prediction for the mixing
probability. The main theoretical uncertainty stems from the
hadronic matrix element of the $\Delta B=2$
four--quark operator generated by the box diagrams:
\bel{eq:DB_op}
\langle\bar B^0\, |\, (\bar b\gamma^\mu(1-\gamma_5)d)\: 
(\bar b\gamma_\mu(1-\gamma_5)d)\, |\, B^0\rangle \,\equiv\,
 {8\over 3} \, M_B^2\, (\sqrt{2}\, \xi_B)^2 \, .
\ee
The size of this matrix element is characterized through
the non--perturbative parameter
$\xi_B\equiv f_B\sqrt{B_B}$, which is rather badly known. 
Present calculations favour
the range \cite{PP:95} 
$\hat\xi_B\equiv\alpha_s(\mu^2)^{-3/23}\xi_B(\mu^2) = 185\pm 50$ MeV
[$\mu$ is the renormalization scale].
With $\overline{m}_t = 173\pm 12$ GeV,
the measured mixing in \eqn{eq:DM_d_x_d} implies
\bel{eq:V_td}
 |\bV_{\!\! td}|\, = \, 0.007\pm 0.002 \, , 
\ee
in good agreement with (but more precise than) the value obtained from the
unitarity constraint in \eqn{eq:CKM_values}.
In terms of the $(\rho,\eta)$ parametrization of 
Eq.~\eqn{eq:wolfenstein},
this gives
\bel{eq:circle_t}
\sqrt{(1-\rho)^2+\eta^2} \, = \,
\left|{\bV_{\!\! td}\over \lambda\bV_{\!\! cb}}\right|
\, = \, 0.8\pm 0.2 \, .
\ee

The same analysis can be applied to the $B^0_s$--$\bar B^0_s$ mixing
probability. The non--perturbative uncertainties can be reduced to the
level of $SU(3)$ breaking corrections through the ratio
\bel{x_ratio}
{\Delta M_{B^0_s}\over \Delta M_{B^0_d}} \approx
{M_{B^0_s}\, \xi^2_{B^0_s}\over M_{B^0_d}\, \xi^2_{B^0_d}}\,
\left|{\bV_{\!\! ts}\over \bV_{\!\! td}}\right|^2 
\approx (1.0\pm 0.2) \times
\left|{\bV_{\!\! ts}\over \bV_{\!\! td}}\right|^2 \, ,
\ee
where we have made
the reasonable assumption 
$(\xi_{B_s}/\xi_{B_d})^2\approx 1.0\pm 0.2$.
The present bounds on $\Delta M_{B^0_s}$ imply then
\bel{eq:Vts_Vtd}
\left|{\bV_{\!\! ts}\over \bV_{\!\! td}}\right| > 1.8
\qquad (95\%\,\mbox{\rm CL})\, .
\ee
This should be compared with the unitarity constraint
$|\bV_{\!\! ts}/ \bV_{\!\! td}| = 4.4\pm 2.6$.

\section{CP--Violation}
\label{sec:CP-Violation}\setcounter{equation}{0}

Since $\delta_{13}$ ($\eta$) is the only possible source of CP violation,
the SM predictions for CP--violating phenomena are quite constrained.
Moreover, the CKM mechanism requires several necessary conditions 
in order to generate an observable CP--violation effect.
With only two fermion generations, the quark--mixing mechanism cannot
give rise to CP violation; therefore,
for CP violation to occur in a particular process,
all 3 generations are required to play an active role.
In the kaon system, for instance, CP--violation effects can only
appear at the one--loop level, where the top quark is present.
In addition, all CKM--matrix elements must be non--zero and the quarks
of a given charge must be non--degenerate in mass. If any of these
conditions were not satisfied, the CKM--phase could be rotated away
by a redefinition of the quark fields. CP--violation effects
are then necessarily proportional to the product of all CKM angles, and
should vanish in the limit where any two (equal--charge) quark masses
are taken to be equal.
All these necessary conditions can be summarized in a very elegant
way as a single requirement \cite{JA:85}
on the original quark--mass matrices
$\bM_u^\prime$ and $\bM_d^\prime$:
\be
\mbox{\rm CP violation} \; \Longleftrightarrow \; 
\mbox{\rm Im}\left\{\det\left[\bM_u^\prime \bM^{\prime\dagger}_u ,
  \bM_d^\prime \bM^{\prime\dagger}_d\right]\right\} \not=0 \, .
\ee

Without performing any detailed calculation, one can make the
following general statements on the implications of the CKM mechanism
of CP violation:
\bi
\item  
Owing to unitarity, for any choice of $i,j,k,l$ (between 1 and 3),
\beqn\label{eq:J_relation}
\mbox{\rm Im}\left[
\bV^{\phantom{*}}_{ij}\bV^*_{ik}\bV^{\phantom{*}}_{lk}\bV^*_{lj}\right]
\, =\, \cJ \sum_{m,n=1}^3 \epsilon_{ilm}\epsilon_{jkn}\, , 
\qquad\quad\\
\cJ \, =\, c_{12} c_{23} c_{13}^2 s_{12} s_{23} s_{13} \sin{\delta_{13}}
\,\approx\, A^2\lambda^6\eta \, < \, 10^{-4}\, .
\eeqn
Any CP--violation observable involves \cite{JA:85} the product
$\cJ$.
Thus, violations of the CP symmetry are necessarily small.
\item In order to have sizeable CP--violating asymmetries
[$(\Gamma - \overline{\Gamma})/(\Gamma + \overline{\Gamma})$], 
one should look
for very suppressed decays, where the decay widths already involve
small CKM matrix elements. 
\item In the SM, CP violation is a low--energy phenomena 
in the sense that any
effect should dissapear when the quark--mass difference $m_c-m_u$ becomes
negligible. 
\item $B$ decays are the optimal place for CP--violation signals to show up.
They involve small CKM matrix elements and are the lowest--mass processes 
where the three quark generations play a direct (tree--level) role.
\ei

The SM mechanism of CP violation is based in the unitarity of the
CKM matrix. Testing the constraints implied by unitarity
is then a way to test the source of CP violation.
Up to now, the only unitarity relation which has been precisely tested
is the one associated with the first row of the CKM matrix;
however, only the moduli of the CKM parameters appear in
Eq.~\eqn{eq:unitarity_test}, while CP violation has to do with their phases.
More interesting are the off--diagonal unitarity conditions:
\be\label{eq:triangles}
\matrix{
\bV^\ast_{\!\! ud}\bV^{\phantom{*}}_{\!\! us} & + & 
\bV^\ast_{\!\! cd}\bV^{\phantom{*}}_{\!\! cs} & + &
\bV^\ast_{\!\! td}\bV^{\phantom{*}}_{\!\! ts} & = & 0 \, ,
\cr 
\bV^\ast_{\!\! us}\bV^{\phantom{*}}_{\!\! ub} & + &
\bV^\ast_{\!\! cs}\bV^{\phantom{*}}_{\!\! cb} & + & 
\bV^\ast_{\!\! ts}\bV^{\phantom{*}}_{\!\! tb} & = & 0 \, ,
\cr
\bV^\ast_{\!\! ub}\bV^{\phantom{*}}_{\!\! ud} & + &
\bV^\ast_{\!\! cb}\bV^{\phantom{*}}_{\!\! cd} & + &
\bV^\ast_{\!\! tb}\bV^{\phantom{*}}_{\!\! td} & = & 0 \, .
\cr}
\ee
These relations can be visualized by triangles in a complex
plane \cite{bj:1} which, owing to
Eq.~\eqn{eq:J_relation}, have the
same area $|\cJ|/2$.
In the absence of CP violation, these triangles would degenerate
into segments along the real axis.

In the first two triangles, one side is much shorter than the other
two (the Cabibbo suppression factors of the three sides are
$\lambda$, $\lambda$ and $\lambda^5$ in the first triangle, 
and $\lambda^4$, $\lambda^2$ and $\lambda^2$ in the second one).
This is the reason why CP effects are so small for $K$ mesons
(first triangle), and why certain  asymmetries in $B_s$ decays are
predicted to be tiny (second triangle).

The third triangle looks more interesting, since the
three sides have a similar size of about $\lambda^3$.
They are small, which means that the relevant $b$--decay branching ratios
are small, but once enough $B$ mesons would be produced, CP--violation
asymmetries are going to be sizeable. 
This triangle is shown in Fig.~\ref{fig:utriangle}, where it has
been scaled by dividing its sides by 
$|\bV^\ast_{\!\! cb}\bV^{\phantom{*}}_{\!\! cd}|$.
In the approximate parametrization \eqn{eq:wolfenstein},
where $\bV^\ast_{\!\! cb}\bV^{\phantom{*}}_{\!\! cd}$ is real,
this aligns one
side of the triangle along the real axis and makes its length equal to
1; the coordinates of the 3 vertices are then
$(0,0)$, $(1,0)$ and $(\rho,\eta)$.
Note that, although the orientation of the triangle in the complex plane
is phase--convention dependent, the triangle itself is a physical
object: the length of the sides and/or the angles can be directly
measured.
In fact, we have already determined its sides from the measured
ratio $\Gamma(b\to u)/\Gamma(b\to c)$ and from $B^0_d$--$\bar B^0_d$
mixing:
\beqn\label{eq:Rb_constraint}
R_b\,\equiv\, 
\left|{\bV_{\!\! ub}^*\bV^{\phantom{*}}_{\!\! ud}
\over\bV_{\!\! cb}^*\bV^{\phantom{*}}_{\!\! cd}}\right|
&\!\!\approx&\!\!
\left|{\bV_{\!\! ub}\over \lambda \bV_{\!\! cb}}\right|
\,\approx\,
\sqrt{\rho^2+\eta^2} 
\, =\, 0.36\pm 0.10 \, ,
\\ \label{eq:Rt_constraint}
R_t\,\equiv\, 
\left|{\bV_{\!\! tb}^*\bV^{\phantom{*}}_{\!\! td}
\over\bV_{\!\! cb}^*\bV^{\phantom{*}}_{\!\! cd}}\right|
&\!\!\approx&\!\!
\left|{\bV_{\!\! td}\over \lambda\bV_{\!\! cb}}\right|
\,\approx\,
\sqrt{(1-\rho)^2+\eta^2} 
\, = \, 0.8\pm 0.2  \, .
\eeqn
In principle, the measurement of these two sides,
performed through CP--conserving observables,
could make possible to establish that CP is violated
(assuming unitarity), by showing that they indeed give rise to a
triangle and not to a straight line.
With the present experimental and theoretical errors, this is however
not possible yet.

%
\begin{figure}[htb]
\centering       
\epsfig{file=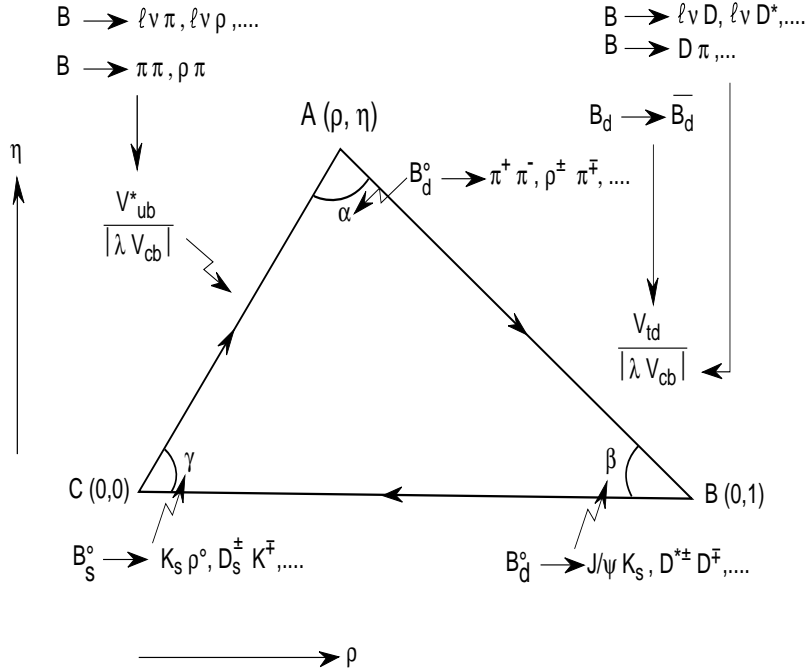,height=9.0cm,width=12.0cm}
\caption{The unitarity triangle. Also shown are various topics in  $B$
physics that allow to measure its sides and angles
\protect\cite{ecfa:93}.}
\label{fig:utriangle}
\end{figure}
%

\subsection{Indirect and Direct CP Violation in the Kaon System}
\label{subsec:CP_kaon}

Any observable CP--violation effect is generated by the interference between
different amplitudes contributing to the same physical transition.
This interference can occur either through meson--antimeson mixing
or via final--state interactions, or by a combination of both effects.

The flavour--specific decays
$K^0\to\pi^- l^+\nu_l$ and $\bar K^0\to\pi^+ l^-\bar\nu_l$
provide a way to measure
the departure of the $K^0$--$\bar K^0$ mixing parameter
$|p/q|$ from unity.
In the SM, 
$|A(\bar K^0\to\pi^+ l^-\bar\nu_l)| = |A(K^0\to\pi^- l^+\nu_l)|$;
therefore,
\be
\delta \equiv
{\Gamma(K_L\to\pi^- l^+\nu_l) - \Gamma(K_L\to\pi^+ l^-\bar\nu_l)\over
\Gamma(K_L\to\pi^- l^+\nu_l) + \Gamma(K_L\to\pi^+ l^-\bar\nu_l)}
 = {|p|^2-|q|^2 \over |p|^2+|q|^2}
= {2\, \mbox{\rm Re}\, (\bar\varepsilon^{\phantom{'}}_K)\over 
(1 + |\bar\varepsilon^{\phantom{'}}_K|^2)} .
\ee
The experimental measurement \cite{pdg:94},
$\delta = (3.27\pm 0.12)\times 10^{-3}$,
implies
\be\label{eq:Repsilon}
\mbox{\rm Re}\, (\bar\varepsilon^{\phantom{'}}_K)\, =\, 
(1.63\pm 0.06)\times 10^{-3}\, ,
\ee
which establishes the presence of {\it indirect} CP--violation
generated by the mixing amplitude.
 
If the flavour of the decaying meson $P$ is known, 
any observed difference between the decay rate
$\Gamma(P\to f)$ and its CP conjugate $\Gamma(\bar P\to \bar f)$
would indicate that CP is directly violated in the decay amplitude.
One could study, for instance,
CP asymmetries in charged--kaon decays, such as $K^\pm\to\pi^\pm\pi^0$, 
where the charge of the final pions clearly identifies the flavour
of the decaying kaon
(these types of decays are often referred to as self-tagging modes).
No positive signal has been reported up to date.

Since at least two interfering amplitudes are needed
to generate a CP--violating effect,
let us write the amplitudes
for the transitions $P \to f$ and
$\bar P \to \bar f$ as
\beqn
\label{eq:direct_b}
A[P \to f] & = & \, M_1 \, e^{i\phi_1}\, e^{i \alpha_1}\,
   +\, M_2 \, e^{i\phi_2}\, e^{i \alpha_2} \, ,
\\
A[\bar P \to \bar f] & = &
  M_1 e^{-i\phi_1} e^{i \alpha_1}\, +\, M_2 e^{-i\phi_2}e^{i \alpha_2} \, ,
\eeqn
where $\phi_1$, $\phi_2$ denote weak phases, $\alpha_1$, $\alpha_2$
strong final--state phases, and $M_1$, $M_2$ the moduli
of the matrix elements. The rate asymmetry is given by
\be
\label{eq:direct_ratediff}
{\Gamma[P \to f] - \Gamma[\bar P \to \bar f] \over
\Gamma[P \to f] + \Gamma[\bar P \to \bar f]}
\,=\, 
{-2 M_1 M_2 \sin{(\phi_1 - \phi_2)}
\sin{(\alpha_1 - \alpha_2)} \over
|M_1|^2 + |M_2|^2 + 2 M_1 M_2 \cos{(\phi_1 - \phi_2)}
\cos{(\alpha_1 - \alpha_2)}} \, .
\ee
Thus, to generate a direct--CP asymmetry one needs:
\begin{enumerate}
\item Two (at least) interfering amplitudes.
\item Two different weak phases 
[$\sin{(\phi_1 - \phi_2)}\not=0$].
\item Two different strong phases
[$\sin{(\alpha_1 - \alpha_2)}\not=0$].
\end{enumerate}
Moreover, in order to get a sizeable asymmetry,
the two amplitudes $M_1$ and $M_2$ should be of comparable size.

In the kaon system, direct CP violation has been searched for in decays of
neutral kaons, where $K^0$--$\bar K^0$ mixing is also involved. Thus,
both direct and indirect CP--violation effects need to be taken into account,
simultaneously.
Since the $\pi^+\pi^-$ and $2\pi^0$ states are even 
under CP, only the $K_1$ state could decay
into $2\pi$ if CP were conserved.
Thus, a CP--violation signal is provided by the ratios:
\beqn\label{eq:etapm}
\eta_{+-} \equiv {A(K_L\to\pi^+\pi^-)\over A(K_S\to\pi^+\pi^-)}
  &\equiv & |\eta_{+-}| \, e^{i\phi_{+-}}
  \,\approx\, \varepsilon_K^{\phantom{'}}
 + {\varepsilon_K'\over 1 + \omega/\sqrt{2}} 
  \, , \\ \label{eq:etazero}
\eta_{00} \,\equiv\, {A(K_L\to\pi^0\pi^0)\over A(K_S\to\pi^0\pi^0)}
  &\equiv & |\eta_{00}|\,  e^{i\phi_{00}}
  \,\approx\, \varepsilon_K^{\phantom{'}} 
- {2\varepsilon_K'\over 1 - \sqrt{2}\omega} 
\, ,\quad
\eeqn
where [terms quadratic in
the small CP--violating quantities have been neglected]
\bel{eq:eps_def}
\varepsilon_K^{\phantom{'}} \equiv\bar\varepsilon_K^{\phantom{'}} 
+ i \xi_0 \, , \quad
\varepsilon_K'\equiv {i\over\sqrt{2}} \,\omega\, (\xi_2 - \xi_0) \, , \quad
\omega\equiv {\mbox{\rm Re}\, (A_2)\over\mbox{\rm Re}\, (A_0)}\,
   e^{i(\delta_2-\delta_0)} \, .\;
\ee
$A_I$ and $\delta_I$ are the decay--amplitudes and strong phase--shifts
of isospin $I=0,2$ (these are the only two values allowed by Bose
symmetry for the final $2\pi$ state),
\be
A[K^0\to(2\pi)_I] \,\equiv\, i A_I\, e^{i\delta_I}\, , \qquad\qquad
A[\bar K^0\to(2\pi)_I] \,\equiv\, -i A_I^*\, e^{i\delta_I} \, ,
\ee
and
\be
\xi_I\,\equiv\, {\mbox{\rm Im}\, (A_I)\over\mbox{\rm Re}\, (A_I)} \, .
\ee

The parameter $\varepsilon_K^{\phantom{'}}$ 
is related to the indirect CP violation. 
Note that
$\varepsilon_K^{\phantom{'}}$ is a physical (measurable) 
phase--convention--independent
quantity, while $\bar\varepsilon_K^{\phantom{'}}$ is not 
[$\varepsilon_K^{\phantom{'}}$ =$\bar\varepsilon_K^{\phantom{'}}$
in the phase convention $\mbox{\rm Im}\, (A_0)=0$; however,
$\mbox{\rm Re}\, (\varepsilon_K^{\phantom{'}}) = 
\mbox{\rm Re}\, (\bar\varepsilon_K^{\phantom{'}})$
in any convention].
Direct CP violation is measured through $\varepsilon'_K$,
which is governed by
the phase--difference between the two isospin amplitudes.
The CP--conserving parameter $\omega$ gives the relative size between
these two amplitudes; experimentally, one finds a very big
enhancement of the $I=0$
channel with respect to the $I=2$ one,
which is known as the $\Delta I = 1/2$ rule:
\be
|\omega| \approx {1\over 22} , \qquad\qquad
\delta_2-\delta_0 = -45^\circ \pm 6^\circ .
\ee
The small size of $|\omega|$ implies a strong suppression of 
$\varepsilon'_K$.  

From the eigenvector equations for $K_S$ and $K_L$ one can easily
obtain the relation
\be
\bar\varepsilon^{\phantom{'}}_K \,\approx\, e^{i\phi_{SW}}\,
{\mbox{\rm Im}(M_{12}) - {i\over 2} \mbox{\rm Im}(\Gamma_{12})\over
\sqrt{\Delta M_{K^0}^2 + {1\over 4} \Delta\Gamma_{K^0}^2}} ,
\ee
where \cite{pdg:94}
$\Delta M_{K^0} \equiv M_{K_L} - M_{K_S} = (3.510\pm0.018)\times 10^{-12}$ MeV,
$\Delta\Gamma_{K^0} \equiv\Gamma_{K_L}-\Gamma_{K_S} =
-(7.361\pm 0.010)\times 10^{-12}$ MeV,
and
\be
\phi_{SW}\equiv\arctan{\left( 
{-2\Delta M_{K^0}\over\Delta\Gamma_{K^0}}\right)}
= 43.64^\circ\pm0.15^\circ 
\ee
is the so--called superweak phase. 
Since $\Delta\Gamma_{K^0}\approx -2\Delta M_{K^0}$, one has 
$\phi_{SW}\approx\pi/4$.
Moreover, 
$\mbox{\rm Im}\, (\Gamma_{12})/ \mbox{\rm Re}\, (\Gamma_{12}) \approx -2\xi_0$
because $\Gamma_{12}$ is dominated by the
$K^0\!\to\! (2\pi)_{I=0}$ decay mode.
Using these relations, one gets the approximate result
\be
\varepsilon^{\phantom{'}}_K\,\approx\, {e^{i\pi/4}\over\sqrt{2}} \,\left\{
{\mbox{\rm Im}\, (M_{12})\over 2 \,\mbox{\rm Re}\, (M_{12})} + \xi_0 \right\} .
\ee
Notice that $\delta_2-\delta_0 + \pi/2\approx \pi/4$, i.e.
\be
\varepsilon'_K\,\approx\, {e^{i\pi/4}\over\sqrt{2}}\,  |\omega | \, 
(\xi_2 - \xi_0 ) .
\ee
Thus, owing to the particular numerical values of the neutral--kaon decay
parameters, the phases of $\varepsilon^{\phantom{'}}_K$ 
and $\varepsilon'_K$ are nearly equal.

The experimental world--averages quoted by the Particle Data Group 
\cite{pdg:94} are
\beqn
|\eta_{+-}| \, = \, (2.269\pm0.023)\times 10^{-3} \, , \qquad
&& \phi_{+-} \, =\, (44.3\pm0.8)^\circ \, , \\
|\eta_{00}| \, = \, (2.259\pm0.023)\times 10^{-3} \, , \qquad
&& \phi_{00} \,\, =\, (43.3\pm1.3)^\circ \, .
\eeqn
The phases are very close to $\pi/4$,
whereas the moduli are equal within errors, showing that indeed
$|\varepsilon'_K|<<|\varepsilon_K|$ as expected from the
$|\omega|$ suppression. Moreover,
these numbers imply 
Re$\, (\varepsilon^{\phantom{'}}_K)\approx 1.63\times 10^{-3}$,
in good agreement with the value
\eqn{eq:Repsilon} extracted from semileptonic decays.

The ratio $\varepsilon'_K/\varepsilon^{\phantom{\prime}}_K$ 
can be determined through the relation
\be
\mbox{\rm Re}\left({\varepsilon'_K\over\varepsilon^{\phantom{'}}_K}\right) 
\approx
{1\over 6} \left\{ 1 - \left| {\eta_{00}\over\eta_{+-}}\right|^2\right\} .
\ee
Two different experiments   
have recently reported a measurement of this
quantity:
\be
\mbox{\rm Re}\left({\varepsilon'_K\over\varepsilon^{\phantom{'}}_K}\right) 
\, = \,
\left\{ 
\begin{array}{ll}
(23.0\pm 6.5)\times 10^{-4}  & \qquad 
   [\mbox{\rm NA31 \protect\cite{NA31:93}}] 
  \\
(7.4\pm5.9)\times 10^{-4} & \qquad 
   [\mbox{\rm E731 \protect\cite{E731:93}}]
\ea \right. .
\ee
The NA31 measurement provides
evidence for a non--zero value of 
$\varepsilon'_K/\varepsilon^{\phantom{\prime}}_K$ (i.e.,
direct CP violation), with a statistical significance of more than three
standard deviations. However, this is not supported by
the E731 result, which is compatible with 
$\varepsilon'_K/\varepsilon^{\phantom{\prime}}_K = 0$, 
thus with no direct CP violation.
The probability for the two results being statistically compatible is
only 7.6\%.

New experiments with a better sensitivity are required in order
to resolve this discrepancy.
A next generation of 
$\varepsilon'_K/\varepsilon^{\phantom{\prime}}_K$ experiments is already
under construction at CERN \cite{cern} and Fermilab \cite{fnal}.
Moreover, a dedicated $\phi$ factory (DA$\Phi$NE), providing
large amounts of tagged $K_S$, $K_L$ and $K^\pm$
($\phi\to K\bar K$), is being built
at Frascati \cite{DAPHNE}. The goal of all these experiments is to
reach sensitivities better than $10^{-4}$.

The CKM mechanism generates CP--violation effects both in the
$\Delta S=2\; $ $K^0$--$\bar K^0$ transition (box--diagrams) and in the
$\Delta S=1$ decay amplitudes (penguin diagrams).
The theoretical analysis of $K^0$--$\bar K^0$ mixing is
quite similar to the one applied to the $B$ system. This time, however,
the charm loop contributions are non--negligible. The main
uncertainty stems from the calculation of the hadronic matrix
element of the four--quark $\Delta S=2$ operator, which is usually
parametrized through the non--perturbative parameter \cite{PP:95}
$\hat B_K\approx 0.4$--0.8.    

\goodbreak

The experimental value of $\varepsilon_K$ specifies a hyperbola in the
$(\rho,\eta)$ plane. 
This is shown in Fig.~\ref{fig:unitarity_constraints},
together with the constraints 
\eqn{eq:Rb_constraint} and \eqn{eq:Rt_constraint}, which result in
the circles centered at $(0,0)$ and $(1,0)$, respectively.
The final allowed range of values for $(\rho,\eta)$  is given by
the intersection of all constraints.

%
\begin{figure}[htb]
\centering       
\epsfig{file=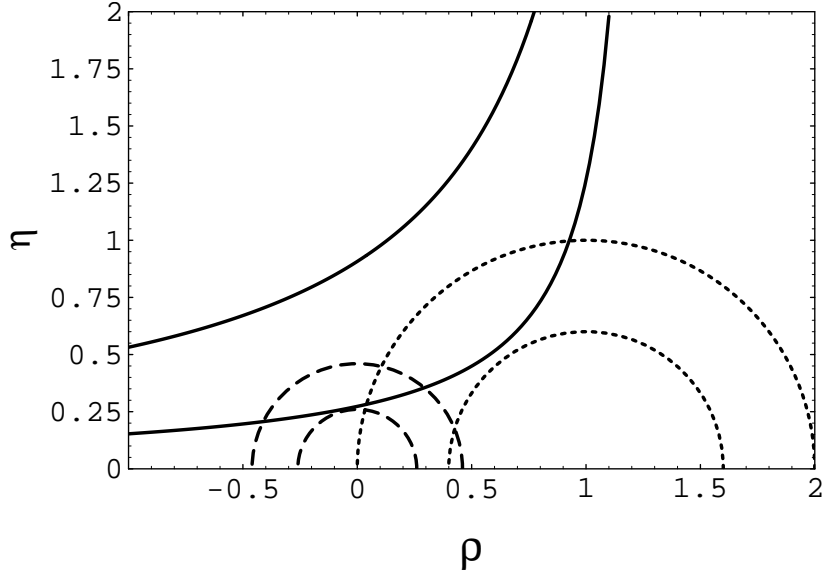,height=7.8cm}   
\caption{Present constraints un the Unitarity Triangle.}
\label{fig:unitarity_constraints}
\end{figure}
%

The theoretical estimate of $\varepsilon'_K/\varepsilon^{\phantom{\prime}}_K$
is much more involved, because ten  
four--quark operators need to be considered
in the analysis and the presence of cancellations between different
contributions tends to amplify the sensitivity to the not
very well controlled long--distance effects.
For large values of the top--mass, the $Z^0$--penguin contributions
strongly suppress the expected value of 
$\varepsilon'_K/\varepsilon^{\phantom{\prime}}_K$,
making the final result very sensitive to $m_t$.
The present theoretical estimates \cite{BJL:93,CI:95} 
range from $-3\times 10^{-4}$ to $10^{-3}$.
More theoretical work is needed in order to get firm predictions.

%
\begin{figure}[htb]
\centering       
\epsfig{file=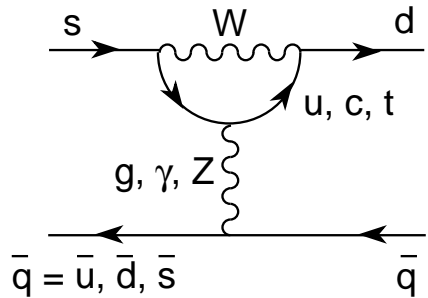,height=4cm}  
\caption{$\Delta S=1$ penguin diagrams.}
\label{fig:penguin}
\end{figure}
%

\goodbreak

\mbox{}\vspace*{-1.1cm}

\subsection{Bottom decays}
\label{subsec:bottom}


The flavour--specific decays 
$B^0\to X l^+\nu_l$ and $\bar B^0\to X l^-\bar\nu_l$
provide the most direct way to measure the amount of CP violation in
the $B^0$--$\bar B^0$ mixing matrix.
The asymmetry between the number of $l^+l^+$ and $l^-l^-$ pairs produced
in the processes $e^+e^-\to B^0\bar B^0\to l^\pm l^\pm X$ 
is easily found to be
\be
\label{eq:a_SL_def}
a_{SL} \equiv {N(l^+l^+) - N(l^-l^-) \over N(l^+l^+) + N(l^-l^-)}
= {\left|p/q\right|^2 - \left|q/p\right|^2 \over
   \left|p/q\right|^2 + \left|q/p\right|^2}
\approx 4 \,\mbox{\rm Re}\, (\bar\varepsilon_B)  .
\ee
Unfortunately, this $\Delta B = 2$ asymmetry is expected to be quite tiny
in the SM, because
$|\Delta\Gamma_{B^0}/\Delta M_{B^0}| \approx |\Gamma_{12}/M_{12}|
\sim m_b^2/m_t^2 << 1 \,\,$
and, moreover, there is an additional GIM suppression
in the phase
$\phi_{\Delta B=2}\equiv\arg{\left(M_{12}/\Gamma_{12}\right)}
\sim (m_c^2-m_u^2)/ m_b^2$, 
implying a value of $|q/p|$  very close to 1.
Thus, 
\be
\label{eq:a_SL_expected}
a_{SL} \leq \left\{ \ba 10^{-3}  \qquad (B^0_d), \\  
10^{-4}  \qquad (B^0_s). \ea
\right.\ee
The observation of an asymmetry $a_{SL}$ at the percent level,
would then be
a clear indication of new physics beyond the SM.


%
\begin{figure}[htb]
\centering       
\epsfig{file=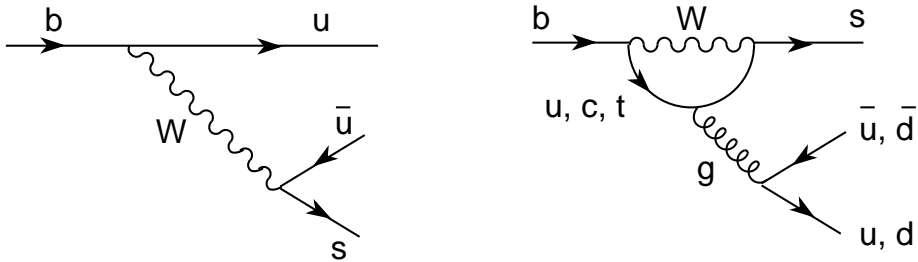,height=3.5cm}
\caption{Feynman diagrams contributing to $B^- \to K^- \rho^0$}
\label{fig:direct_cp_diagrams}
\end{figure}
%

Direct CP violation could be established by measuring a non--zero rate
asymmetry in $B^\pm$ decays.
One example is the decay $B^\pm \to K^\pm \rho^0$ which proceeds via
a tree and a penguin diagram
the weak couplings of which are given
by $\bV^{\phantom{*}}_{\!\! ub} \bV_{\!\! us}^*\approx A\lambda^4(\rho-i\eta)$
and $\bV^{\phantom{*}}_{\!\! tb} \bV_{\!\! ts}^*\approx -A\lambda^2$,
respectively\footnote{
Since $m_u, m_c << M_W$, we can neglect the small quark--mass corrections
in the up and charm penguin contributions. These two diagrams
then differ in their CKM factors only, and their sum is regulated by the
same CKM factor than the top--quark loop, due to the unitarity
of $\protect\bV$.}.
Although the penguin contribution is of higher--order in the strong
coupling, and suppressed by the loop factor $1/(16\pi^2)$,
one could expect both amplitudes to be of comparable size, owing
to the additional $\lambda^2$ suppression factor of the tree diagram.
The needed strong--phase difference can be generated through the
absorptive part of the penguin diagram, corresponding to
on--shell intermediate particle rescattering \cite{BSS:79}.
Therefore, one could expect a sizeable asymmetry, provided the
strong--phase difference is not too small.
However, a very large number of $B^{\pm}$ is required, because
the branching ratio is quite suppressed ($\sim 10^{-5}$).
Other decay modes such as \cite{GH:91} $B^\pm\to K^\pm K_S,K^\pm K^{*0}$ involve
the interference between penguin diagrams only and might show
sizeable CP--violating asymmetries as well, but the corresponding
branching fractions are expected to be even smaller.

The two interfering amplitudes can also be generated through
other mechanisms. For instance, one can have an interplay between two
different cascade processes \cite{CS:81,BS:81} like 
$B^-\to D^0 X^-\to K_S Y^0 X^-$
and  $B^-\to\bar D^0 X^-\to K_S Y^0 X^-$.
Another possibility would be an interference between
two  tree diagrams corresponding to two different decay mechanisms like
direct decay (spectator) and weak annihilation \cite{BJ:81}.
Direct CP violation could also be studied in decays of bottom
baryons \cite{Baryon}, 
where it could show up as a rate asymmetry and in various decay
parameters.

Note that, for all these flavour--specific decays,
the necessary presence of strong phases makes very difficult to extract
useful information on the CKM  factors from their measured
CP asymmetries.
Nevertheless, the experimental observation of a non--zero  CP--violating
asymmetry in any of these decay modes would be a major milestone in our
understanding of CP--violation phenomena, as it would clearly establish
the existence of direct CP violation in the decay amplitudes.


The large $B^0$--$\bar B^0$ mixing provides a different way to generate the
required CP--violating interference.
There are quite a few non--leptonic final states which are reachable
both from a $B^0$ and a $\bar B^0$. For these flavour non--specific decays
the $B^0$ (or $\bar B^0$) can decay directly to the given final state $f$,
or do it after the meson has been changed to its antiparticle via the
mixing process; i.e., there are two different amplitudes,
$A(B^0\to f)$ and $A(B^0\to\bar B^0\to f)$, corresponding to two possible
decay paths. CP--violating effects can then result from the interference
of these two contributions.

  The time--dependent decay probabilities for the decay of a neutral
$B$ meson created at the time $t_0=0$ as a pure $B^0$
($\bar B^0$) into the final state $f$ ($\bar f\equiv CP\, f$) are
(we neglect the tiny $\Delta\Gamma_{B^0}$ corrections):
\beqn
\label{eq:decay_b}
\Gamma[B^0(t)\to f] &\!\!\propto &\!\! 
{1\over 2}\, e^{-\Gamma_{B^0} t}\, |A_f|^2\,
        \Biggl\{
[1 + |\bar\rho_f|^2] +  [1 - |\bar\rho_f|^2] \cos{(\Delta M_{B^0} t)}
\qquad\no\\  & & \qquad\qquad\qquad
  - 2 \,\mbox{\rm Im}\left( {q\over p} \bar\rho_f\right) 
\sin{(\Delta M_{B^0} t)}
    \Biggr\} ,
\\
\label{eq:decay_bbar}
\Gamma[\bar B^0(t)\to \bar f] &\!\!\propto &\!\!
{1\over 2}\, e^{-\Gamma_{B^0} t}\,
|\bar A_{\bar f}|^2 \,       \Biggl\{
[1 + |\rho_{\bar f}|^2] +  [1 - |\rho_{\bar f}|^2] \cos{(\Delta M_{B^0} t)}
 \qquad\no\\ && \qquad\qquad\qquad
  - 2 \,\mbox{\rm Im}\left( {p\over q} \rho_{\bar f}\right)
\sin{(\Delta M_{B^0} t)}
        \Biggr\} ,
 \eeqn
where we have introduced the notation
\be
\begin{array}{lll}
A_f \equiv A[B^0\to f] , \qquad & \bar A_f \equiv -A[\bar B^0\to f] , \qquad &
\bar\rho_f\equiv \bar A_f / A_f , \quad
\\
A_{\bar f} \equiv A[B^0\to \bar f], \qquad &
\bar A_{\bar f} \equiv -A[\bar B^0\to \bar f] , \qquad &
\rho_{\bar f}\equiv A_{\bar f} / \bar A_{\bar f} . \quad
\ea
\ee

CP invariance demands the probabilities of CP conjugate processes to be
identical.
Thus, CP conservation requires
$A_f = \bar A_{\bar f}$, $A_{\bar f} = \bar A_f$,
$\bar\rho_f = \rho_{\bar f}$ and
$\mbox{\rm Im}({q\over p} \bar\rho_f) = 
\mbox{\rm Im}({p\over q} \rho_{\bar f})$.
Violation of any of the first three equalities would be a signal of
direct CP violation. The fourth equality tests CP violation generated
by the interference of the direct decay $B^0\to f$ and the
mixing--induced decay $B^0\to\bar B^0\to f$.

To observe any CP--violating asymmetry,
one needs to distinguish between $B^0$ and $\bar B^0$ decays.
However, a final state $f$ that is common to both $B^0$ and $\bar B^0$ decays
cannot reveal by itself whether it came from a $B^0$ or a $\bar B^0$.
Therefore, one needs independent information
on the flavour identity of the decaying neutral $B$ meson.
Since beauty hadrons are always produced in pairs, one can use for
instance the flavour--specific decays of one $B$ to {\it tag} the
flavour of the companion $B$.

An obvious example of final states $f$ which can be reached both from the
$B^0$ and the $\bar B^0$ are CP eigenstates; i.e., states such that
$\bar f = \zeta_f f$  ($\zeta_f = \pm 1$).
The ratios $\bar\rho_f$ and $\rho_{\bar f}$ depend in general on the
underlying strong dynamics.
However,
for CP self--conjugate final states, all dependence on the
strong interaction disappears \cite{CS:81,BS:81}
if only one weak amplitude contributes to
the $B^0\to f$ and $\bar B^0\to f$ transitions.
In this case, we can write the decay amplitude as
$A_f = M e^{i \phi_D} e^{i \delta_s}$, where $M = M^*$, $\phi_D$ is the phase
of the weak decay amplitude and $\delta_s$ is the strong phase associated with
final--state interactions.
It is easy to check that the ratios $\bar\rho_f$ and $\rho_{\bar f}$
are then given  by
($A_{\bar f} = M \zeta_f e^{i\phi_D} e^{i\delta_s}$,
 $\bar A_{f} = M  \zeta_f e^{-i\phi_D} e^{i\delta_s}$,
 $\bar A_{\bar f} = M  e^{-i\phi_D} e^{i\delta_s}$)
\be
\rho_{\bar f} = \bar\rho_f^* = \zeta_f e^{2i\phi_D} .
\ee
The unwanted effect of final--state interactions cancels out completely
from these two ratios.
Moreover, $\rho_{\bar f}$ and $\bar\rho_f$ simplify in this case to
a single weak phase, associated with the underlying weak quark transition.

Since $|\rho_{\bar f}| = |\bar\rho_f| = 1$,
the time-dependent decay probabilities
become much simpler. In particular, there is no
longer any dependence on $\cos{(\Delta M_{B^0} t)}$.
Moreover, for $B$ mesons
$|\Gamma_{12}/M_{12}|<<1$, implying
\be
{q\over p} \,\approx\, \sqrt{{M_{12}^*\over M_{12}}} \,\approx\,
{\bV_{\!\! tb}^* \bV_{\!\! tq}^{\phantom{*}} 
\over \bV_{\!\! tb}^{\phantom{*}} \bV_{\!\! tq}^*}
\,\equiv\, e^{-2 i \phi_M} .
\ee
Here $q \equiv d, s$ stands for $B^0_d$, $B^0_s$.
Therefore, the mixing ratio $q/p$ is also given by a known weak phase,
and the coefficients of the sinusoidal terms in the time--dependent decay
amplitudes are then fully known in terms of CKM mixing angles only:
\be
\label{eq:im_coeff}
\mbox{\rm Im}\left( {p\over q} \rho_{\bar f}\right) \,\approx\,
-\mbox{\rm Im}\left( {q\over p} \bar\rho_f\right) \,\approx\,
\zeta_f\sin{[2(\phi_M + \phi_D)]}
\,\equiv\, \zeta_f\sin{(2\Phi)}.
\ee

The time--dependent decay rates are finally given by
\beqn
\label{eq:rate_b}
\Gamma[B^0(t)\to f] & = & \Gamma[B^0\to f] \, e^{-\Gamma_{B^0} t} \,
       \{ 1 + \zeta_f\sin{(2 \Phi)} \sin{(\Delta M_{B^0} t)} \} , \\
\label{eq:rate_bbar}
\Gamma[\bar B^0(t)\to \bar f] & = & \Gamma[\bar B^0\to \bar f] 
   \, e^{-\Gamma_{B^0} t} \,
       \{ 1 - \zeta_f\sin{(2 \Phi)} \sin{(\Delta M_{B^0} t)} \} .
\eeqn

\noindent In this ideal case,
the time-dependent CP--violating decay asymmetry
\be
{\Gamma[B^0(t)\to f] - \Gamma[\bar B^0(t)\to\bar f] \over
 \Gamma[B^0(t)\to f] + \Gamma[\bar B^0(t)\to\bar f]} \, = \,
 \zeta_f\sin{(2 \Phi)} \, \sin{(\Delta M_{B^0} t)}
\ee
provides a direct and clean measurement of  the CKM parameters
\cite{KLPS:88}.
Integrating over all decay times yields
\be
\label{eq:integ_rate}
\int_0^\infty dt \,\Gamma[\BBB(t)\to\ffb] \,\propto\,
1 \mp\zeta_f\,\sin{(2 \Phi)}
  \, {x \over 1 + x^2} .
\ee
For $B^0_d$ mesons the mixing
term $x_{B^0_d} /(1+x_{B^0_d}^2)$ only suppresses the observable asymmetry 
by a factor of about two. 
For $B^0_s$ mesons, however,
the large $B^0_s$--$\bar B^0_s$ mixing would
lead to a huge dilution of the CP asymmetry.
The measurement of the time--dependence is then a crucial requirement for
observing CP--violating asymmetries with $B^0_s$ mesons.

In $e^+e^-$ machines, running near the $B^0\bar B^0$ production threshold,
one needs to take also into account the oscillation of the
{\it tagging} meson.
The observable time--dependent
asymmetry takes then the form
\beqn
\lefteqn{{\Gamma[(B^0\bar B^0)_{C=\mp}\to f + (l^-\bar\nu_l X^+)]
 - \Gamma[(B^0\bar B^0)_{C=\mp}\to f + (l^+\nu_l X^-)] \over
\Gamma[(B^0\bar B^0)_{C=\mp}\to f + (l^-\bar\nu_l X^+)]
+\Gamma[(B^0\bar B^0)_{C=\mp}\to f + (l^+\nu_l X^-)]} \, = }
 && \qquad\qquad\qquad\qquad\qquad\qquad\qquad\qquad\qquad
 \qquad\qquad\qquad\qquad\qquad\qquad\qquad\qquad\mbox{}
\no\\ && \qquad\qquad\qquad\qquad\qquad\qquad\qquad\qquad\qquad
 \zeta_f \sin{(2\Phi)} \sin{[\Delta M_{B^0} (t\mp \bar t\, )]}\, ,
 \qquad
\eeqn
where the $B$ flavour has been assumed to be {\it tagged} through
the semileptonic decay, and
$t$ ($\bar t\, $) denotes the time of decay into $f$ ($l^\pm$).
Note that for $C=-1$ the asymmetry vanishes if $t$ and $\bar t$ are
treated symmetrically.
A measurement of at least the sign of $\Delta t \equiv t - \bar t$
is necessary to detect CP violation in this case.
This is the main reason for building asymmetric $B$ factories.

\begin{table}[tbh]
\centering
\hphantom{}
\begin{tabular}{|c|c|c|l|c|}
\hline
Decay & CKM factor & CKM factor & \,\,\,\, Exclusive channels &
 $\quad \Phi \quad$ \\
           & (Direct) & (Penguin) & & \\ \hline
$\bar b \to \bar c  c \bar s$ & $A \lambda^2$ & $-A \lambda^2$ & 
$ B^0_d\to J/\psi K_S ,
J/\psi K_L$ & $\beta$ \\
&&& $ B^0_s\to D_s^+ D_s^-, J/\psi\eta$ & 0 \\ \hline
$\bar b\to\bar s s \bar s$ & -- & $-A \lambda^2$ & 
$ B^0_d\to K_S\phi, K_L\phi$ &
$\beta$ \\ \hline
$\bar b\to\bar d d\bar s$ & -- & $-A \lambda^2$ & 
$ B^0_s\to K_S K_S, K_L K_L$ &
0 \\ \hline
$\bar b\to\bar c c\bar d$ & $-A\lambda^3$ & $A\lambda^3 (1-\rho - i \eta)$ &
$ B^0_d\to D^+ D^- , J/\psi\pi^0$ & $\approx \beta$
\\ &&& $ B^0_s\to J/\psi K_S, J/\psi K_L$ & 0 \\ \hline
$\bar b\to\bar u u\bar d$ & $A \lambda^3 (\rho + i \eta)$ & 
$A\lambda^3 (1 - \rho - i \eta)$ & 
$ B^0_d\to\pi^+\pi^- , \rho^0\pi^0 ,\omega\pi^0$
& $\approx \beta+\gamma$ \\
 & & & $ B^0_s\to\rho^0 K_S ,\omega K_S ,\pi^0 K_S$,
  & $\approx \gamma$ \\
 & & & $\phantom{ B^0_s\to} \rho^0 K_L, \omega K_L, \pi^0 K_L$ &
\\ \hline
$\bar b\to\bar s s\bar d$ & -- & $A\lambda^3 (1 - \rho - i \eta)$ &
$ B^0_d\to K_S K_S, K_L K_L$ & 0
\\ &&& $ B^0_s\to K_S\phi, K_L\phi$ & $-\beta$
\\ \hline
\end{tabular}
\caption{CKM factors and relevant angle $\Phi$ for some $B$--decays into
CP--eigenstates.}
\label{tab:decays}
\end{table}
 
We have assumed up to now that there is only one amplitude contributing
to the given decay process.
Unfortunately, this is usually not the case.
If several decay amplitudes
with different weak and strong phases
contribute, $|\bar{\rho}_f|\not=1$, and the interference term will
depend both on the CKM mixing parameters and on the strong dynamics embodied
in the ratio $\bar{\rho}_f$.
 
The leading contributions to
$\bar b\to\bar q' q'\bar q$ decay amplitudes are either
{\it direct} (Fermi) or  generated by gluon exchange (penguin).
Although of higher order in the strong coupling constant,  penguin
amplitudes are logarithmically enhanced, due to the virtual $W$--loop, and
are  potentially competitive. Table~\ref{tab:decays} contains the CKM
factors associated with the direct and penguin diagrams for
different $B$--decay modes into CP--eigenstates.
Also shown is the relevant angle $\Phi$. 
In terms of CKM elements, the angles $\alpha$, $\beta$
and $\gamma$ are:
\be\label{eq:angles}
\alpha\equiv\arg{\left[
   -{\bV^{\phantom{*}}_{\!\! td}\bV^*_{\!\! tb}\over 
   \bV^{\phantom{*}}_{\!\! ud}\bV^*_{\!\! ub}}
  \right]} , \quad
\beta\equiv\arg{\left[
   -{\bV^{\phantom{*}}_{\!\! cd}\bV^*_{\!\! cb}\over 
   \bV^{\phantom{*}}_{\!\! td}\bV^*_{\!\! tb}}
  \right]} , \quad
\gamma\equiv\arg{\left[
   -{\bV^{\phantom{*}}_{\!\! ud}\bV^*_{\!\! ub}\over 
   \bV^{\phantom{*}}_{\!\! cd}\bV^*_{\!\! cb}}
  \right]} , \quad
\ee
which correspond to the angles of the unitarity triangle
in Fig.~\ref{fig:utriangle}
($\alpha + \beta + \gamma = \pi$). 

The $\bar b\to\bar c c\bar s$ quark decays are theoretically unambiguous
\cite{LP:89}:
the direct and penguin amplitudes have the same
weak phase $\Phi = \beta $ ($0$), for $ B^0_d $ ($ B^0_s$). Ditto for
$\bar b\to\bar s s\bar s$ and $\bar b\to\bar d d\bar s$,
where only the penguin mechanism is possible.
The same is true for the Cabibbo--suppressed $\bar b\to\bar s s \bar d$ mode,
which only gets contribution from the penguin diagram;
the $ B^0_d$ ($ B^0_s$) phases are 0 ($-\beta$) in this case.
The $\bar b\to\bar c c\bar d$ and $\bar b\to\bar u u\bar d$ decay modes 
are not so
simple; the two decay mechanisms
have the same Cabibbo suppression ($\lambda^3$) and
different weak phases,
but the
penguin amplitudes are  down by
${(\alpha_s / 6 \pi}) \ln(m_W / m_b) \approx 3 \% $:
these decay modes can be used
as approximate measurements of the CKM factors.
We have  not considered doubly Cabibbo--suppressed
decay amplitudes, such as
$\bar b\to\bar u u\bar s$, for which
penguin effects can be important and
spoil the simple estimates based on the direct decay mechanism.
 
Presumably  the most realistic channels for the measurement of the angles
  $\Phi=(\beta ,\,\alpha ,\,\gamma)$ are
   $ B^0_d\to J/\psi K_S$, $ B^0_d\to\pi^+\pi^-$ 
($\beta + \gamma = \pi - \alpha$)
and $ B^0_s\to\rho^0 K_S$,
respectively. The first of these processes is no doubt
the one with the cleanest signature and the most tractable background
\cite{nakada}.
The last process has the disadvantage of requiring a $B^0_s$ meson and,
moreover,
its branching ratio is expected to be very small because the
{\it direct} decay amplitude is colour suppressed, leading presumably to
a much larger penguin contamination;
thus, the determination
of $\gamma$, through this decay mode looks a quite formidable task.
 
 The decay modes where $\Phi = 0$ are useless for making
a determination of the CKM factors.
However, some of them provide a very interesting test
of the SM, because the prediction
that no CP--asymmetry should be seen is very clean.
Any detected CP--violating signal would be a clear indication of new physics.

Many other decay modes of $B$ mesons can be used to get information on
the CKM factors responsible for CP violation phenomena. A
summary, including alternative ways of measuring $\gamma$, can be found
in Ref.~\citenum{ecfa:93}. 

\section{Rare Decays}

Rare decays of $K$ and $B$ mesons are a useful tool to improve our
understanding of the interplay among electromagnetic, weak and strong
interactions. 
Decays such as $K\to\pi\nu\bar\nu$ or $B\to X_s\nu\bar\nu$, where
QCD corrections can be easily estimated,
could provide clean measurements of the relevant CKM factors.
CP-violating signals can be looked for in the decays
$K_L\to\pi^0\nu\bar\nu$ and $K_L\to\pi^0 l^+l^-$.
Other higher--order weak decays like $K_L\to\mu^+\mu^-$, 
$K_L\to\pi^0\gamma\gamma$,
$B\to X_s\gamma$, $B\to X_s l^+ l^-$ or $B\to l^+l^-$
can be used to make interesting tests of the SM.
A detailed discussion of rare decays can be found in Refs.
\citenum{chpt:95}, \citenum{EC:95}, \citenum{DEIN:95}, \citenum{DR:95}
and \citenum{BBL:95}.

\section{Summary}

The flavour structure of the SM is one of the main pending questions
in our undertanding of weak interactions.
Although we do not know the reason of the observed family replication,
we have learn experimentally that the number of SM generations is
just three (and no more). Therefore, we must study as precisely
as possible the few existing flavours, to get some hints on the
dynamics responsible for their observed structure.

The SM imposes two basic constraints on flavour--changing
transitions: the universality of the charged--current interactions
(the same gauge coupling $g$ for all fermions) and the unitarity of
the quark--mixing matrix $\bV$.
The empirical verification of these two properties is
one of the main motivations to perform a precise
experimental investigation of flavour--changing processes.

Since quarks are confined within hadrons, the theoretical analysis 
of hadronic weak decays requires a good understanding of 
strong interaction effects.
In these lectures, we have discussed a few selected processes
where our control on the QCD interplay is good enough to allow
a meaningful determination of CKM parameters.
Many more weak decays are available for a comprehensive
phenomenological study, which could bring precious additional
information on the underlying quark couplings, provided
our present understanding of strong interactions is improved in 
a significative way.
Obviously, a good sample of measured decays would help to
discriminate among different theoretical models and obtain
more reliable predictions.
Thus, accurate experimental analyses of weak transitions offer
the possibility to test both the electroweak and strong interactions.

The SM incorporates a mechanism to generate CP violation, through the
single phase naturally occurring in the CKM matrix.
This mechanism, deeply rooted into the unitarity structure of $\bV$,
implies very specific requirements for CP violation
to show up, which should be tested in appropriate
experiments.
The tiny violation of the CP symmetry observed in the kaon system,
can be parametrized through the CKM phase; however, we do not have yet
an experimental verification of the CKM mechanism.
Moreover, a fundamental explanation of the origin of this phenomena is
still lacking.

In the SM, CP violation is associated with a charged--current interaction
with changes the quark flavour in a very definite way:
$u_i\to d_j W^+$, $d_j\to u_i W^-$. Therefore, CP should be directly
violated in many ($\Delta S=1$, $\Delta D=1$, $\Delta B=1$) decay
processes without any relation with meson--antimeson mixing.
Although the quantitative predictions are often uncertain, owing to
the not so--well understood long--distance strong--interaction
dynamics, the experimental observation of a non-zero asymmetry in any
self-tagging decay mode would be a major achivement, as it would
clearly establish the existence of direct CP violation in the decay
amplitudes.

The observation of CP--violating asymmetries with neutral $B$ mesons,
would allow to independently measure the angles of the unitarity triangle,
providing an overconstrained determination of the CKM matrix.
If the measured sides and angles turn out to be consistent with a
geometrical triangle, we would have a beautiful test of the CKM unitarity,
providing strong support to the SM mechanism of CP violation.
On the contrary, any deviation from a triangular shape would be a clear
proof that new physics is needed to understand CP--violating phenomena.

The dynamics of flavour is a broad and fascinating subject, which is
closely related to the so far untested scalar sector of the SM.
The experimental verification of the SM predictions is a very
important challenge for future experiments.
Large surprises may well be discovered, probably giving the first
hints of new physics and offering clues to the problems
of fermion--mass generation, quark mixing and family replication.

\vspace*{0.6cm}\noindent {\normalsize\bf Acknowledgements}
\par\vspace*{0.4cm}
I would like to thank the organizers for the charming atmosphere of this 
meeting. I am also grateful to
V. Gim\'enez, J. Prades and E. de Rafael
for reading the manuscript.
This work has been supported by CICYT (Spain) under grant
No. AEN-93-0234.

\vspace*{0.6cm}\noindent {\normalsize\bf References}
\vspace*{0.4cm}

\end{document}